\documentclass[a4paper,11pt]{article}
\usepackage{jcappub} 
\usepackage{amsmath}

\arxivnumber{2502.02636} 

\title{Bounds on Velocity-Dependent Dark Matter-Baryon Scattering from Large-Scale Structure}

\author[a]{Adam He,}
\author[b,c]{Mikhail M. Ivanov,}
\author[a]{Rui An,}
\author[a]{Trey Driskell,}
\author[a]{Vera Gluscevic}
\affiliation[a]{Department of Physics and Astronomy, University of Southern California,\\Los Angeles, CA 90089, USA}
\affiliation[b]{Center for Theoretical Physics, Massachusetts Institute of Technology,\\Cambridge, MA 02139, USA}
\affiliation[c]{The NSF AI Institute for Artificial Intelligence and Fundamental Interactions,\\Cambridge, MA 02139, USA}

\emailAdd{adamhe@usc.edu}

\abstract{We explore interacting dark matter (DM) models that allow DM and baryons to scatter off of each other with a cross section that scales with relative particle velocity. 
Using the effective field theory of large-scale structure, we perform the first analysis of BOSS full-shape galaxy clustering data for velocity-dependent DM-baryon interactions. 
We determine that while the addition of BOSS full-shape data visibly modifies the shape of the posterior 
distribution, it 
does not significantly alter the 95\% confidence level intervals for the interaction cross section obtained from an analysis of the cosmic microwave (CMB) anisotropy from \textit{Planck} measurements alone. 
Moreover, in agreement with previous findings, we note that the DM-baryon interacting model presents a good fit to both large-scale structure (LSS) data and CMB data and alleviates the $S_8$ tension between the two data sets. After combining LSS and CMB data with weak lensing data from the Dark Energy Survey, we find a $\gtrsim2\sigma$ preference for non-zero interactions between DM and baryons in a velocity-independent model. We also explore a scenario where only a fraction of DM undergoes scattering with baryons; we find a similar $\gtrsim2\sigma$ preference for the presence of interactions. 
Our results suggest that a suppression of the linear matter power spectrum at small scales may be needed to resolve certain discrepancies between LSS and CMB data that are found in the cold DM (CDM) scenario.}

\begin{document}
\maketitle
\flushbottom



\section{\label{sec:level1}Introduction\protect}

Late-time and early-time cosmological probes disagree considerably on the current expansion rate of the universe $H_0$ and the amplitude of density fluctuations at late times, quantified by the $S_8$ parameter, when assuming the $\Lambda$CDM model \cite{Abdalla_2022}. This could be due to unknown systematic
errors \cite{Bernal_2016, Di_Valentino_2021_S}, or it may indicate that new physics beyond $\Lambda$CDM is needed to correctly model the early and late universe simultaneously \cite{Di_Valentino_2021_H, Abdalla_2022}.

A myriad of compelling dark matter (DM) models beyond the cold colisionless DM (CDM) have been proposed, including the weakly-interacting massive particles (WIMPs) and other interacting DM (IDM) scenarios that feature DM-baryon scattering \cite{Snowmass_2013, battaglieri2017cosmic, akerib2022snowmass2021,Sigurdson2004,Dvorkin2014,Gluscevic_2018, Boddy_2018, Boddy2018, Boehm_2005, Xu_2021, Nguyen2021, Maamari_2021, Rogers_2022, Becker_2021, Nadler_2019, Nadler_2021, Li_2022, Gluscevic_2019, Slatyer_2018, Buen_Abad_2022, Hooper_2022}. In this study, we explore one such scenario, in which DM features elastic scattering with baryons \cite{Boddy2018}, exchanging heat and momentum between the two cosmological fluids, and resulting in a collisional damping of matter perturbations and a scale-dependent suppression of structure \cite{B_hm_2001, Boehm_2005}; see Fig.~\ref{fig:Pk for all models}. Recent studies have shown that a scale-dependent suppression of the linear matter power spectrum $P(k)$ might be able to resolve the $S_8$ tension~\cite{Amon_2022, Poulin_2022,Rogers:2023ezo,he2023s8}. In CDM cosmology, this $S_8$ tension between large-scale structure (LSS) data and the cosmic microwave background (CMB) anisotropy measurements from \textit{Planck} is nearing 3$\sigma$ \cite{Di_Valentino_2021_S,Ghirardini_2024,DESKiDS2023}. With the power suppression that is characteristic to DM-baryon scattering scenarios, this tension may be alleviated \cite{he2023s8}.

Observational data that have been used to constrain DM-baryon interacting models include \textit{Planck} CMB measurements \cite{Boddy2018, Nguyen2021}, Lyman-$\alpha$ forest measurements \cite{Becker_2021, Hooper_2022, Rogers_2022}, and Milky Way satellite measurements \cite{Nadler_2019}; only recently has this model been studied in the context of galaxy clustering and lensing \cite{he2023s8}. This is because modeling accuracy up to $k \sim 0.2 \ h$/Mpc is needed to analyze LSS data, including the Baryon Oscillation Spectroscopic Survey (BOSS) \cite{Alam_2017}. The HALOFIT suite is one way of modeling CDM-like cosmologies up to these mildly-nonlinear scales; however, HALOFIT is calibrated against N-body simulations that assume a CDM universe, and therefore breaks down in an IDM context \cite{Smith_2003, Takahashi_2012}. Alternatively, one may use the effective field theory of large-scale structure (EFT)~\cite{Baumann:2010tm,Carrasco_2012,Ivanov:2022mrd} to model IDM up to these mildly non-linear scales \cite{he2023s8}. EFT uses perturbation theory to evolve modes at the mildly-nonlinear scale and has been used extensively to test CDM-like as well as non-CDM cosmologies~\cite{Ivanov_2020,DAmico:2019fhj,Chen:2021wdi,Lagu__2022, Xu_2022, Nunes_2022, Rubira_2023, rogers2023ultralight, mosbech2025desiforecastdarkmatterneutrino,he2023s8}.

In this paper, we use EFT to perform a joint analysis of \textit{Planck} CMB data and BOSS full-shape galaxy clustering data and look for evidence of velocity-dependent DM-proton elastic scattering. We first focus on scenarios in which all of the DM experiences interactions with baryons; we find that the CMB+LSS bounds are weaker than the bounds obtained from Lyman-$\alpha$ forest measurements and Milky Way satellite measurements, which restrict the suppression of power from $0.2\lesssim k \lesssim2 \ h/\mathrm{Mpc}$ to be $\lesssim$ 25\% \cite{Chabanier_2019, Nadler_2019, Nadler_2021,Ivanov:2024jtl}. We further also consider unconstrained scenarios where a fraction of DM interacts with baryons, and the impact on $P(k)$ is not sufficiently prominent to alter substructure in the Milky Way; we assume that the rest of DM is collisionless.
The effect of interactions is shown in Fig.~\ref{fig:fractions}, for several different IDM fractions. 

We determine that the addition of BOSS data does not significantly alter the 95\% confidence level intervals for the interaction cross section from \textit{Planck} data alone. However, we find that the velocity-independent scattering fits both BOSS data and \textit{Planck} data, alleviating the $S_8$ tension. After combining BOSS and \textit{Planck} with a $S_8$ prior from the weak lensing data from the Dark Energy Survey (DES) \cite{Abbott_2022}, we find a $\gtrsim2\sigma$ preference for non-vanishing interaction cross section in a velocity-independent case, consistent with previous analyses \cite{he2023s8}. We further find a $\gtrsim2\sigma$ preference for scattering in scenarios where only a fraction of the DM exchanges heat and momentum with baryons. Our results, in line with other proposed solutions to the $S_8$ tension, imply a suppression of the linear matter power spectrum at small scales, which can resolve the mild discrepancy between LSS and CMB data found in CDM \cite{Amon_2022, Ye_2021, Poulin_2022, he2023s8}.

This paper is organized as follows. Sec.~\ref{sec:formalism} describes IDM cosmology and the EFT of LSS, in context of IDM. 
We describe our analysis method in Sec.~\ref{sec:methodology}.
Sec.~\ref{sec:results} presents our results. We discuss and conclude in Sec.~\ref{sec:discussion}.

\section{\label{sec:formalism}Evolution of structure and IDM\protect}

\subsection{\label{sec:linear}Linear evolution}

In the presence of DM-proton scattering, the standard Boltzmann equations contain additional interaction terms that encapsulate the momentum transfer occurring  between DM and baryons \cite{Gluscevic_2018, Boddy2018},
\begin{equation}\label{boltzmann}
     \begin{split}
        \dot{\delta}_{\chi} = &-\theta_{\chi} - \frac{\dot{h}}{2}, \qquad \dot{\delta}_{\mathrm{b}} = -\theta_{\mathrm{b}} - \frac{\dot{h}}{2}, \\ \dot{\theta}_{\chi} = &-\frac{\dot{a}}{a}\theta_{\chi}+c^{2}_{\chi}k^{2}\delta_{\chi} + R_{\chi}\left(\theta_{\mathrm{b}}-\theta_{\chi}\right), \\
        \dot{\theta}_{\mathrm{b}} = &-\frac{\dot{a}}{a}\theta_{\mathrm{b}}+c^{2}_{\mathrm{b}}k^{2}\delta_{\mathrm{b}} + \frac{\rho_{\chi}}{\rho_{\mathrm{b}}}R_{\chi}\left(\theta_{\chi}-\theta_{\mathrm{b}}\right) \\ 
        &+ R_{\gamma}\left(\theta_{\gamma}-\theta_{\mathrm{b}}\right),
    \end{split}
\end{equation}
where subscripts ${\chi}$ and ${\mathrm{b}}$ denote DM and baryons, respectively\footnote{We ignore helium fraction in our analysis, which is shown to have a minimal impact on the relevant observables for DM-baryon scattering models \cite{Boddy2018}.}; $\delta$ denotes density perturbations and $\theta$ represents velocity divergence; $h$ is the trace of the scalar metric perturbation; $c$ represents the sound speeds in respective fluids; $R_{\gamma}$ is the momentum transfer rate between baryons and photons from Compton scattering; and $R_{\chi}$ is the momentum transfer rate between DM and baryons from their non-gravitational interaction, 
\begin{equation}\label{Rchi}
    R_{\chi} = \frac{ac_n\rho_{\mathrm{b}}\sigma_{0}}{m_{\chi}+m_\mathrm{b}}\left(\frac{T_{\chi}}{m_{\chi}}+\frac{T_{\mathrm{b}}}{m_{\mathrm{b}}}+\frac{V^{2}_{\mathrm{RMS}}}{3}\right)^{-\frac{n+1}{2}},
\end{equation}

\noindent where $c_n = \frac{2^{\frac{n+5}{2}}\Gamma(3+\frac{n}{2})}{3\sqrt{\pi}}$, $m_\chi$ is the DM particle mass, $m_\mathrm{b}$ is the mean baryon mass, $T$ denotes fluid temperatures, and $n$ is the power index that dictates the velocity dependence in the interaction cross section:
\begin{equation}\label{eq:sigma}
    \sigma=\sigma_0 v^n,
\end{equation}

\begin{figure}[!t]
\centering
\includegraphics[scale=0.95]{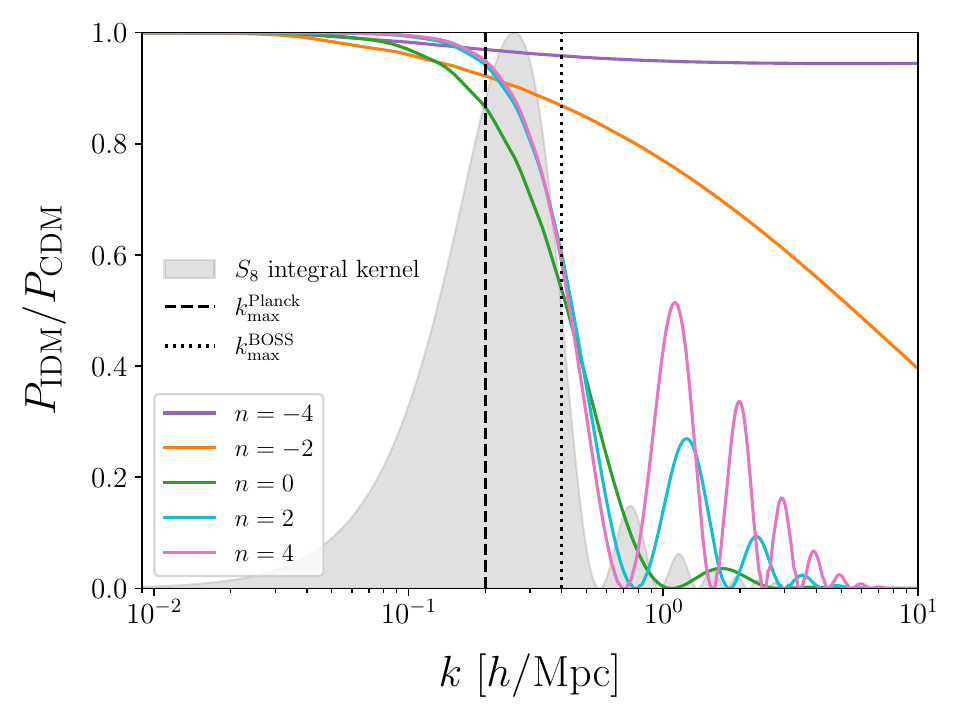}
\caption{\label{fig:Pk for all models} Fractional difference between the linear matter power spectrum for IDM versus that of CDM cosmology; note that different lines correspond to different powers of velocity dependence in the interaction cross section $n$. In the presence of DM-baryon interactions, there is a scale-dependent suppression at small scales. The illustrated models correspond to the cross section at the 95\% C.L. upper limit derived from \textit{Planck} for each $n$ \cite{Nguyen2021}. The shaded region displays the Fourier-space filter from the integral calculation of the matter clustering amplitude $S_8$, indicating the level at which different wavenumbers contribute to $S_8$. The dashed line indicates the maximum wavenumber that the \textit{Planck} likelihood probes, and the dotted line indicates the maximum wavenumber that the BOSS likelihood probes.
}
\end{figure}

\noindent where $\sigma$ is the interaction cross section, $\sigma_0$ is the coefficient of the momentum transfer cross section, and $v$ is the bulk relative velocity between the DM and baryons. The root-mean-square bulk relative velocity between DM and baryons is approximated by \cite{Dvorkin2014},
\begin{equation}\label{RMS}
    V_{\mathrm{RMS}}^{2} = \left< \vec{V}_{\chi}^{2}\right>_{\xi} = \int \frac{dk}{k}\Delta_{\xi} \left(\frac{\theta_{\mathrm{b}}-\theta_{\chi}}{k^{2}}\right)^{2},
\end{equation}
\noindent where $\Delta_{\xi}$ is the primordial curvature variance per log wavenumber $k$. Integrating over $k$ in $V_{\mathrm{RMS}}^{2}$ disrupts the linearity of the Boltzmann equations and mixes modes; however, the bulk relative velocity may be approximated analytically with a function that remains constant for $z > 10^{3}$ and scales linearly with $z$ for $z\lesssim 10^{3}$. Following previous literature, we use this approach to capture the effect of $V_{\mathrm{RMS}}^{2}$ on $R_{\chi}$ \cite{Tseliakhovich2010, Dvorkin2014}.
To solve the modified Boltzmann equations, we use an altered version of the Boltzmann solver \texttt{CLASS} that allows for DM-baryon interactions parameterized by a momentum transfer cross section $\sigma=\sigma_0 v^n$, presented in Refs.~\cite{Gluscevic_2018, Boddy2018} \footnote{\url{https://github.com/kboddy/class_public/tree/dmeff}}\textsuperscript{,}\footnote{We also implement a tight-coupling scheme between baryons and DM for positive values of $n$, shown in Appendix~\ref{Appendix:tight-coupling}.}. 

\subsection{\label{sec:nonlinear}Non-linear evolution}

To model the late-time evolution of the matter power spectrum on scales associated with galaxy clustering, weak lensing, and other LSS observables, we merge the modified IDM \texttt{CLASS} code with \texttt{CLASS-PT}~\cite{Chudaykin2020}\footnote{\url{https://github.com/Michalychforever/CLASS-PT}}. \texttt{CLASS-PT} is a perturbation theory extension of \texttt{CLASS} that calculates non-linear 1-loop corrections to the linear matter power spectrum for a given model, as \cite{Baumann:2010tm,Carrasco_2012,Chudaykin:2020hbf,Cabass:2022avo,Ivanov:2022mrd}
\begin{equation}\label{total Pk}
    P(z,k) = P_{\mathrm{lin}}(z,k) + P_{\mathrm{1-loop}}(z,k) + P_{\mathrm{ctr}}(z,k)
\end{equation}

\noindent where $P$ is the total matter power spectrum, $P_{\mathrm{lin}}$ is the linear matter power spectrum, $P_{\mathrm{1-loop}}$ is the first-order correction to the matter power spectrum from standard perturbation theory (SPT), and $P_{\mathrm{ctr}}$ is the counterterm that compensates for the overshooting of SPT \cite{Carrasco_2012, Chudaykin2020}. Each term in the expression above is a function of $P_{\mathrm{lin}}$; $P_{\mathrm{1-loop}}$ may be expressed as
\begin{equation}\label{1-loop}
    P_{\mathrm{1-loop}}(z,k) =  D^4(z)(P_{13}(k) + P_{22}(k))
\end{equation}

\noindent where $D(z)$ is the linear growth factor, and $P_{22}(k)$ and $P_{13}(k)$ are given by
\begin{equation}\label{P22}
    P_{22}(k) = 2 \int_{\textbf{q}}F_2^2(\textbf{q}, \textbf{k} - \textbf{q})P_{\mathrm{lin}}(q)P_{\mathrm{lin}}(|\textbf{k}-\textbf{q}|)
\end{equation}

\noindent and
\begin{equation}\label{P13}
    P_{13}(k) = 6P_{\mathrm{lin}}(k)\int_{\textbf{q}}F_3(\textbf{k}, -\textbf{q}, \textbf{q})P_{\mathrm{lin}}(q)
\end{equation}

\noindent where $F_2$ and $F_3$ are the perturbation theory kernels, defined e.g. in Ref.~\cite{Blas_2016}. $P_{\mathrm{ctr}}$ is expressed as
\begin{equation}\label{Pctr}
    P_{\mathrm{ctr}}(z,k) = -2c_s^2(z)k^2P_{\mathrm{lin}}(z,k)
\end{equation}

\noindent where $c_s^2(z)$ is an effective sound speed that is treated as a nuisance parameter in our analyses.

\texttt{CLASS-PT} calculates the redshift-space galaxy power spectrum as~\cite{Chudaykin:2020hbf}
\begin{equation}\label{galaxy galaxy Pk}
    P_{\mathrm{gg}}(z,k) = P_{\mathrm{lin}}(z,k) + P_{\mathrm{1-loop}}^\mathrm{gg}(z,k) + P_{\mathrm{ctr}}^\mathrm{gg}(z,k) + 
    P_{\mathrm{shot}}(z)
\end{equation}

\noindent where $P_{\mathrm{gg}}$ is the redshift-space galaxy power spectrum and $P_{\mathrm{shot}}$ is the scale-independent shot noise contribution to the power spectrum. $P_{\mathrm{1-loop}}^\mathrm{gg}$ and $P_{\mathrm{ctr}}^\mathrm{gg}$ are given in Eqs.~\ref{1-loop} and~\ref{Pctr}, with nuisance parameters added to them to account for galaxy bias, denoted $b_1$ (linear bias), $b_2$ (quadratic bias), and $b_{\mathcal{G}_2}$ (tidal bias). As is standard in the literature, we express $P_{\mathrm{gg}}$ as a multipole expansion using Legendre polynomials and focus on the monopole ($l=0$), quadrupole ($l=2$), and hexadecapole ($l=4$)~\cite{Chudaykin2020}.

\begin{figure}[!tb]
\includegraphics[scale=0.95]{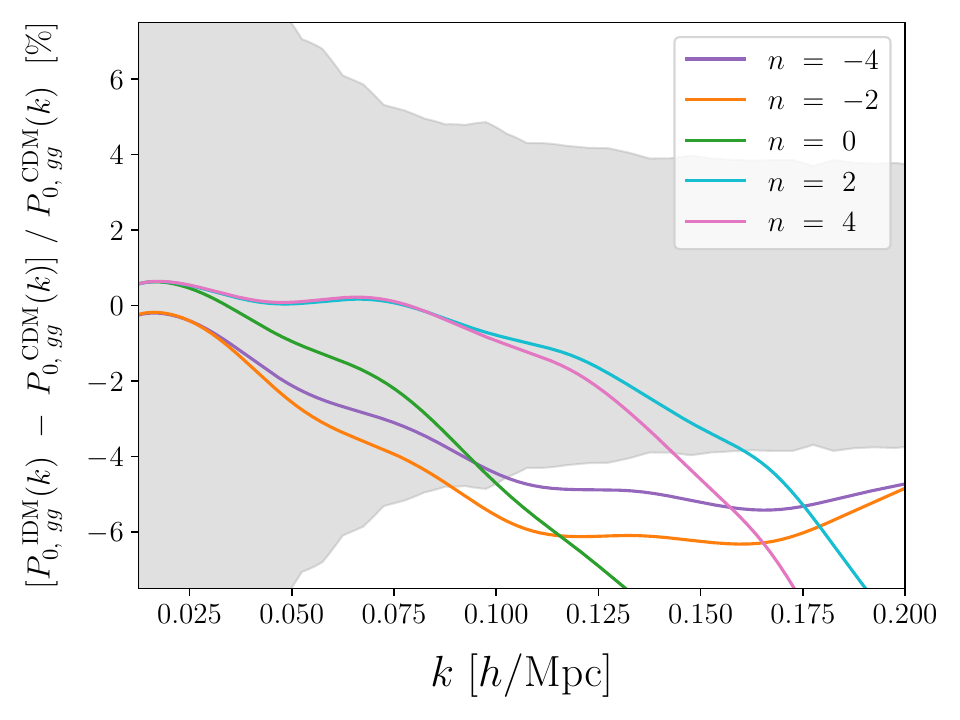}
\caption{\label{fig:galaxy galaxy monopole} Residuals of the redshift space galaxy power spectrum monopole for different DM-baryon interacting models, compared to the $\Lambda$CDM case. The blue shaded region indicates $1\sigma$ uncertainty on the BOSS data, derived from Patchy mock galaxy catalogs~\cite{Ivanov_2020}. 
Each curve is associated with a value of the momentum-transfer cross section that is excluded by the BOSS data, for a given value of $n$.}
\end{figure}

\texttt{CLASS-PT} uses EFT to model the redshift-space galaxy power spectrum in the mildly-nonlinear regime; in the context of non-gravitational interactions between baryons and DM, the EFT should in principle be modified to account for such a scenario.
However, non-linear effects are entirely negligible at the high redshifts where DM-baryon scattering impacts the evolution of matter perturbations. 
Conversely, DM-baryon interactions are negligible at the low redshifts relevant for galaxy surveys, and the matter perturbations evolve as in $\Lambda$CDM but with an altered initial power spectrum, shown in Fig.~\ref{fig:Pk for all models}.\footnote{We show that DM-baryon interactions only impact matter perturbations at redshifts before recombination in Appendix~\ref{Appendix:structure evolution}, for all values of $n$ considered here.}
Thus, the standard version of \texttt{CLASS-PT} is appropriate for predicting late-time LSS observables in the context of IDM\footnote{We make a minor modification to the standard \texttt{CLASS-PT} code to account for IDM scenarios in which the power spectrum vanishes; this modification is explained in Appendix~\ref{Appendix:IR}.}.
Let us finally note that the relative velocity 
between baryons and dark matter source 
additional terms in the galaxy bias~\cite{Tseliakhovich2010,Schmidt:2016coo}.
This effect is suppressed for low redshift galaxies
which we use in our analysis, and therefore we will
ignore it in what follows.

We show the redshift space galaxy power spectrum monopole calculated for different DM-baryon interacting models in Fig.~\ref{fig:galaxy galaxy monopole}. Note that the residuals exceed the measurement uncertainty, indicating that these models are excluded by the data.

\section{\label{sec:methodology}Data and Methodology}

We analyze a combination of CMB and LSS data:

\begin{itemize}
    \item \textbf{\textit{Planck}}: full TT, TE, EE, and lensing power spectra from \textit{Planck} 2018 \cite{Planck2018_V}
    \item \textbf{BOSS}: anisotropic galaxy clustering data from BOSS DR12 at $z=0.38$ and 0.61 \cite{Alam_2017,Ivanov_2020,Ivanov:2019hqk}. As in~\cite{Chudaykin:2020ghx,Philcox:2021kcw}, we perform our analysis up to $k_\mathrm{max} = $ 0.2~$h/\mathrm{Mpc}$ for the galaxy power spectrum multipoles, 
    from $ 0.2<k<  0.4$~$h/\mathrm{Mpc}$ for the real-space
    power spectrum proxy $Q_0$~\citep{Ivanov:2021fbu}, and up to 
    $k_\mathrm{max} = $ 0.08~$h/\mathrm{Mpc}$
    for the bispectrum monopole~\citep{Ivanov:2021kcd,Philcox:2021kcw}.\footnote{The BOSS full-shape likelihood that we use is 
    available at~\mbox{\url{https://github.com/oliverphilcox/full_shape_likelihoods}}. }
    We also add post-reconstructed BOSS DR12 BAO data following~\cite{Philcox:2020vvt}.
    \item \textbf{DES}: weak lensing data from the DES Year 3 data release (DES-Y3), in the form of a prior on $S_8$: 0.776 $\pm$ 0.017 \cite{Abbott_2022}.
    
\end{itemize}

We use the following data set combinations in our analysis: `\textit{Planck}', `BOSS', `\textit{Planck} + BOSS', `\textit{Planck} + BOSS + DES', and `BOSS + DES'.

Imposing a prior on $S_8$ is equivalent to adding the complete DES-Y3 dataset to our analysis, as DES provides a largely model-independent measurement of $S_8$; this argument is substantiated by the observation that the inferred value of $S_8$ is found to be the same under several models, including $\Lambda$CDM, WDM, and $\Lambda$CDM with early dark energy (EDE) \cite{Abbott_2022, DES_extensions_2021, Hill_2020,Ivanov:2020ril}. The value of $S_8$ is robust to the choice of the cosmological model, as long as the late-time growth of structure is unmodified. Moreover, $S_8$ is the primary directly 
observed principle component of the weak lensing data, and, as such, is close to being model independent. We thus safely leave the full calculation of the DES-Y3 likelihood for DM-baryon scattering for future work.

\begin{figure}[!tb]
\includegraphics[scale=0.95]{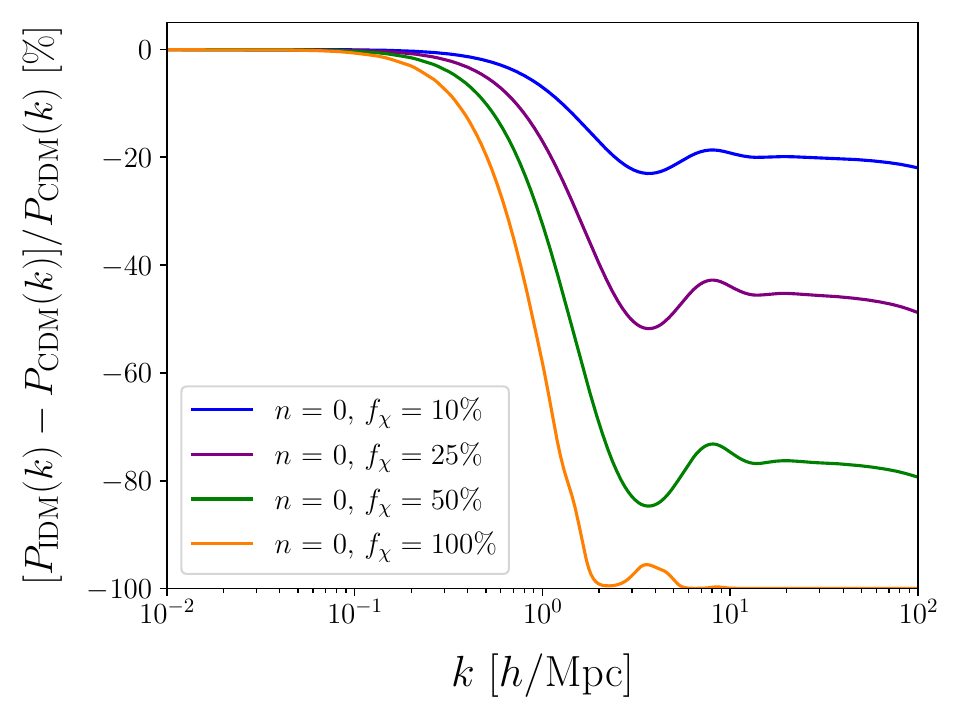}
\caption{Effect on the linear matter power spectrum for different interacting DM fractions $f_\chi$ in a DM-baryon scattering cosmology. The residuals of the linear spectra with respect to CDM are displayed for fractions $f_\chi = 10\%$, 25\%, 50\%, and 100\%. These spectra are generated with the best-fit parameter values from a \textit{Planck} + BOSS + DES analysis of the $n = 0$, $f_\chi=100\%$ model for a DM mass of $m_\chi=1$ MeV.
\label{fig:fractions}}
\end{figure}

For our EFT-based full-shape analysis, we consistently 
marginalize over necessary 
nuisance parameters which capture galaxy bias, baryonic feedback, non-linear redshift space-distortions, etc.~\citep{Philcox:2021kcw}.\footnote{The priors we use using are sufficiently wide 
to account for the 
dependence of EFT parameters
on DM physics~\cite{Ivanov:2024hgq,Ivanov:2024xgb,Akitsu:2024lyt,Ivanov:2024dgv}, which is especially
important for non-minimal DM models.
Our priors are also consistent with 
the physics of BOSS 
red luminous galaxies. 
For the discussion of the role of
the priors in the EFT full-shape 
analysis based on \texttt{CLASS-PT}
see e.g.~the original works~\cite{Ivanov_2020,Chudaykin:2020ghx,Philcox:2021kcw}
and recent detailed analyses of
\cite{Ivanov:2024xgb,Chudaykin:2024wlw}.
}
Our analysis is thus independent of 
the details of galaxy formation.
We also apply our BOSS galaxy 
clustering data to the IDM scenario without any $\Lambda$CDM assumptions; more specifically,
we do not use the compressed BOSS likelihood containing BAO and RSD parameters that are derived with a
fixed \textit{Planck}-like $\Lambda$CDM template~\citep{Ivanov_2020,eBOSS:2020yzd}. Finally, as in \cite{Ivanov_2020}, our EFT-based 
likelihood includes galaxy power 
spectrum 
shape information 
that the standard BOSS
likelihood does not have \cite{Alam_2017}.

We use our modified \texttt{CLASS} code\footnote{\url{https://github.com/ash2223/class-dmeff-EFT}} and the MCMC sampler \texttt{MontePython} to obtain bounds on the IDM models \cite{Brinckmann_2018, Audren_2012}. We assume flat priors on $\{\omega_{\mathrm{b}}$, $\omega_{\mathrm{DM}}$, 100$\theta_\mathrm{s}$, $\tau_\mathrm{reio}$, $\mathrm{ln}(10^{10}A_{\mathrm{s}})$, $n_{\mathrm{s}}$$\} + \sigma_{0}$.
We consider interacting models for which $n=-4$, $-2$, 0, 2, and 4.
The additional free parameter is the IDM particle mass $m_\chi$; following \cite{Gluscevic_2018}, we fix the mass in each MCMC fit to be 1 MeV. We choose this mass because of the strict constraints on IDM from direct detection above 1 GeV, and constraints on $N_\mathrm{eff}$ that rule out masses lower than $\sim1$ MeV \cite{Lewin1995}, \cite{An_2022}.
We set the fraction of DM that interacts with baryons $f_\chi$ to be 100\% for the first part of our analysis. We then also analyse data with a fixed IDM fraction $f_\chi$, setting it to 10\% for models $n=-2$, 0, 2, and 4, and we use the following benchmark DM masses: 1 MeV, 1 GeV, and 10 GeV.\footnote{The $n=-2$, $f_\chi = 10\%$ model impacts perturbations down to redshift $z\sim8$ (see Appendix~\ref{Appendix:structure evolution}). Thus, our results for this particular model are approximate; we hope our findings encourage a more detailed analysis in future work.}\textsuperscript{,}\footnote{We do not explore the $n=-4$, $f_\chi = 10\%$ model as it impacts perturbations down to $z\sim0$, so the EFT cannot be used in this case without modification. Additionally, this model exhibits small levels of suppression in the power spectrum, and changing the interacting fraction would thus have a minimal effect.}
We model free-streaming neutrinos as two massless species and one massive species for which $m_\nu$ = 0.06 eV, in line with the \textit{Planck} convention \cite{Planck2018_VI}. 
A chain is deemed converged if the Gelman-Rubin convergence criterium $|R-1|$ is less than 0.03. 

\section{\label{sec:results}Results\protect}

\subsection{\label{sec:parameter estimation} Parameter estimation}

\begin{figure*}[!tb]
\includegraphics[scale=0.375]{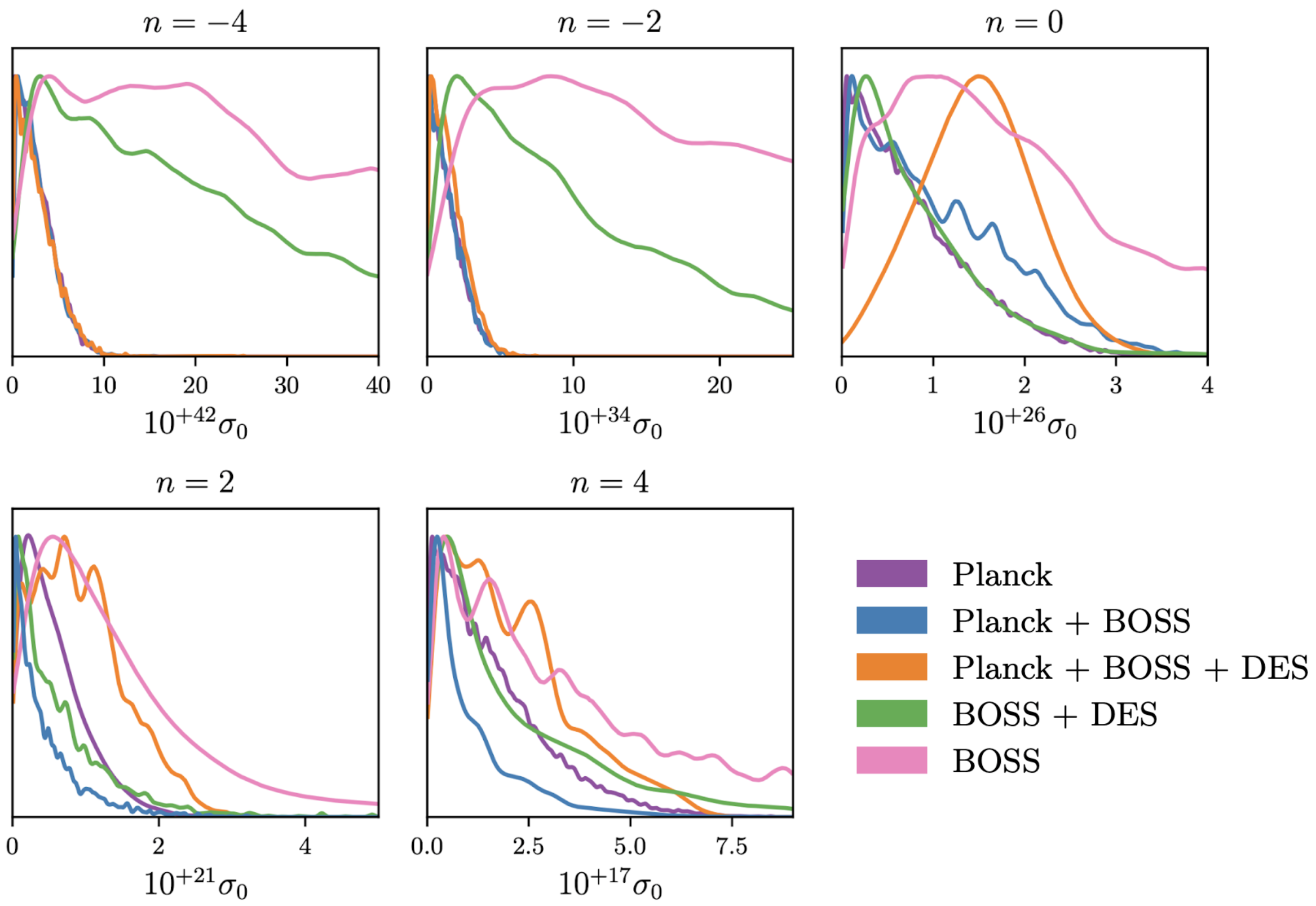}
\caption{\label{fig:sigma0} Marginalized posterior probability distributions for the coefficient of the momentum-transfer cross section for DM-proton elastic scattering, for each interaction model; different interaction models correspond to different power-law indices $n$; see Eq.~\ref{eq:sigma}. 
All cross sections are presented in units of cm$^{2}$.
Different colors correspond to analyses involving different combinations of data. We find that most of the constraining power is in \textit{Planck} CMB anisotropy measurements (purple), while BOSS and DES alone provide a weaker bound, regardless of the model (pink and green). The combination of BOSS with \textit{Planck} data only marginally improves the constraints in the cases with strong relative velocity-dependence in the momentum transfer cross section, for $n=2, 4$ cases (blue). The addition of the DES $S_8$ prior (orange) leads to a mild preference for non-vanishing interaction cross section in the velocity-independent scattering case, consistent with previous analyses \cite{he2023s8}.}
\end{figure*}

If all of DM is assumed to interact with baryons, a joint analysis of \textit{Planck} and BOSS data for cases $n=-4$, $-2$, 0, 2, and 4 shows no evidence of interactions, and all marginalized probability distributions for the momentum-transfer cross section coefficient $\sigma_0$ are consistent with zero. Moreover, the addition of BOSS data does not alter the 95\% C.L. upper limit on $\sigma_0$ as compared to a \textit{Planck}-only analysis; this is the case for all scattering models we considered here. In fact, a \textit{Planck} + BOSS analysis of the $n=0$ model slightly broadens the allowed range of $\sigma_0$ values as compared to a \textit{Planck}--only analysis of the same model; we examine this case in Sec.~\ref{sec:tensions}.

We further find that a joint likelihood analysis of \textit{Planck}, BOSS, and DES data yields a $\gtrsim2\sigma$ preference for a velocity-independent elastic scattering between DM and protons. Fig.~\ref{fig:sigma0} shows 1D posterior probability distributions for each value of $n$, with the top right panel corresponding to $n=0$. The orange curve for the joint analysis of \textit{Planck}, BOSS, and DES data displays a maximum at $\sigma_0 = 1.47\cdot10^{-26}$  $\mathrm{cm}^{2}$. The $\gtrsim2\sigma$ preference for a non-vanishing interaction cross section is consistent with the $\Delta\chi^2_\mathrm{min}=-6.02$  value found for this model; see Table~\ref{table:Chi squared}.

\begin{table*}[!htb]
\centering
\caption{\label{table:Chi squared}$\Delta\chi_{\mathrm{min}}^2$ values for all models and data combinations tested in our analysis for 100\% of DM scattering with protons, for a given interaction model (denoted by the power law index $n$ describing the velocity dependence of the momentum-transfer cross section; see Eq.~\ref{eq:sigma}). DM mass is set to 1 MeV. Each $\Delta\chi_{\mathrm{min}}^2$ is computed as a difference with respect to the CDM case, for a given data set; negative values correspond to the cases where the IDM model leads to improvement in fit.}
\begin{tabular}{|c | c | c | c | c | c|}
\hline
 Model
 &$n=-4$
 &$n=-2$
 &$n=0$
 &$n=2$
 &$n=4$ \\ [0.5ex] 
 \hline\hline
 \textit{Planck}
 &$+0.02$
 &$+1.46$
 &$+1.42$
 &$+0.64$
 &$+0.28$ \\ [0.5ex]

 BOSS
 & +0.92
 & $-0.034$
 & $-0.61$
 & $+0.896$
 & $+1.702$ \\ [0.5ex]

 \textit{Planck} + BOSS
 &$-2.2$
 &$-2.56$
 &$-1.84$
 &$-1.98$
 &$-2.14$ \\ [0.5ex]

 \textit{Planck} + BOSS + DES
 &$+0.58$
 &$-0.18$
 &$-6.02$
 &$-0.02$
 &$+0.42$\\ [0.5ex]

 BOSS + DES
 & $+0.812$
 & $+0.954$
 & $+0.408$
 & $+1.376$
 & $+1.838$ \\ [0.5ex]
 \hline
\end{tabular}
\end{table*}

\begin{table*}[!htb]\footnotesize
\centering
\caption{\label{table:cross section}Constraints on the DM-baryon scattering momentum-transfer cross section $\sigma_0$ from different models and data combinations in our analysis for 100\% of DM interacting with protons. DM mass is set to 1 MeV. ``Best fit'' denotes the maximum of the full posterior, while ``Marginalized max'' denotes the maxima of the marginalized posteriors.}
\begin{tabular}{|c | c | c | c | c | c | c|}
\hline
 Model & Dataset & Best-fit & Marginalized max $\pm \ \sigma$ & 95\% lower & 95\% upper \\ \hline
& \textit{Planck} &$0.1563$ & $2.64^{+0.77}_{-2.6}$ & $>0$ & $6.45$ \\
$n=-4 \ [10^{-42} \ \rm{cm}^2]$& \textit{Planck}+BOSS &$2.767$ & $2.67^{+0.75}_{-2.5}$ & $>0$ & $6.738$ \\
& \textit{Planck}+BOSS+DES &$1.117$ & $2.73^{+0.74}_{-2.7}$ & $>0$ & $6.834$ \\
\hline
& \textit{Planck} &$2.01$ & $1.38^{+0.46}_{-1.4}$ & $>0$ & $3.56$ \\
$n=-2\ [10^{-34} \ \rm{cm}^2]$& \textit{Planck}+BOSS &$0.5662$ & $1.39^{+0.39}_{-1.4}$ & $>0$ & $3.544$ \\
& \textit{Planck}+BOSS+DES &$1.942$ & $1.52^{+0.44}_{-1.5}$ & $>0$ & $3.721$ \\
\hline
& \textit{Planck} &$0.8792$ & $0.80^{+0.20}_{-0.79}$ & $>0$ & $2.07$ \\
$n=0\ [10^{-26} \ \rm{cm}^2]$& \textit{Planck}+BOSS &$0.2136$ & $1.07^{+0.59}_{-1.1}$ & $>0$ & $2.6$ \\
& \textit{Planck}+BOSS+DES &$1.606$ & $1.47\pm 0.63$ & $0.2336$ & $2.664$ \\
\hline
& \textit{Planck} &$0.2277$ & $0.691^{+0.077}_{-0.67}$ & $>0$ & $1.535$ \\
$n=2\ [10^{-21} \ \rm{cm}^2]$& \textit{Planck}+BOSS &$0.1614$ & $0.462^{+0.070}_{-0.47}$ & $>0$ & $1.591$ \\
& \textit{Planck}+BOSS+DES &$1.091$ & $0.93^{+0.33}_{-0.86}$ & $>0$ & $2.108$ \\
\hline
& \textit{Planck} &$0.3951$ & $1.64^{+0.38}_{-1.6}$ & $>0$ & $4.429$ \\
$n=4\ [10^{-17} \ \rm{cm}^2]$& \textit{Planck}+BOSS &$0.02186$ & $1.08^{+0.28}_{-1.1}$ & $>0$ & $3.714$ \\
& \textit{Planck}+BOSS+DES &$1.428$ & $2.08^{+0.8}_{-1.9}$ & $>0$ & $5.247$ \\
 \hline
\end{tabular}
\end{table*}

The full set of marginalized posterior probability distributions from our analyses are shown in Appendix \ref{Appendix:posteriors_100}, and constraints on all relevant cosmological parameters are shown in Appendix \ref{Appendix:constraints}. 
Table~\ref{table:Chi squared} displays $\Delta\chi^2_\mathrm{min}$ values for each DM-baryon interacting model, as compared to $\Lambda$CDM. Table~\ref{table:cross section} displays constraints on the momentum-transfer cross section for each DM-baryon interacting model. Appendix \ref{Appendix:bias} displays posterior probability distributions for the EFT bias parameters referenced in Sec.~\ref{sec:nonlinear}. We show that all best-fit values for EFT bias parameters are similar to those found in $\Lambda$CDM. 

A comment is in order on the role of the BOSS data, which we find to depend on the power law index $n$.
For $n<0$ models, the 
addition of the BOSS 
does not lead to any noticeable effect
on the posteriors for $\sigma_0$. 
For $n=0$, the constraints become slightly
worse because the IDM model starts 
accounting for the $S_8$ tension between 
\textit{Planck} and BOSS (see Sec.~\ref{sec:tensions}).
For $n>0$, the BOSS data 
changes the $68\%$ CLs of \textit{Planck} quite significantly, but the $95\%$ CLs
are not strongly modified, as seen in Table~\ref{table:cross section}. In particular,
we find a $20\%$ improvement in the
$n=4$ case. This peculiar behavior happens because the addition
of the BOSS data makes the posterior
distribution of $\sigma_0$
more non-Gaussian for $n>0$.

\begin{figure}[!htb]
\includegraphics[scale=0.95]{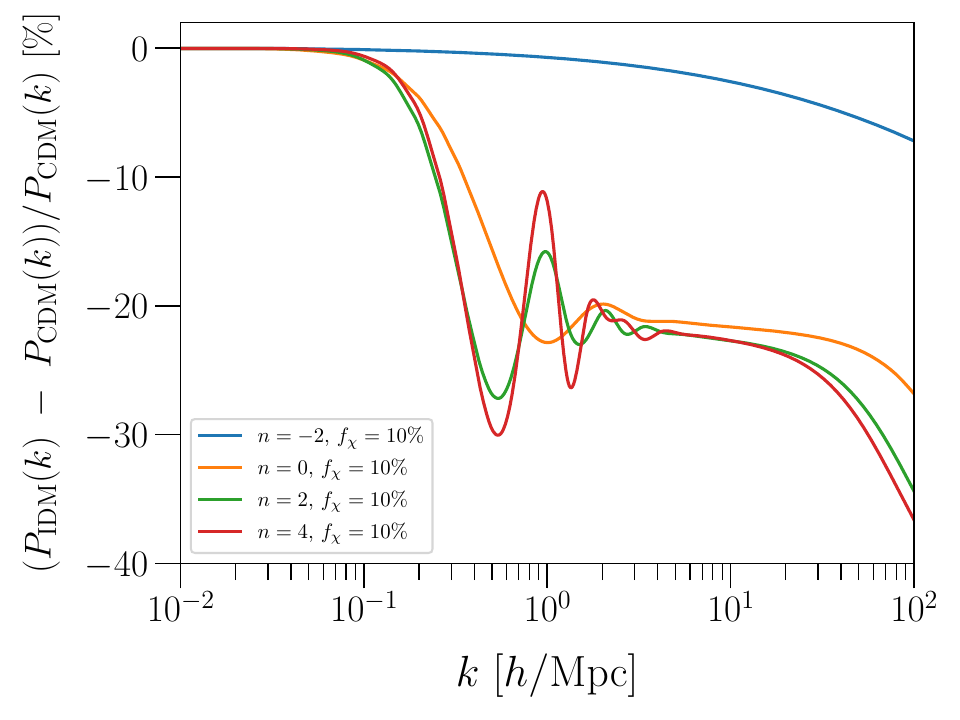}
\caption{\label{fig:fractional models} Percent difference between the linear matter power spectrum for different powers of $n$ in a fractional IDM cosmology and the linear matter power spectrum for $\Lambda$CDM. DM mass is set to 1 MeV.}
\end{figure}

We further consider cases where only a fraction of the total DM abundance is coupled to protons through $n = -2$, 0, 2, and 4 interaction models. We specifically consider cases where 10\% of the DM interacts with baryons, following \cite{he2023s8}, and the DM mass $m_\chi$ is 1 MeV, 1 GeV, and 10 GeV. We analyze these scenarios with the  combination of \textit{Planck} + BOSS + DES data. The resulting marginalized posterior probability distributions are shown in Appendix \ref{Appendix:posteriors_10}. Table~\ref{table:Chi squared fractional} displays $\Delta\chi^2_\mathrm{min}$ values for each fractional IDM model, compared to $\Lambda$CDM. We display curves for the best-fit fractional $n = -2$, 0, 2, and 4 models from our analysis in Fig.~\ref{fig:fractional models}.

\begin{table}[!htb]
\centering
\caption{\label{table:Chi squared fractional}$\Delta\chi_{\mathrm{min}}^2$ values for different models and masses under a $Planck$ + BOSS + DES analysis. In each case, $f_\chi=10\%$. Each $\Delta\chi_{\mathrm{min}}^2$ value is given with respect to the CDM $\chi^2$ value for a $Planck$ + BOSS + DES analysis.}
\begin{tabular}{|c | c | c | c | c | c | c|}
\hline
 Model
 &$n=-2$
 &$n=0$
 &$n=2$
 &$n=4$ \\ [0.5ex] 
 \hline\hline
 1 MeV
 & $-1.6$
 & $-6.7$
 & $-2.16$
 & $-3.08$ \\ [0.5ex]

 1 GeV
 & $+0.86$
 & $-3.98$
 & $-2$
 & $+0.42$ \\ [0.5ex]

 10 GeV
 & $-0.18$
 & $-4.8$
 & $-4.28$
 & $-1.16$ \\ [0.5ex]

 \hline
\end{tabular}
\end{table}

We find that in some cases, a non-zero cross section is mildly preferred at a level of $\gtrsim$ 2$\sigma$. We also note that this preference reaches a level of $\sim3\sigma$ for the $n=0$ model, for $m_\chi=1$ MeV, as reported in \cite{he2023s8}. While the preference is mild, these fractional models are currently consistent with both large-scale and small-scale structure observations, and as such, their consistent preference over $\Lambda$CDM may have interesting implications for cosmology; we discuss this point in more detail in Sec.~\ref{sec:discussion}.

\subsection{\label{sec:tensions} Cosmological tensions}

\begin{figure*}[!tb]
\includegraphics[scale=0.41]{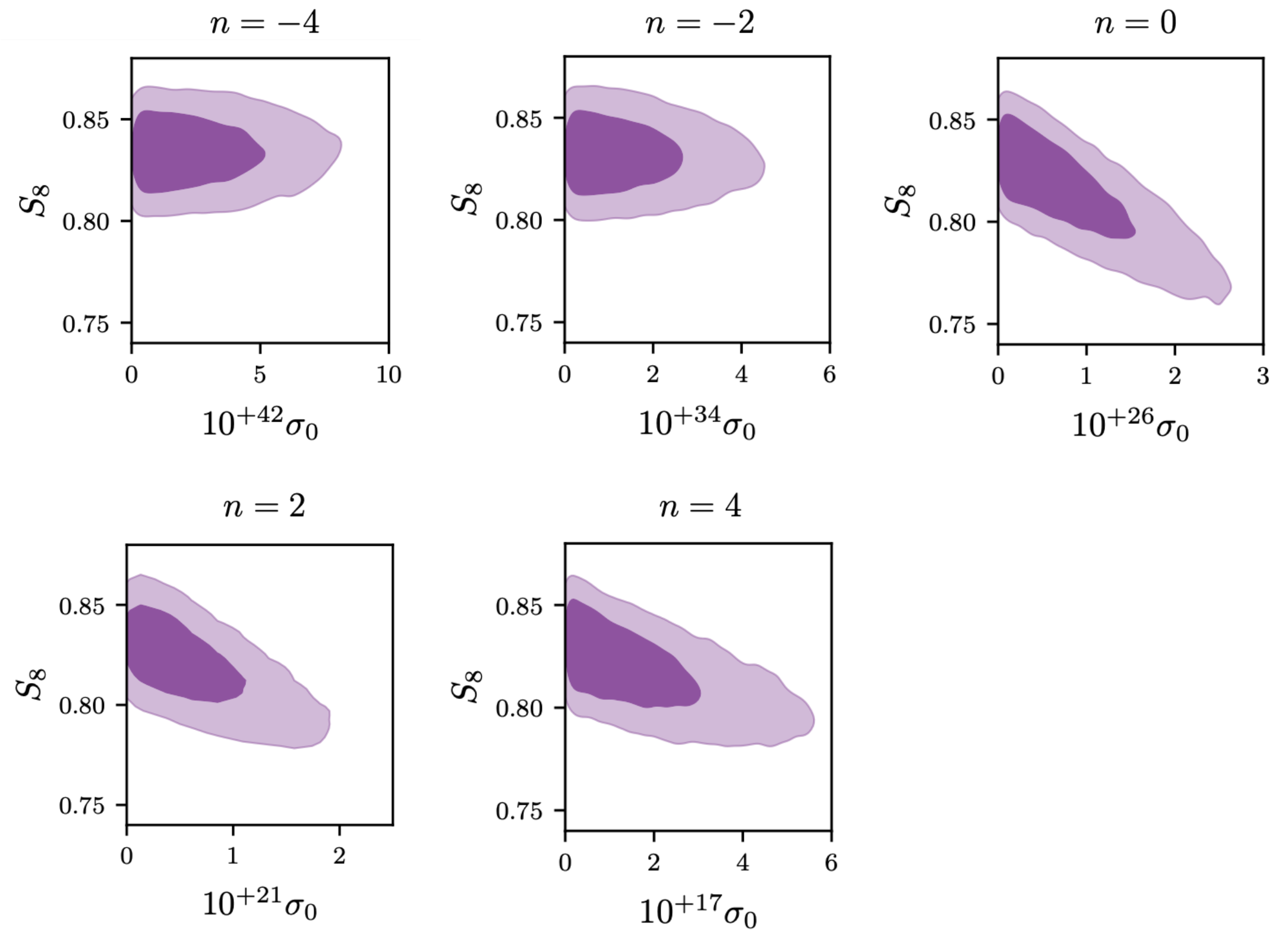}
\caption{\label{fig:degeneracy} 68\% and 95\% confidence level marginalized posterior distributions for $S_8$ and the coefficient of the cross section (in units of cm$^{2}$) of momentum transfer between DM and protons are shown for different values of $n$ when analyzed with \textit{Planck} data. Note the strong degeneracy between $S_8$ and $\sigma_0$ that begins to take form as $n$ is increased. Models
with negative powers of $n$ do not experience enough suppression to lead to a lower value of $S_8$.
}
\end{figure*}

In Fig.~\ref{fig:degeneracy}, we show the 2D marginalized posterior probability distribution for $S_8$ and $\sigma_0$, for all values of $n$ in models where all of DM interacts with baryons. We find that $\sigma_0$ is strongly degenerate with $S_8$ for $n$ = 0, 2, and 4 models. This degeneracy is expected for non-negative powers of velocity dependence in the cross section: a higher value of $\sigma_0$ leads to stronger coupling between DM and baryons, and an increased suppression in the power spectrum. For non-negative powers of $n$, this suppression is so steep that a shift in any other cosmological parameter cannot compensate for the sharp reduction in power; hence, higher values of $\sigma_0$ lead to a corresponding decrease in $S_8$. Models with negative powers of $n$ are unable to achieve the level of suppression needed to observe a lower value of $S_8$, and do not lead to this degeneracy. This is precisely why we see a preference for non-zero interaction cross section when we analyze these models with BOSS + DES data: the LSS probes prefer a lower value for $S_8$, and therefore disfavor lower cross sections that do not produce this lower value. 
As noted in Sec.~\ref{sec:parameter estimation}, a \textit{Planck} + BOSS analysis of the $n=0$ model slightly increases the allowed range of $\sigma_0$ values as compared to a \textit{Planck}--only analysis; again, this is because BOSS prefers a lower value for $S_8$, and higher cross sections in the $n=0$ model can suppress the power spectrum enough to achieve this lower value.

Fig.~\ref{fig:degeneracy} also demonstrates that the $n=0$ model most effectively alleviates the $S_8$ tension, with the degeneracy between $\sigma_0$ and $S_8$ being most notable in the $n$ = 0 case. 
We explicitly quantify the $S_8$ tension for each model using a Gaussian tension metric and display these values in Table~\ref{table:stdev}. As expected, we find the tension is minimized in the $n=0$ case (1.47$\sigma$), which is 40\% lower than the $S_8$ tension in $\Lambda$CDM (2.59$\sigma$). 

The $n$ = 0, $f_\chi$ = 100\% model leads to a preference for non-zero interaction when CMB and LSS data are considered; in addition, this model alleviates the $S_8$ tension. Although this scenario is already constrained by Milky Way substructure, this result indicates the shape of the transfer function that is necessary to fit both the LSS and CMB data. As previously noted in the literature, the scale-dependent power suppression we see in this model is the key feature that allows certain beyond-CDM models to alleviate cosmological tensions \cite{Amon_2022,Poulin_2022}.

\begin{table}[!htb]
\centering
\caption{\label{table:stdev}$S_8$ tension for $\Lambda$CDM and IDM models where 100\% of DM elastically scatters with protons. Note that the $S_8$ tension is minimized for the $n=0$ model, which is $\sim 50\%$ lower than the standard $S_8$ tension in $\Lambda$CDM.}

$S_8$ tension between \textit{Planck} and DES \\ [0.5ex]

\begin{tabular}{|c | c | c | c | c | c|}
\hline
 $\Lambda$CDM
 &$n=-4$
 &$n=-2$
 &$n=0$
 &$n=2$
 &$n=4$ \\ [0.5ex] 
 \hline\hline
 2.59$\sigma$
 & 2.7$\sigma$
 & 2.61$\sigma$
 & 1.47$\sigma$
 & 1.81$\sigma$
 & 1.83$\sigma$ \\ [0.5ex]
 
 \hline
\end{tabular}
\end{table}

Because the qualitative picture for fractional IDM models is very similar, we do not discuss these models in detail here. For a more detailed discussion of the $S_8$ tension in the context of fractional IDM models, see \cite{he2023s8}.

The $H_0$ tension is neither alleviated nor exacerbated when comparing a \textit{Planck}-only analysis of IDM to $\Lambda$CDM. The inclusion of LSS data shifts the mean of $H_0$ to higher values, but decreases the width of the associated posterior probability distribution, keeping the $H_0$ tension at the same level as in CDM case. 
$H_0$ is typically affected when either the size of the sound horizon at recombination $r_\mathrm{s}^\star$ or the angular diameter distance to the surface of last scattering $D_\mathrm{A}^\star$ is altered \cite{Bernal_2016}, where

\begin{equation}\label{r_s}
    r_\mathrm{s}^\star = \int_{z^\star}^{\infty} \frac{dz}{H(z)}c_s(t)
\end{equation}

\noindent where $z^\star$ is the redshift at recombination, $H$ is the Hubble parameter, and $c_\mathrm{s}$ is the sound speed of the photon-baryon fluid before recombination, typically defined as

\begin{equation}\label{c_s}
    c_\mathrm{s} = \sqrt{\frac{1}{3(1+R)}}
\end{equation}

\noindent where $R$ is the baryon-to-photon energy ratio. We find that this ratio is unaffected by DM-baryon interactions; even though DM and baryons are seemingly indistinguishable when they are strongly coupled, the photons in the fluid are still able to distinguish between the two particles, so the DM does not contribute to the overall density of baryons. Neither $r_\mathrm{s}^\star$ or $D_\mathrm{A}^\star$ change in a DM-baryon interacting scenario, and therefore $H_0$ stays the same between analyses of IDM and CDM.

\section{\label{sec:discussion}Discussion and Summary\protect}

This study explores DM interactions with the Standard Model within a range of models that feature velocity-dependent DM-proton elastic scattering, in the context of CMB and LSS data. In these models, DM and baryons exchange heat and momentum, leading to a scale-dependent suppression of matter perturbations in the early universe, which in turn affects CMB anisotropy and galaxy populations throughout cosmic history. Following previous literature, we model the momentum-transfer cross section for the interactions as a power law of the relative particle velocity, leaving the amplitude as a free parameter of the model. The power law index $n$ then captures the velocity dependence of the interaction for a specific interaction model at hand; we explore $n\in\{-4,-2, 0, 2, 4\}$.

To model the effects of interactions on cosmological observables beyond linear theory, we apply the effective field theory of LSS. We then use this modeling approach to analyze BOSS galaxy clustering data and \textit{Planck} CMB data in a joint likelihood analysis, and assess the validity of interacting DM models. Assuming that all of DM exchanges momentum with protons, we find that the addition of BOSS data does not alter the constraints on the momentum-transfer cross section inferred from CMB data alone. In contrast, we find that the inclusion of DES weak lensing data leads to a $\gtrsim$ 2$\sigma$ preference for velocity-independent DM-baryon scattering with a momentum-transfer cross section of $\sigma_0 = 1.47\cdot10^{-26}$ $\mathrm{cm}^{-2}$. We further explore scenarios where only a fraction of DM features interactions with protons, and find a $\gtrsim$ 2$\sigma$ preference for interactions in scenarios where $n=0$ and 2, consistent with \cite{he2023s8}. 

We note that the scenarios in which all of DM interacts with protons lead to the most severe suppression of perturbations on small scales; these models are thus found to be in significant tension with the existence of known satellite galaxies within the Milky Way \cite{Nadler_2019, Nadler_2021}. At the same time, we find that these scenarios tend to relieve the $S_8$ tension between cosmological data sets, indicating that the scale-dependent suppression they feature may be preferred by the data. 
Interestingly, and consistent with \cite{he2023s8}, the fractional interacting cases $n=0$ and $n=2$ which likewise restore the consistency of data on cosmological scales are currently unconstrained by the Milky Way substructure. 
The consistency in data is restored without significant changes to other standard cosmological parameters.
Indeed, the shape of the IDM linear power spectrum is similar to those associated with other proposed solutions to $S_8$ that also find a preference when jointly analyzing data from early-universe and late-universe observations \cite{Amon_2022, Poulin_2022}. 

The fractional IDM scenarios are distinct from other potential solutions to the $S_8$ tension in several interesting ways. First, IDM does not exacerbate the $H_0$ tension; this is a common pitfall of many models that attempt to address the $S_8$ tension \cite{Abdalla_2022}. Second, IDM is based on new DM physics that was extensively explored in contexts of direct detection, rather than as an a posteriori proposal designed solely to resolve cosmological tensions. Third, the preferred range of cross sections we find in the fractional cases is unconstrained by small-scale structure observations and other analyses \cite{Maamari_2021, Nadler_2019, Nadler_2021, Xu_2018, Rogers_2022, Becker_2021, Hooper_2022}, but may become accessible to detection with upcoming surveys, and is therefore imminently falsifiable. Indeed, DESI and the Vera C. Rubin Observatory will probe the Lyman-$\alpha$ forest and Milky Way substructure in detail, allowing new constraints on the preferred parameter space for interacting DM. Similarly, Stage 3 and Stage 4 CMB data will further refine measurements of the sub-degree-scale CMB primary and secondary anisotropy, further putting pressure on these scenarios within the coming decade \cite{SPT-3G:2014dbx, ACT:2020gnv,SimonsObservatory:2018koc,CMB-S4:2016ple,CMB-HD:2022bsz}. Finally, further full-shape analyses of similar models with existing 
and upcoming LSS datasets, e.g.,
eBOSS~\cite{eBOSS:2020yzd,Ivanov:2021zmi,Chudaykin:2022nru}, DESI~\cite{DESI:2024hhd},
KiDS \cite{KiDS:2020suj}, 
HSC \cite{HSC:2018mrq} are underway. We also acknowledge that recent studies of cluster counts from eROSITA \cite{Ghirardini_2024} and combined KiDS/DES data \cite{DESKiDS2023} showed no $S_8$ tension, and could thus challenge the results of this work.

We note that in our analysis, the DM mass $m_\chi$ and interacting fraction $f_\chi$ are set to fixed values, rather than free parameters. We choose to explore $f_\chi=10\%$ because this model has the highest chance of having an impact on the $S_8$ tension, and is also unconstrained by Milky Way substructure data \cite{he2023s8}. Moreover, in \cite{he2023s8} we found that varying $m_\chi$ qualitatively produces the same results, and results for different values of $f_\chi$ do not change considerably either when perturbed near 10\%. Indeed, a full model selection in which fraction and mass are both free parameters should be performed to evaluate whether the data truly prefer IDM over a vanilla CDM cosmology; however, this analysis exceeds the scope of this work.

More generally, our results indicate that a specific modification of the linear matter power spectrum (i.e., a power cutoff at mildly non-linear $k$) restores consistency between LSS and CMB data (Figure \ref{fig:fractions}).\footnote{Note that the BOSS data alone has a significant 
tension with \textit{Planck}~\cite{Chen:2024vuf,Ivanov:2024xgb}
on large scales, $k\lesssim 0.1~h$Mpc$^{-1}$.
It will be interesting to see if this tension 
can be accounted for by beyond-$\Lambda$CDM models, 
such as e.g.~\cite{Fuss:2022zyt,Fuss:2024dam}. 
} Other models with similar scale-dependent power suppression have also shown preference when analyzed with CMB and LSS data \cite{Amon_2022,Poulin_2022}; this consistent finding of a preference for these models over $\Lambda$CDM when combining LSS and CMB data motivates further exploration of cosmological models that uniquely affect the matter power spectrum at different points in cosmic history. 

\section*{Acknowledgements}
RA and VG acknowledge the support from NASA through the Astrophysics Theory Program, Award Number 21-ATP21-0135. VG additionally acknowledges the support from the National Science Foundation (NSF) CAREER Grant No. PHY-2239205 and from the Research Corporation for Science Advancement under the Cottrell Scholar Program.  

\clearpage
\appendix

\section{Matter perturbations at low redshifts}
\label{Appendix:structure evolution}

\begin{figure*}[ht!]
\includegraphics[scale=0.75]{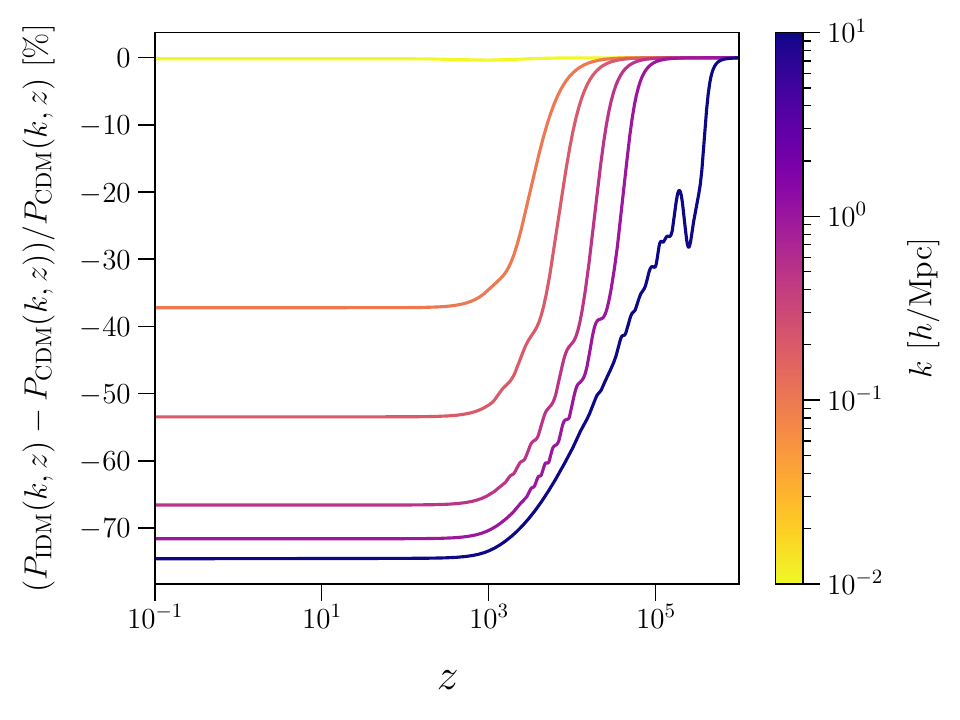}
\centering
\caption{Residual between the power spectrum for the $n=-4$ IDM model and the power spectrum for $\Lambda$CDM as a function of redshift, for different values of $k$. The curves are static for $z < 10^3$, indicating that there is no evolution on these scales past recombination. This plot is generated with best-fit cosmological parameters from a \textit{Planck} analysis of the $n=-4$ IDM model, with a DM particle mass $m_\chi=1$ MeV and cross section at its 5$\sigma$ limit from BOSS, $\sigma_0=1\cdot10^{-40} \ \mathrm{cm}^2$.
\label{fig:transfer}}
\end{figure*}

\begin{figure*}[ht!]
\includegraphics[scale=0.75]{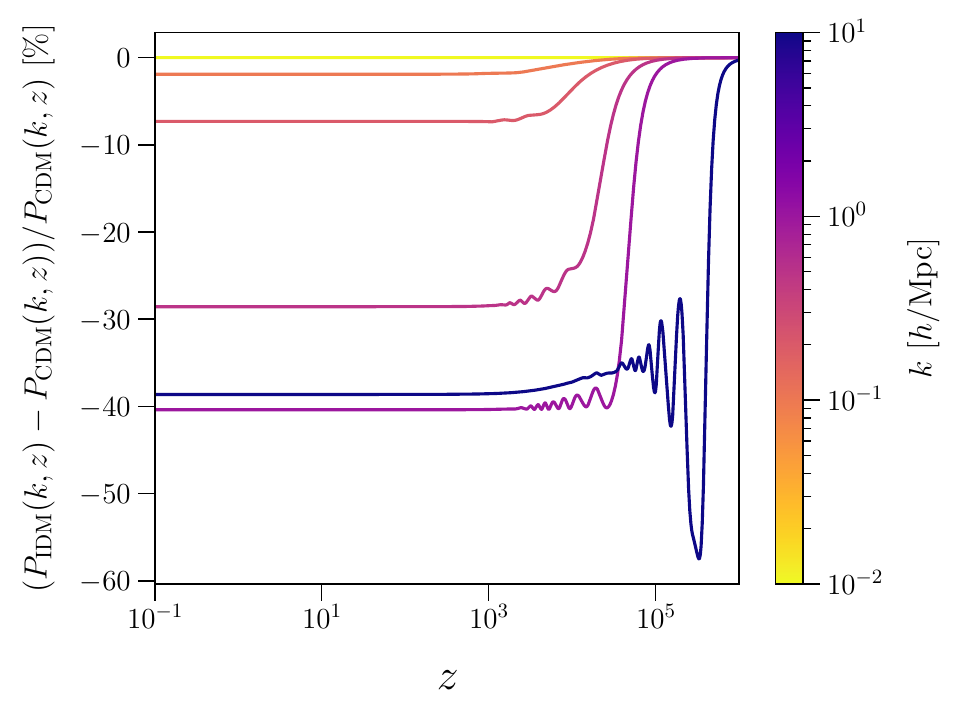}
\centering
\caption{Residual between the power spectrum for the $n=0$, $f_\chi = 10\%$ IDM model and the power spectrum for $\Lambda$CDM as a function of redshift, for different values of $k$. The curves are static for $z < 10^3$, indicating that there is no evolution on these scales past recombination. This plot is generated with best-fit cosmological parameters from a \textit{Planck} + BOSS + DES analysis of the $n=0$, $f_\chi = 10\%$ IDM model, with a DM particle mass $m_\chi=1$ MeV and cross section $\sigma_0=5.16\cdot10^{-26} \ \mathrm{cm}^2$.
\label{fig:transfer fractional n=0}}
\end{figure*}

\begin{figure*}[ht!]
\includegraphics[scale=0.75]{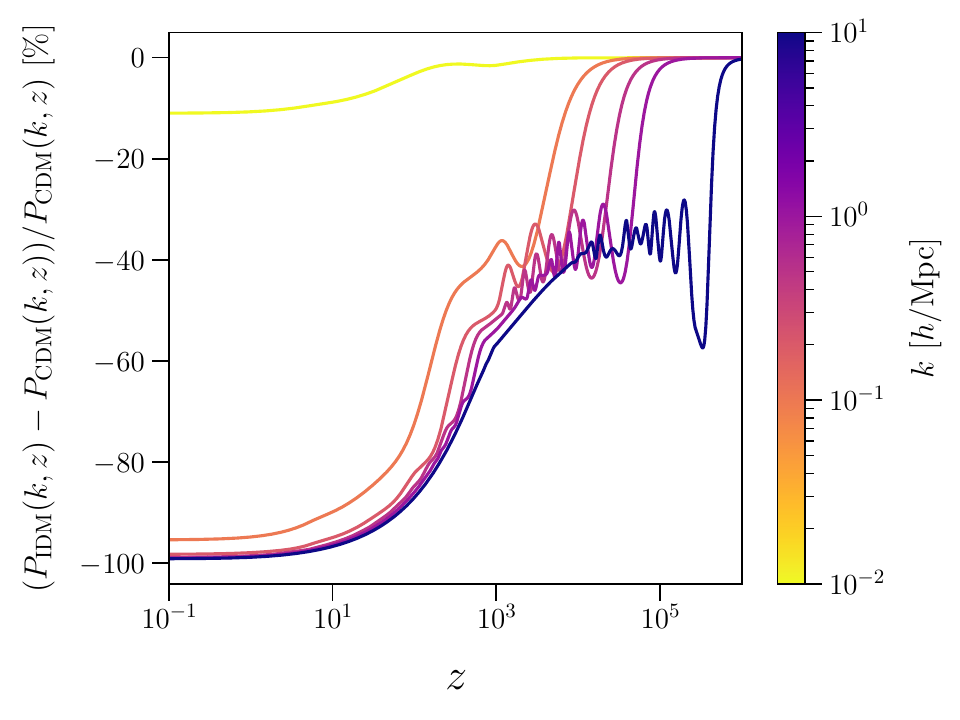}
\centering
\caption{Residual between the power spectrum for the $n=-2$, $f_\chi = 10\%$ IDM model and the power spectrum for $\Lambda$CDM as a function of redshift, for different values of $k$. The curves continue oscillating up to $z \sim 8$, meaning that the EFT should be modified to account for this model's impact on late-time growth. We thus treat our results for this model as approximate. This plot is generated with best-fit cosmological parameters from a \textit{Planck} analysis of the $n=-2$, $f_\chi = 10\%$ IDM model, with a DM particle mass $m_\chi=1$ MeV and cross section $\sigma_0=1\cdot10^{-32} \ \mathrm{cm}^2$.
\label{fig:transfer fractional n=-2}}
\end{figure*}

We check that there are no alternations in structure evolution on any modes after recombination for IDM models in which all of DM interacts with baryons by plotting the residual of the $n=-4$ power spectrum with respect to $\Lambda$CDM as a function of redshift for different $k$ (Fig.~\ref{fig:transfer}). Therefore, we may take the linear power spectrum generated for these models and pass it to the standard non-linear CDM pipeline implemented by \texttt{CLASS-PT}, without introducing additional counterterms to the non-linear power spectrum calculation. For IDM models in which only a fraction of DM interacts with baryons, we plot the residual of the $n=0$, $f_\chi = 10\%$ power spectrum with respect to $\Lambda$CDM as a function of redshift for different $k$ (Fig.~\ref{fig:transfer fractional n=0}), showing that the EFT does not need to be modified for fractional $n=0$, 2, 4 models. As noted in Sec.~\ref{sec:methodology}, the $n=-2$, $f_\chi = 10\%$ model impacts perturbations down to redshifts $z\sim8$; we show this in Fig.~\ref{fig:transfer fractional n=-2}. Our results for this particular model can thus be treated as \textit{approximate}; we leave a more detailed analysis of this model to future work.

\clearpage

\section{Cosmological parameter constraints}
\label{Appendix:constraints}

We show constraints on relevant cosmological parameters for models in which all of DM is interacting. Tables~\ref{tab:n-4},~\ref{tab:n-2},~\ref{tab:n0},~\ref{tab:n2},~\ref{tab:n4} correspond with models $n=-4$, $-2$, 0, 2, 4 respectively, all with $f_\chi=100\%$ and $m_\chi=1$ MeV. We also show constraints on relevant cosmological parameters for models in which 10\% of DM is interacting. For this case, Tables~\ref{tab:n-2f},~\ref{tab:n0f},~\ref{tab:n2f},~\ref{tab:n4f} correspond with models $n=-2$, 0, 2, 4 respectively, all with $f_\chi=10\%$ and analyzed with \textit{Planck}+BOSS+DES. In all tables, ``Best fit'' denotes the maximum of the full posterior, while ``Marginalized max'' denotes the maxima of the marginalized posteriors.

\begin{table*}[htb!]
\centering
\caption{Constraints on $n=-4$, $f_\chi=100\%$, $m_\chi=1$ MeV
} 
\label{tab:n-4}
\begin{tabular}{|l|c|c|c|c|c|}
 \hline
Dataset & Parameter & Best-fit & Marginalized max $\pm \ \sigma$ & 95\% lower & 95\% upper \\ \hline
& $10^{+42}\sigma{}_0$ &$0.1563$ & $2.64^{+0.77}_{-2.6}$ & $>0$ & $6.45$ \\
\textit{Planck}& $\sigma_8$ &$0.8132$ & $0.8113\pm 0.006$ & $0.7995$ & $0.8232$ \\
& $S_8$ &$0.8376$ & $0.8339_{-0.0128}^{+0.0129}$ & $0.8082$ & $0.8594$ \\
\hline
& $10^{+42}\sigma{}_0$ &$2.767$ & $2.67^{+0.75}_{-2.5}$ & $>0$ & $6.738$ \\
\textit{Planck}+BOSS& $\sigma_8$ &$0.8099$ & $0.8077\pm 0.0058$ & $0.7964$ & $0.8191$ \\
& $S_8$ &$0.8309$ & $0.8259\pm{0.0102}$ & $0.8055$ & $0.8463$ \\
\hline
& $10^{+42}\sigma{}_0$ &$1.117$ & $2.73^{+0.74}_{-2.7}$ & $>0$ & $6.834$ \\
\textit{Planck}+BOSS+DES& $\sigma_8$ &$0.803$ & $0.802\pm 0.0054$ & $0.7913$ & $0.8129$ \\
& $S_8$ &$0.808$ & $0.8123_{-0.0089}^{+0.0088}$ & $0.7951$ & $0.8297$ \\
\hline
\end{tabular}
\end{table*}

\begin{table*}[htb!]
\centering
\caption{Constraints on $n=-2$, $f_\chi=100\%$, $m_\chi=1$ MeV
} 
\label{tab:n-2}
\begin{tabular}{|l|c|c|c|c|c|}
 \hline
Dataset & Parameter & Best-fit & Marginalized max $\pm \ \sigma$ & 95\% lower & 95\% upper \\ \hline
& $10^{+34}\sigma{}_0$ &$2.01$ & $1.38^{+0.46}_{-1.4}$ & $>0$ & $3.56$ \\
\textit{Planck}& $\sigma_8$ &$0.8102$ & $0.8102\pm 0.0061$ & $0.7983$ & $0.8223$ \\
& $S_8$ &$0.8374$ & $0.8318\pm{0.013}$ & $0.8061$ & $0.8574$ \\
\hline
& $10^{+34}\sigma{}_0$ &$0.5662$ & $1.39^{+0.39}_{-1.4}$ & $>0$ & $3.544$ \\
\textit{Planck}+BOSS& $\sigma_8$ &$0.81$ & $0.8065\pm 0.0059$ & $0.7949$ & $0.818$ \\
& $S_8$ &$0.8219$ & $0.8243_{-0.0103}^{+0.01}$ & $0.8041$ & $0.8448$ \\
\hline
& $10^{+34}\sigma{}_0$ &$1.942$ & $1.52^{+0.44}_{-1.5}$ & $>0$ & $3.721$ \\
\textit{Planck}+BOSS+DES& $\sigma_8$ &$0.7999$ & $0.8009\pm 0.0055$ & $0.790$ & $0.8117$ \\
& $S_8$ &$0.814$ & $0.8113_{-0.0089}^{+0.009}$ & $0.764$ & $0.7967$ \\
\hline
\end{tabular}
\end{table*}

\begin{table*}[htb!]
\centering
\caption{Constraints on $n=0$, $f_\chi=100\%$, $m_\chi=1$ MeV
} 
\label{tab:n0}
\begin{tabular}{|l|c|c|c|c|c|}
 \hline
Dataset & Parameter & Best-fit & Marginalized max $\pm \ \sigma$ & 95\% lower & 95\% upper \\ \hline
& $10^{+26}\sigma{}_0$ &$0.8792$ & $0.80^{+0.20}_{-0.79}$ & $>0$ & $2.07$ \\
\textit{Planck}& $\sigma_8$ &$0.7924$ & $0.793^{+0.022}_{-0.0094}$ & $0.7549$ & $0.8208$ \\
& $S_8$ &$0.8214$ & $0.8154^{+0.0238}_{-0.0156}$ & $0.7728$ & $0.8529$ \\
\hline
& $10^{+26}\sigma{}_0$ &$0.2136$ & $1.07^{+0.59}_{-1.1}$ & $>0$ & $2.6$ \\
\textit{Planck}+BOSS& $\sigma_8$ &$0.8023$ & $0.781^{+0.028}_{-0.014}$ & $0.7359$ & $0.8158$ \\
& $S_8$ &$0.8242$ & $0.7983_{-0.0172}^{+0.0286}$ & $0.7513$ & $0.8386$ \\
\hline
& $10^{+26}\sigma{}_0$ &$1.606$ & $1.47\pm 0.63$ & $0.2336$ & $2.664$ \\
\textit{Planck}+BOSS+DES& $\sigma_8$ &$0.7659$ & $0.768\pm 0.016$ & $0.7374$ & $0.7997$ \\
& $S_8$ &$0.7828$ & $0.7846_{-0.0149}^{+0.015}$ & $0.7554$ & $0.8138$ \\
\hline
\end{tabular}
\end{table*}

\begin{table*}[htb!]
\centering
\caption{Constraints on $n=2$, $f_\chi=100\%$, $m_\chi=1$ MeV
} 
\label{tab:n2}
\begin{tabular}{|l|c|c|c|c|c|}
 \hline
Dataset & Parameter & Best-fit & Marginalized max $\pm \ \sigma$ & 95\% lower & 95\% upper \\ \hline
& $10^{+21}\sigma{}_0$ &$0.2277$ & $0.691^{+0.077}_{-0.67}$ & $>0$ & $1.535$ \\
\textit{Planck}& $\sigma_8$ &$0.8102$ & $0.797^{+0.017}_{-0.0068}$ & $0.7698$ & $0.8208$ \\
& $S_8$ &$0.8326$ & $0.8195_{-0.0132}^{+0.0204}$ & $0.7855$ & $0.8542$ \\
\hline
& $10^{+21}\sigma{}_0$ &$0.1614$ & $0.462^{+0.070}_{-0.47}$ & $>0$ & $1.591$ \\
\textit{Planck}+BOSS& $\sigma_8$ &$0.8039$ & $0.796^{+0.015}_{-0.0074}$ & $0.7665$ & $0.8181$ \\
& $S_8$ &$0.8305$ & $0.8134_{-0.0123}^{+0.0174}$ & $0.7803$ & $0.8436$ \\
\hline
& $10^{+21}\sigma{}_0$ &$1.091$ & $0.93^{+0.33}_{-0.86}$ & $>0$ & $2.108$ \\
\textit{Planck}+BOSS+DES& $\sigma_8$ &$0.777$ & $0.781^{+0.018}_{-0.012}$ & $0.7517$ & $0.807$ \\
& $S_8$ &$0.7911$ & $0.7954_{-0.0124}^{+0.0159}$ & $0.7663$ & $0.8222$ \\
\hline
\end{tabular}
\end{table*}

\begin{table*}[htb!]
\centering
\caption{Constraints on $n=4$, $f_\chi=100\%$, $m_\chi=1$ MeV
} 
\label{tab:n4}
\begin{tabular}{|l|c|c|c|c|c|}
 \hline
Dataset & Parameter & Best-fit & Marginalized max $\pm \ \sigma$ & 95\% lower & 95\% upper \\ \hline
& $10^{+17}\sigma{}_0$ &$0.3951$ & $1.64^{+0.38}_{-1.6}$ & $>0$ & $4.429$ \\
\textit{Planck}& $\sigma_8$ &$0.8053$ & $0.796^{+0.015}_{-0.0095}$ & $0.7708$ & $0.818$ \\
& $S_8$ &$0.8243$ & $0.8193_{-0.0153}^{+0.0176}$ & $0.7867$ & $0.8505$ \\
\hline
& $10^{+17}\sigma{}_0$ &$0.02186$ & $1.08^{+0.28}_{-1.1}$ & $>0$ & $3.714$ \\
\textit{Planck}+BOSS& $\sigma_8$ &$0.8099$ & $0.796^{+0.013}_{-0.0078}$ & $0.7725$ & $0.8164$ \\
& $S_8$ &$0.8364$ & $0.8148_{-0.0121}^{+0.015}$ & $0.7868$ & $0.8408$ \\
\hline
& $10^{+17}\sigma{}_0$ &$1.428$ & $2.08^{+0.8}_{-1.9}$ & $>0$ & $5.247$ \\
\textit{Planck}+BOSS+DES& $\sigma_8$ &$0.7907$ & $0.785^{+0.013}_{-0.011}$ & $0.7609$ & $0.8067$ \\
& $S_8$ &$0.8097$ & $0.7988_{-0.0115}^{+0.013}$ & $0.7744$ & $0.8222$ \\
\hline
\end{tabular}
\end{table*}

\begin{table*}[htb!]
\centering
\caption{Constraints on $n=-2$, $f_\chi=10\%$, \textit{Planck}+BOSS+DES
} 
\label{tab:n-2f}
\begin{tabular}{|l|c|c|c|c|c|}
 \hline
Mass & Parameter & Best-fit & Marginalized max $\pm \ \sigma$ & 95\% lower & 95\% upper \\ \hline
& $10^{+34}\sigma{}_0$ &$2.111$ & $15.9^{+4.3}_{-16}$ & $>0$ & $40.52$ \\
1 MeV & $\sigma_8$ &$0.8035$ & $0.8015\pm 0.0054$ & $0.7908$ & $0.8122$ \\
& $S_8$ &$0.8113$ & $0.8118_{-0.0085}^{+0.0087}$ & $0.7945$ & $0.829$ \\
\hline
& $10^{+34}\sigma{}_0$ &$15.28$ & $41^{+10}_{-40}$ & $>0$ & $102.99$ \\
1 GeV & $\sigma_8$ &$0.8039$ & $0.8014\pm 0.0055$ & $0.7907$ & $0.8122$ \\
& $S_8$ &$0.8089$ & $0.8115_{-0.0089}^{+0.0087}$ & $0.7944$ & $0.8288$ \\
\hline
& $10^{+34}\sigma{}_0$ &$119.96$ & $244_{-200}^{+70}$ & $>0$ & $643.82$ \\
10 GeV & $\sigma_8$ &$0.8044$ & $0.8016\pm 0.0055$ & $0.7908$ & $0.8124$ \\
& $S_8$ &$0.8121$ & $0.8118_{-0.0086}^{+0.0089}$ & $0.7944$ & $0.8288$ \\
\hline
\end{tabular}
\end{table*}

\begin{table*}[htb!]
\centering
\caption{Constraints on $n=0$, $f_\chi=10\%$, \textit{Planck}+BOSS+DES}
\label{tab:n0f}
\begin{tabular}{|l|c|c|c|c|c|}
 \hline
Mass & Parameter & Best-fit & Marginalized max $\pm \ \sigma$ & 95\% lower & 95\% upper \\ \hline
& $10^{+26}\sigma{}_0$ &$5.163$ & $13.23_{-6.5}^{+5.2}$ & $1.55$ & $24.57$ \\
1 MeV & $\sigma_8$ &$0.7921$ & $0.7796_{-0.0094}^{+0.0068}$ & $0.764$ & $0.7967$ \\
& $S_8$ &$0.8039$ & $0.7939_{-0.0104}^{+0.0094}$ & $0.7744$ & $0.8141$ \\
\hline
& $10^{+26}\sigma{}_0$ &$99.64$ & $165^{+70}_{-90}$ & $11.17$ & $326.59$ \\
1 GeV & $\sigma_8$ &$0.785$ & $0.7842^{+0.0062}_{-0.0084}$ & $0.7699$ & $0.7999$ \\
& $S_8$ &$0.7952$ & $0.7984_{-0.0099}^{+0.0093}$ & $0.7795$ & $0.8178$ \\
\hline
& $10^{+26}\sigma{}_0$ &$1067.18$ & $1155\pm 700$ & $>0$ & $2326.7$ \\
10 GeV & $\sigma_8$ &$0.7837$ & $0.7858^{+0.0063}_{-0.009}$ & $0.7711$ & $0.8025$ \\
& $S_8$ &$0.7972$ & $0.7997_{-0.01}^{+0.0092}$ & $0.7808$ & $0.819$ \\
\hline
\end{tabular}
\end{table*}

\begin{table*}[htb!] 
\centering
\caption{Constraints on $n=2$, $f_\chi=10\%$, \textit{Planck}+BOSS+DES
} 
\label{tab:n2f}
\begin{tabular}{|l|c|c|c|c|c|}
 \hline
Mass & Parameter & Best-fit & Marginalized max $\pm \ \sigma$ & 95\% lower & 95\% upper \\ \hline
& $10^{+21}\sigma{}_0$ &$19.48$ & $15.8^{+6.5}_{-9.7}$ & $1.321$ & $32.34$ \\
1 MeV & $\sigma_8$ &$0.7684$ & $0.7769^{+0.0093}_{-0.012}$ & $0.7578$ & $0.7975$ \\
& $S_8$ &$0.7785$ & $0.7916_{-0.0115}^{+0.0113}$ & $0.7697$ & $0.8137$ \\
\hline
& $10^{+21}\sigma{}_0$ &$13660.35$ & $5931^{+2000}_{-6000}$ & $>0$ & $40586.35$ \\
1 GeV & $\sigma_8$ &$0.786$ & $0.796^{+0.011}_{-0.0055}$ & $0.7745$ & $0.8128$ \\
& $S_8$ &$0.8035$ & $0.8075_{-0.0091}^{+0.0122}$ & $0.7847$ & $0.8283$ \\
\hline
& $10^{+21}\sigma{}_0$ &$225061.1$ & $\left(\,2^{+1}_{-2}\,\right)\cdot 10^{5}$ & $>0$ & $707963.01$ \\
10 GeV & $\sigma_8$ &$0.7842$ & $0.79^{+0.016}_{-0.011}$ & $0.7651$ & $0.8109$ \\
& $S_8$ & $0.7957$ & $0.8019_{-0.0123}^{+0.014}$ & $0.7772$ & $0.8252$ \\
\hline
\end{tabular}
\end{table*}

\begin{table*}[t]
\centering
\caption{Constraints on $n=4$, $f_\chi=10\%$, \textit{Planck}+BOSS+DES
} 
\label{tab:n4f}
\begin{tabular}{|l|c|c|c|c|c|}
 \hline
Mass & Parameter & Best-fit & Marginalized max $\pm \ \sigma$ & 95\% lower & 95\% upper \\ \hline
& $10^{+17}\sigma{}_0$ &$61.62$ & $48\pm20$ & $2.667$ & $91.57$ \\
1 MeV & $\sigma_8$ &$0.7763$ & $0.7827^{+0.0074}_{-0.0093}$ & $0.7663$ & $0.8002$ \\
& $S_8$ &$0.7914$ & $0.7971_{-0.0107}^{+0.0096}$ & $0.7763$ & $0.8183$ \\
\hline
& $10^{+17}\sigma{}_0$ &$666.43$ & $10080^{+5000}_{-10000}$ & $>0$ & $48546.2$ \\
1 GeV & $\sigma_8$ &$0.8033$ & $0.8028\pm 0.0055$ & $0.7918$ & $0.8135$ \\
& $S_8$ & $0.8125$ & $0.8132_{-0.0092}^{+0.0089}$ & $0.7955$ & $0.8308$ \\
\hline
& $10^{+17}\sigma{}_0$ &$33350.65$ & $15280^{-5200}_{-16000}$ & $>0$ & $110734.9$ \\
10 GeV & $\sigma_8$ &$0.8059$ & $0.8027\pm 0.0054$ & $0.7921$ & $0.8134$ \\
& $S_8$ &$0.8184$ & $0.8129\pm{0.0088}$ & $0.7956$ & $0.8299$ \\
\hline
\end{tabular}
\end{table*}

\clearpage

\section{Posterior probability distributions for models in which all of DM interacts}
\label{Appendix:posteriors_100}

\subsection{$n=-4$}

\begin{figure*}[!htb]
\includegraphics[scale=0.575]{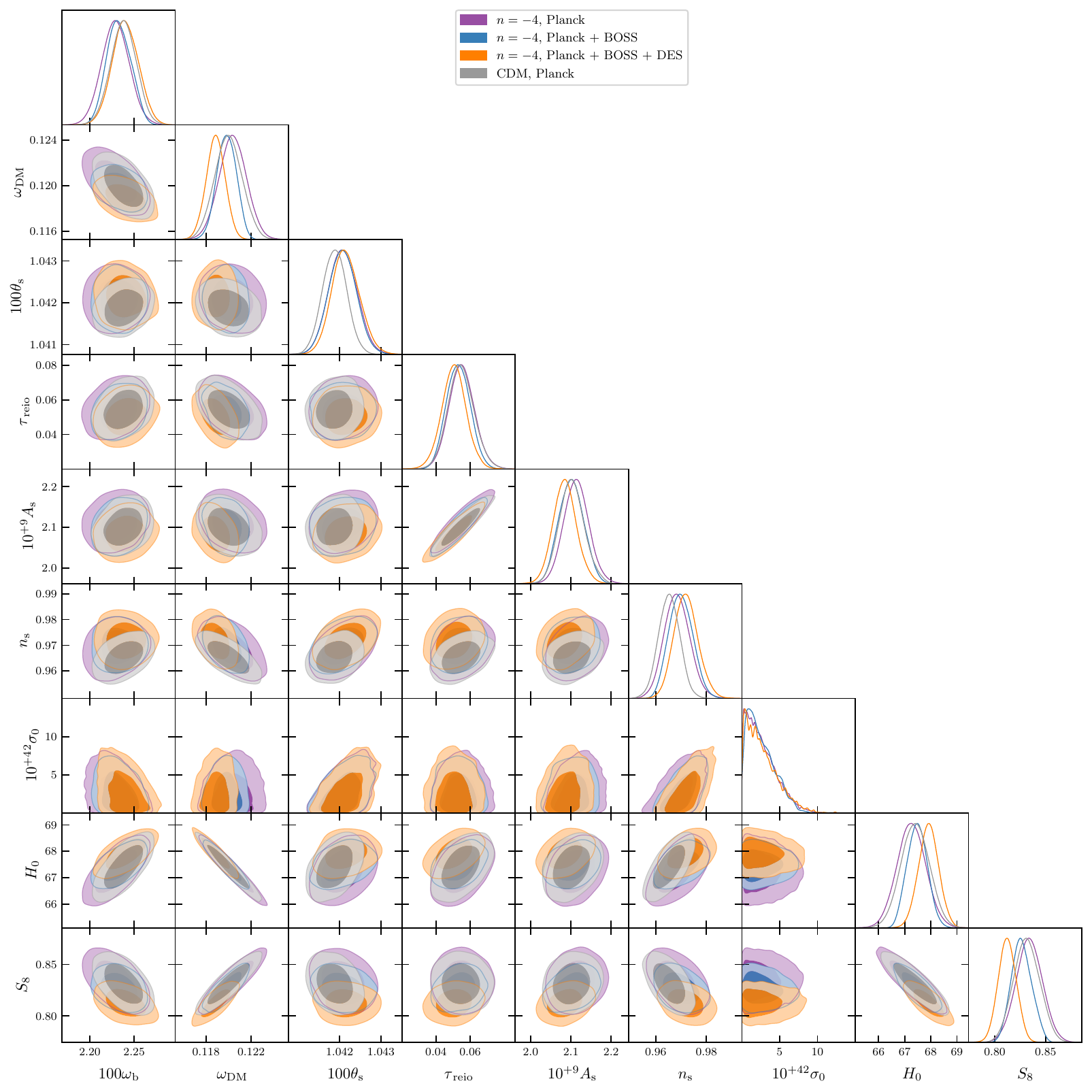}
\caption{\label{fig:n=-4 triangle plot} 68\% and 95\% confidence level marginalized posterior distributions for the $n=-4$ DM-baryon interacting model from different combinations of \textit{Planck}, BOSS, and DES data (colored), compared with posteriors for $\Lambda$CDM from \textit{Planck} (gray).}
\end{figure*}

\begin{figure*}[!htb]
\includegraphics[scale=0.575]{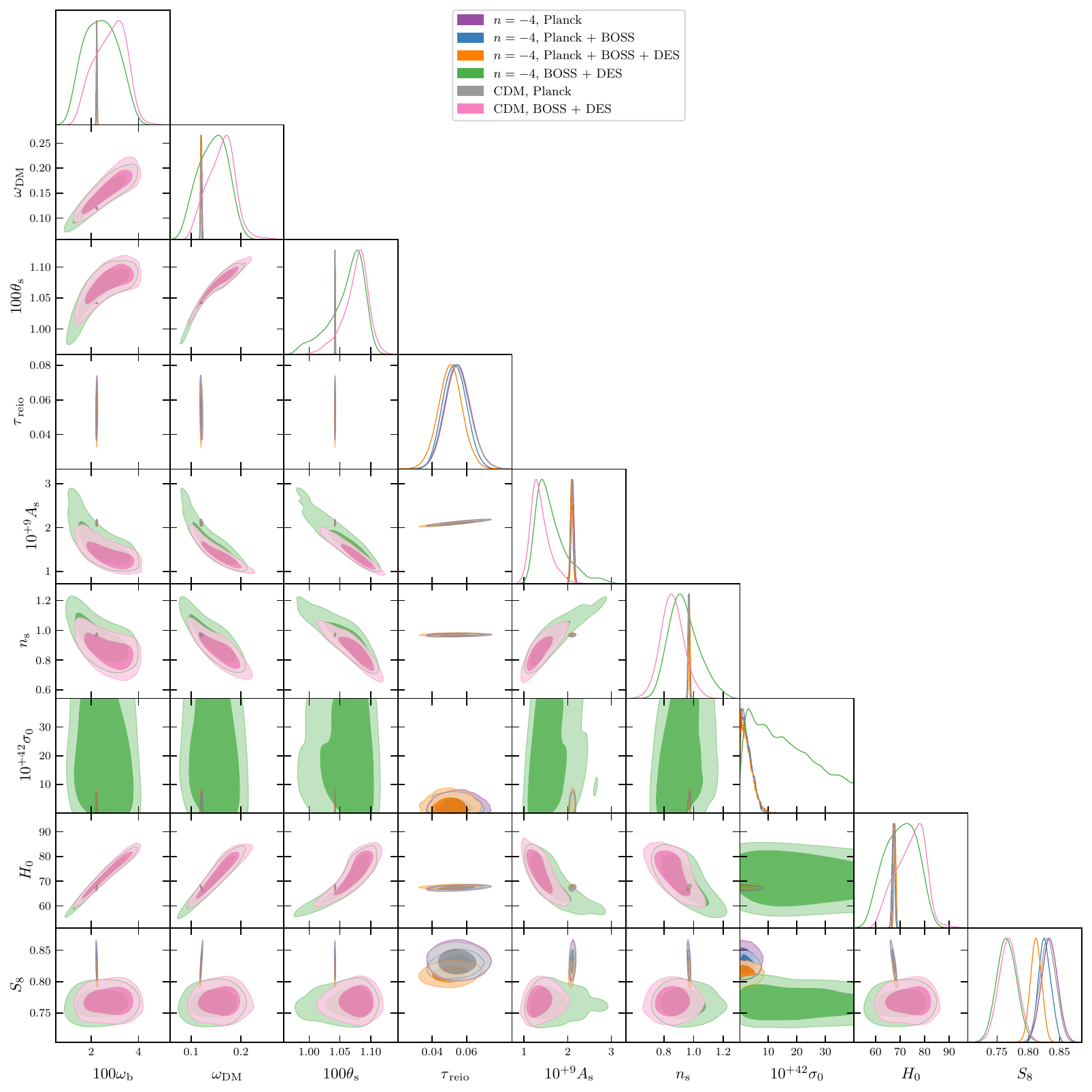}
\caption{\label{fig:n=-4 triangle plot BOSS + DES} 68\% and 95\% confidence level marginalized posterior distributions for the $n=-4$ DM-baryon interacting model from different combinations of \textit{Planck}, BOSS, and DES data (colored), compared with posteriors for $\Lambda$CDM from \textit{Planck} (gray). Same plot as Fig.~\ref{fig:n=-4 triangle plot} but with BOSS + DES posteriors added.}
\end{figure*}

\begin{figure*}[!htb]
\includegraphics[scale=0.575]{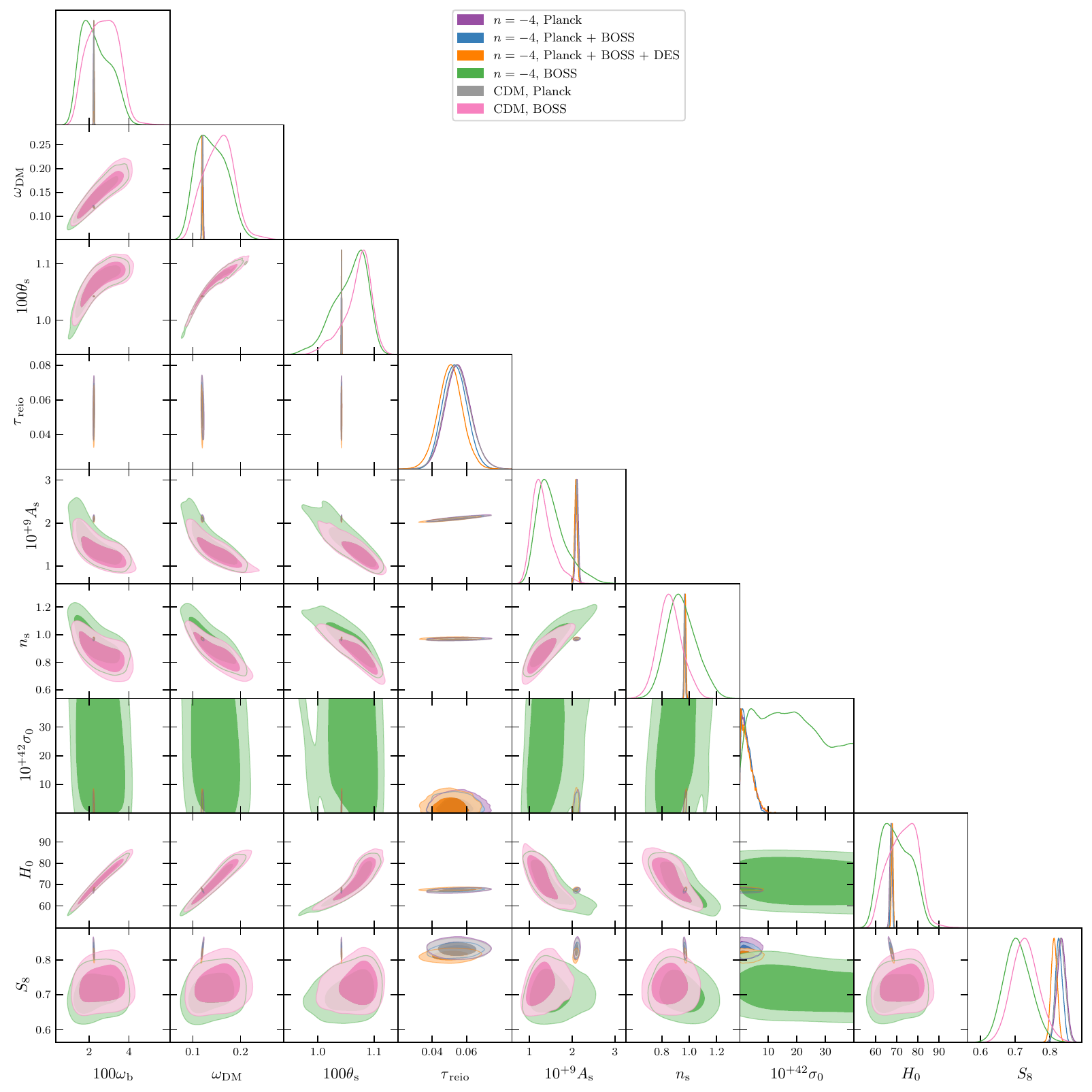}
\caption{\label{fig:n=-4 triangle plot BOSS} 68\% and 95\% confidence level marginalized posterior distributions for the $n=-4$ DM-baryon interacting model from different combinations of \textit{Planck}, BOSS, and DES data (colored), compared with posteriors for $\Lambda$CDM from \textit{Planck} (gray). Same plot as Fig.~\ref{fig:n=-4 triangle plot} but with BOSS posteriors added.}
\end{figure*}

We display full marginalized posterior distributions for all relevant parameters in our analysis of the $n=-4$, $f_\chi$ = 100\%, and $m_\chi$ = 1 MeV model in Fig.~\ref{fig:n=-4 triangle plot}. We show the same posterior distributions along with the BOSS + DES posteriors in Fig.~\ref{fig:n=-4 triangle plot BOSS + DES}, and the same posterior distributions along with the BOSS posteriors in Fig.~\ref{fig:n=-4 triangle plot BOSS}.

\clearpage

\subsection{$n=-2$}

\begin{figure*}[!htb]
\includegraphics[scale=0.575]{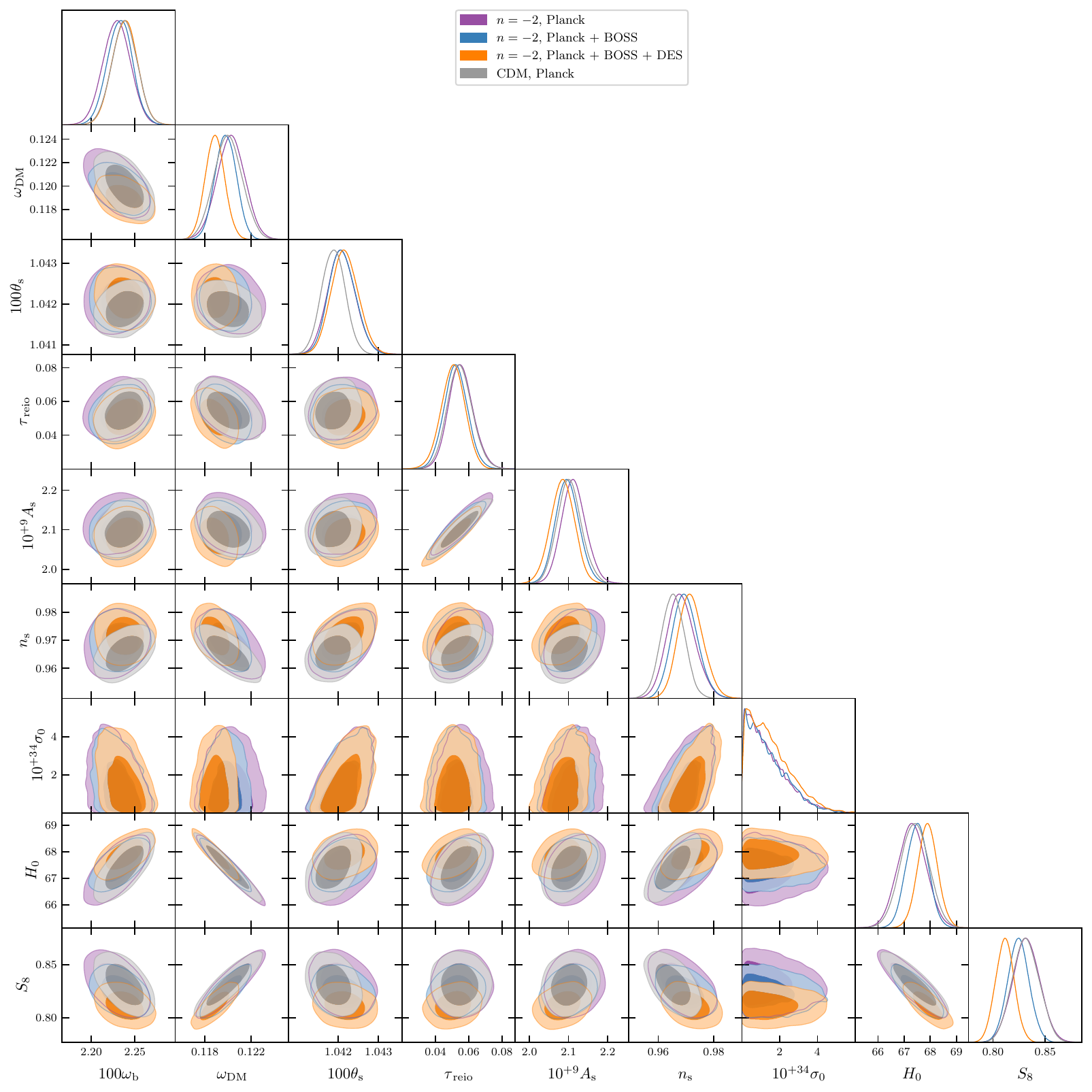}
\caption{\label{fig:n=-2 triangle plot} 68\% and 95\% confidence level marginalized posterior distributions for the $n=-2$ DM-baryon interacting model from different combinations of \textit{Planck}, BOSS, and DES data (colored), compared with posteriors for $\Lambda$CDM from \textit{Planck} (gray).}
\end{figure*}

\begin{figure*}[!htb]
\includegraphics[scale=0.575]{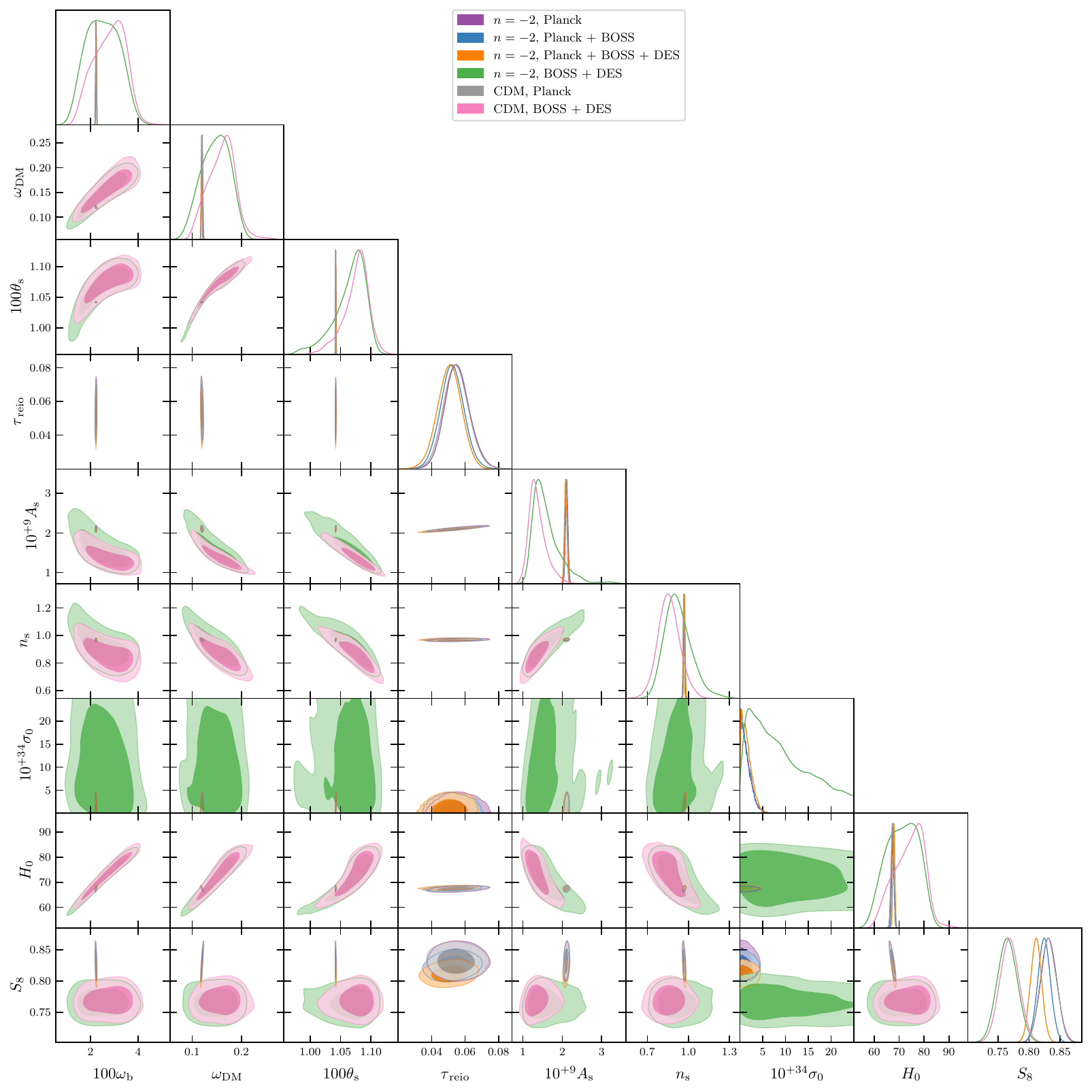}
\caption{\label{fig:n=-2 triangle plot BOSS + DES} 68\% and 95\% confidence level marginalized posterior distributions for the $n=-2$ DM-baryon interacting model from different combinations of \textit{Planck}, BOSS, and DES data (colored), compared with posteriors for $\Lambda$CDM from \textit{Planck} (gray). Same plot as Fig.~\ref{fig:n=-2 triangle plot} but with BOSS + DES posteriors added.}
\end{figure*}

\begin{figure*}[!htb]
\includegraphics[scale=0.575]{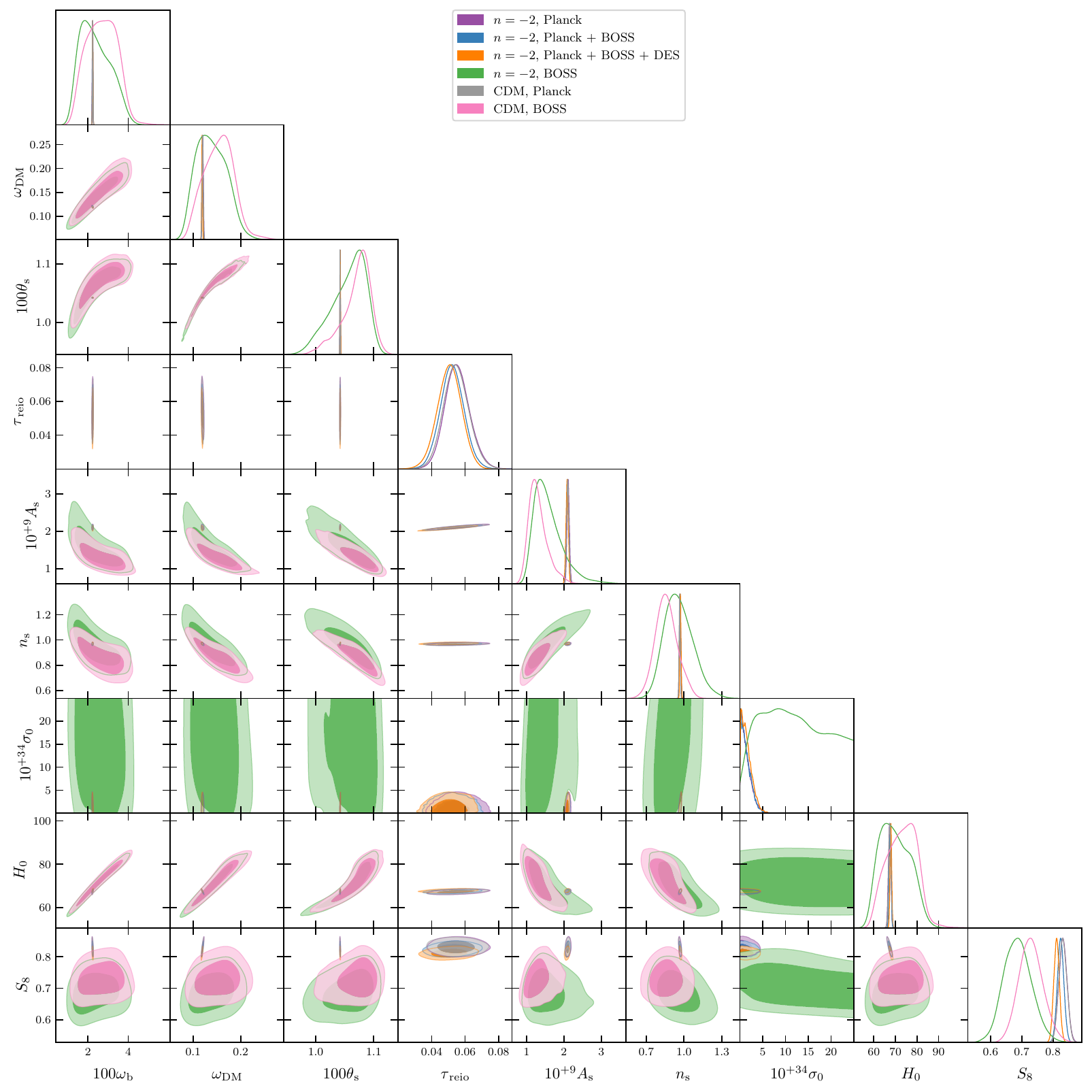}
\caption{\label{fig:n=-2 triangle plot BOSS} 68\% and 95\% confidence level marginalized posterior distributions for the $n=-2$ DM-baryon interacting model from different combinations of \textit{Planck}, BOSS, and DES data (colored), compared with posteriors for $\Lambda$CDM from \textit{Planck} (gray). Same plot as Fig.~\ref{fig:n=-2 triangle plot} but with BOSS posteriors added.}
\end{figure*}

We display full marginalized posterior distributions for all relevant parameters in our analysis of the $n=-2$, $f_\chi$ = 100\%, and $m_\chi$ = 1 MeV model in Fig.~\ref{fig:n=-2 triangle plot}. We show the same posterior distributions along with the BOSS + DES posteriors in Fig.~\ref{fig:n=-2 triangle plot BOSS + DES}, and the same posterior distributions along with the BOSS posteriors in Fig.~\ref{fig:n=-2 triangle plot BOSS}.

\clearpage

\subsection{$n=0$}

\begin{figure*}[!htb]
\includegraphics[scale=0.575]{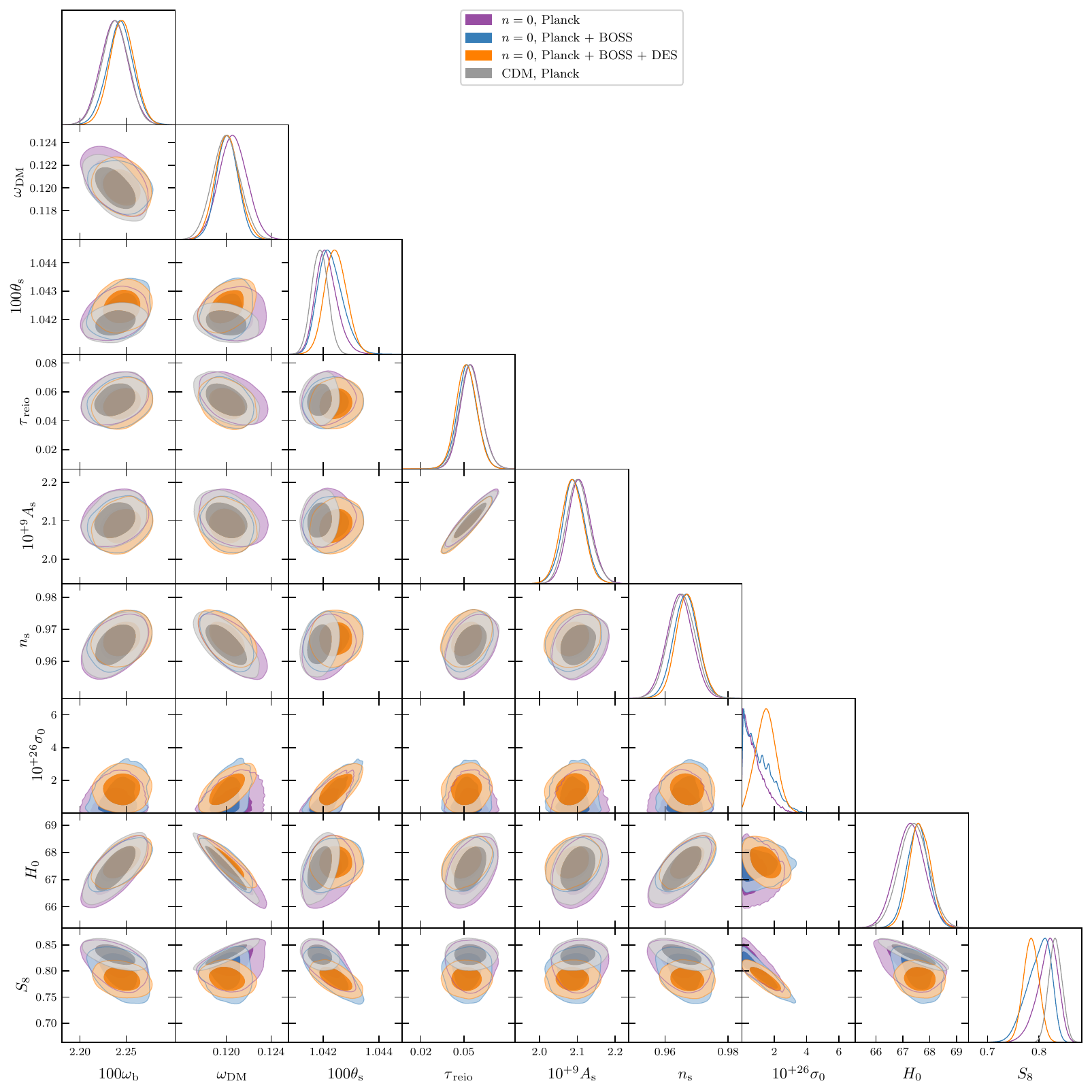}
\caption{\label{fig:n=0 triangle plot} 68\% and 95\% confidence level marginalized posterior distributions for the $n=0$ DM-baryon interacting model from different combinations of \textit{Planck}, BOSS, and DES data (colored), compared with posteriors for $\Lambda$CDM from \textit{Planck} (gray).}
\end{figure*}

\begin{figure*}[!htb]
\includegraphics[scale=0.575]{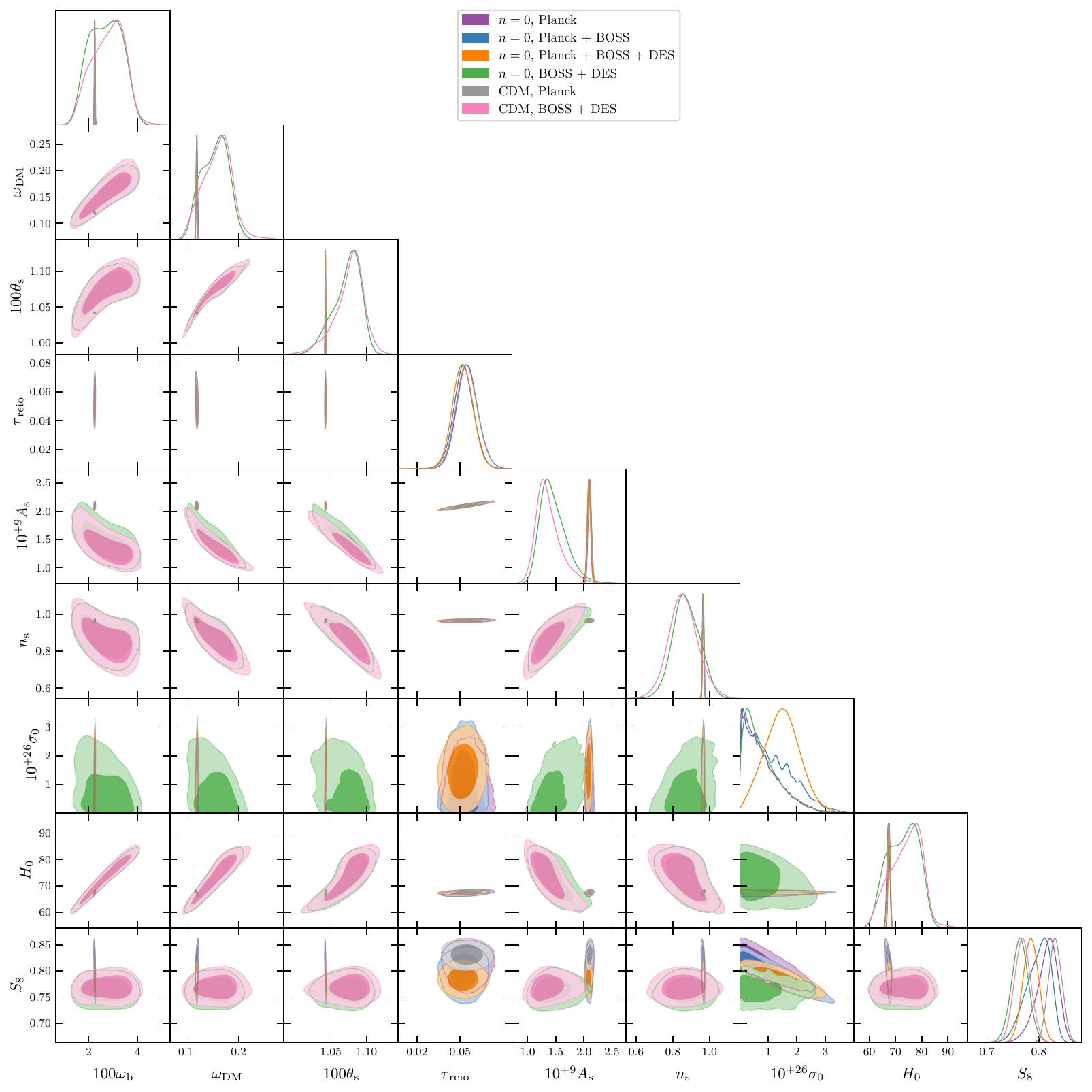}
\caption{\label{fig:n=0 triangle plot BOSS + DES} 68\% and 95\% confidence level marginalized posterior distributions for the $n=0$ DM-baryon interacting model from different combinations of \textit{Planck}, BOSS, and DES data (colored), compared with posteriors for $\Lambda$CDM from \textit{Planck} (gray). Same plot as Fig.~\ref{fig:n=0 triangle plot} but with BOSS + DES posteriors added.}
\end{figure*}

\begin{figure*}[!htb]
\includegraphics[scale=0.575]{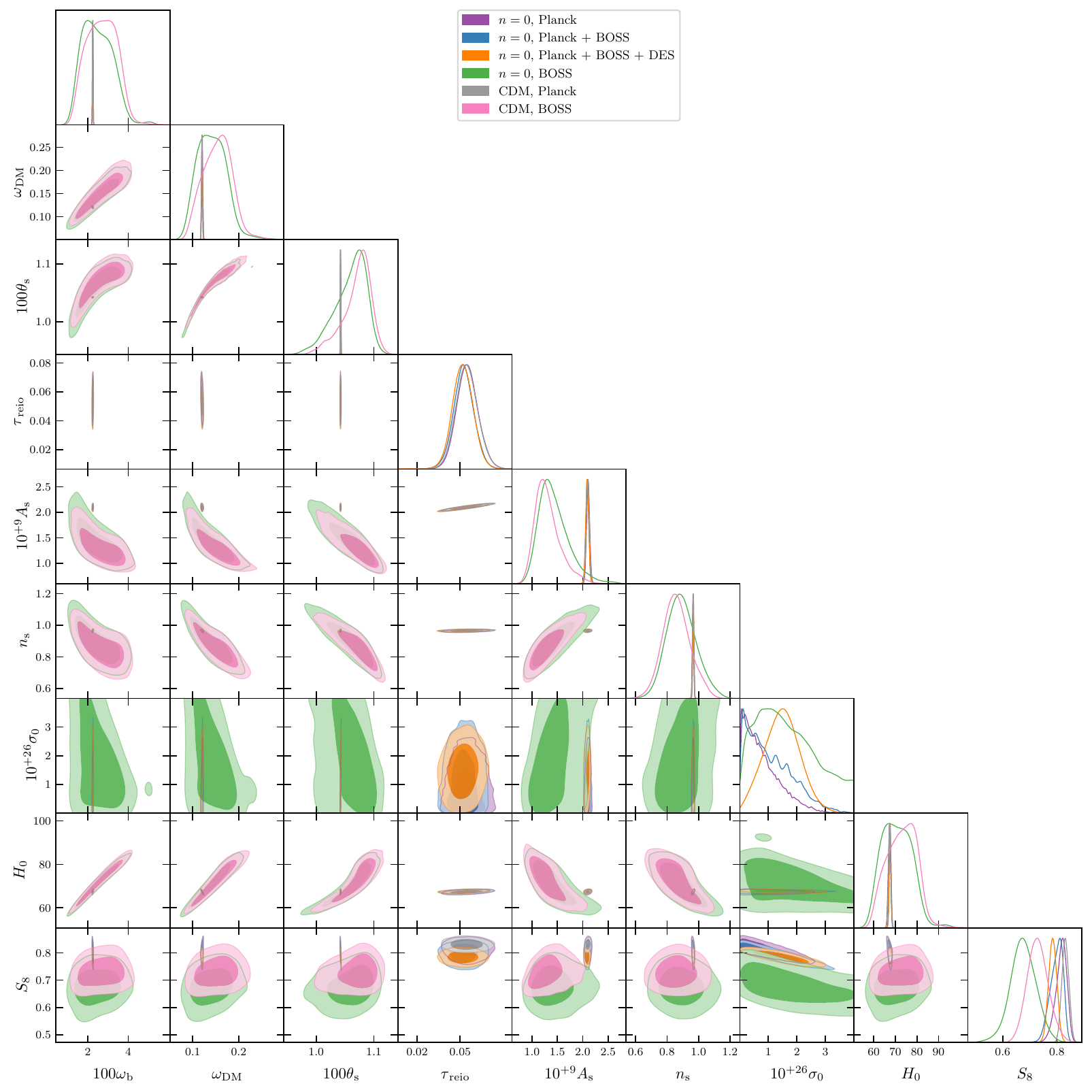}
\caption{\label{fig:n=0 triangle plot BOSS} 68\% and 95\% confidence level marginalized posterior distributions for the $n=0$ DM-baryon interacting model from different combinations of \textit{Planck}, BOSS, and DES data (colored), compared with posteriors for $\Lambda$CDM from \textit{Planck} (gray). Same plot as Fig.~\ref{fig:n=0 triangle plot} but with BOSS posteriors added.}
\end{figure*}

We display full marginalized posterior distributions for all relevant parameters in our analysis of the $n=0$, $f_\chi$ = 100\%, and $m_\chi$ = 1 MeV model in Fig.~\ref{fig:n=0 triangle plot}. We show the same posterior distributions along with the BOSS + DES posteriors in Fig.~\ref{fig:n=0 triangle plot BOSS + DES}, and the same posterior distributions along with the BOSS posteriors in Fig.~\ref{fig:n=0 triangle plot BOSS}.

\clearpage

\subsection{$n=2$}

\begin{figure*}[!htb]
\includegraphics[scale=0.575]{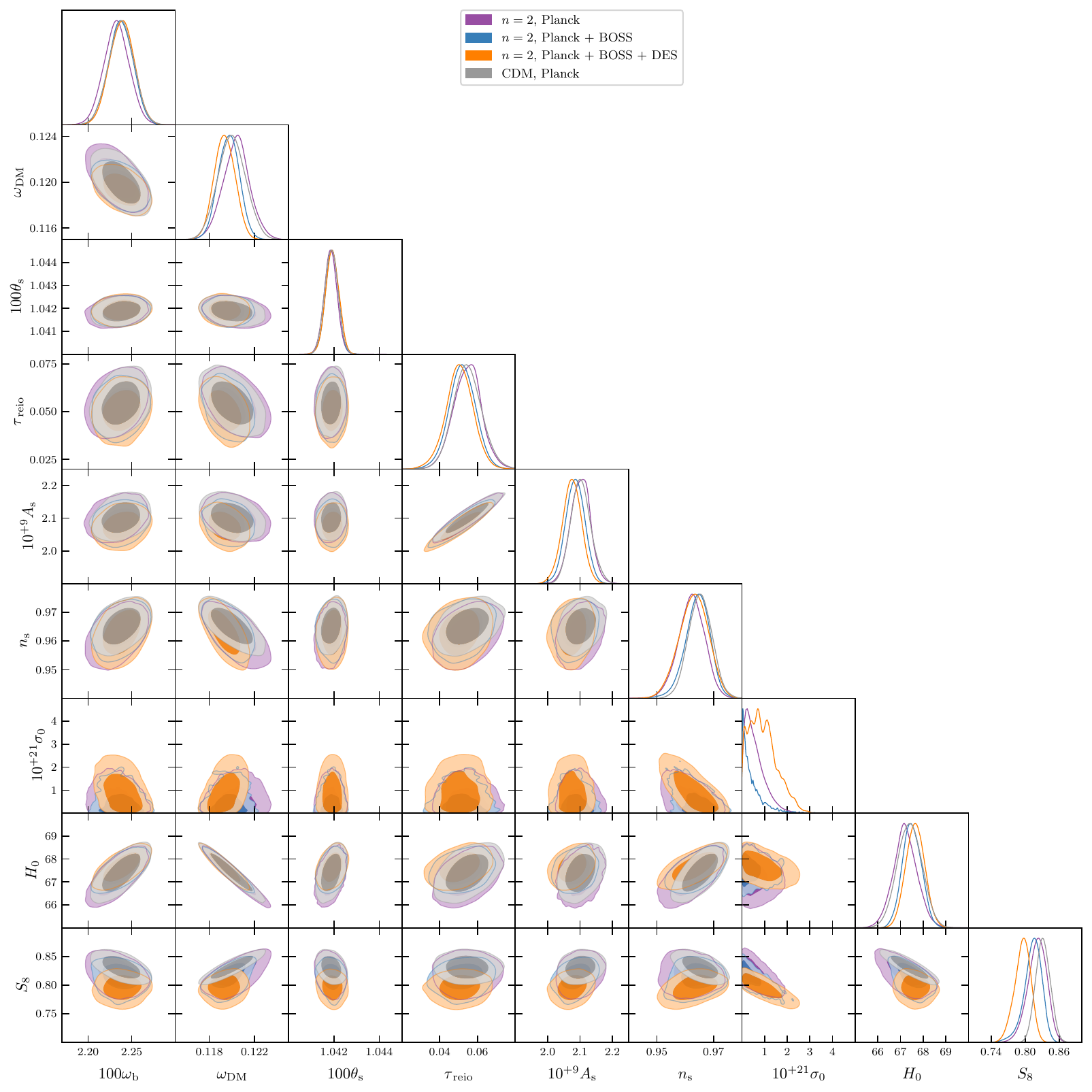}
\caption{\label{fig:n=2 triangle plot} 68\% and 95\% confidence level marginalized posterior distributions for the $n=2$ DM-baryon interacting model from different combinations of \textit{Planck}, BOSS, and DES data (colored), compared with posteriors for $\Lambda$CDM from \textit{Planck} (gray).}
\end{figure*}

\begin{figure*}[!htb]
\includegraphics[scale=0.575]{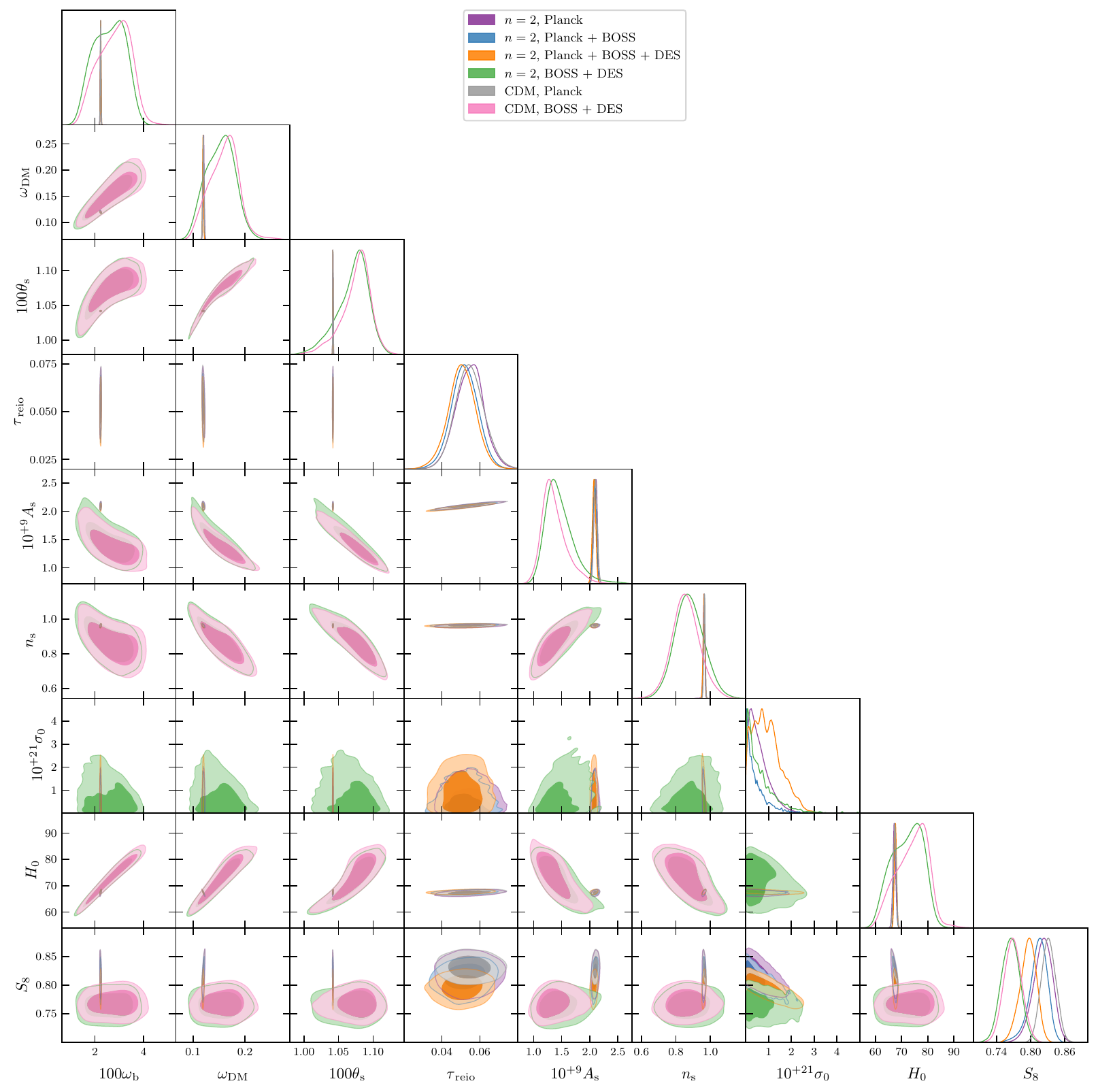}
\caption{\label{fig:n=2 triangle plot BOSS + DES} 68\% and 95\% confidence level marginalized posterior distributions for the $n=2$ DM-baryon interacting model from different combinations of \textit{Planck}, BOSS, and DES data (colored), compared with posteriors for $\Lambda$CDM from \textit{Planck} (gray). Same plot as Fig.~\ref{fig:n=2 triangle plot} but with BOSS + DES posteriors added.}
\end{figure*}

\begin{figure*}[!htb]
\includegraphics[scale=0.575]{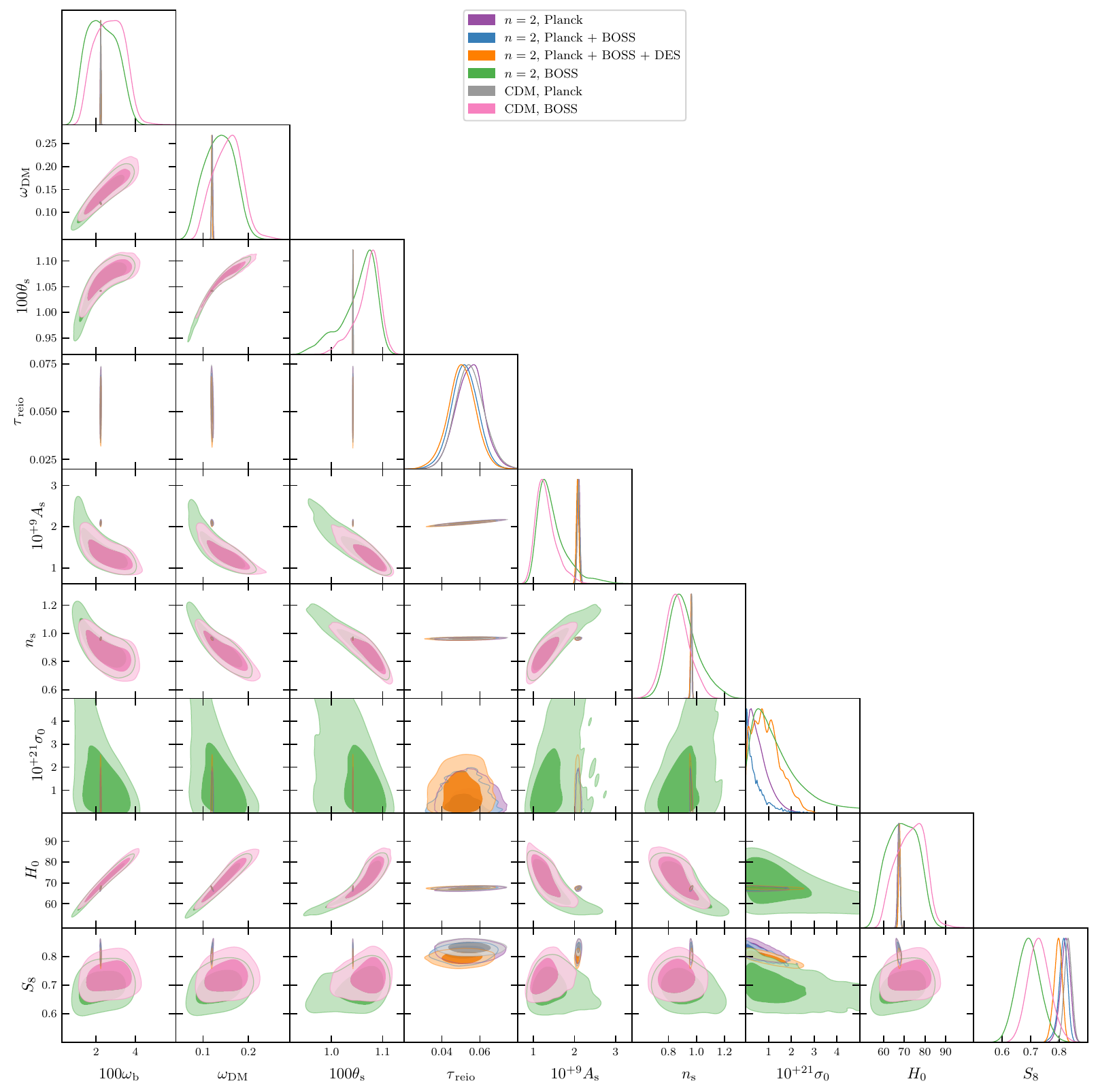}
\caption{\label{fig:n=2 triangle plot BOSS} 68\% and 95\% confidence level marginalized posterior distributions for the $n=2$ DM-baryon interacting model from different combinations of \textit{Planck}, BOSS, and DES data (colored), compared with posteriors for $\Lambda$CDM from \textit{Planck} (gray). Same plot as Fig.~\ref{fig:n=2 triangle plot} but with BOSS posteriors added.}
\end{figure*}

We display full marginalized posterior distributions for all relevant parameters in our analysis of the $n=2$, $f_\chi$ = 100\%, and $m_\chi$ = 1 MeV model in Fig.~\ref{fig:n=2 triangle plot}. We show the same posterior distributions along with the BOSS + DES posteriors in Fig.~\ref{fig:n=2 triangle plot BOSS + DES}, and the same posterior distributions along with the BOSS posteriors in Fig.~\ref{fig:n=2 triangle plot BOSS}.

\clearpage

\subsection{$n=4$}

\begin{figure*}[!htb]
\includegraphics[scale=0.575]{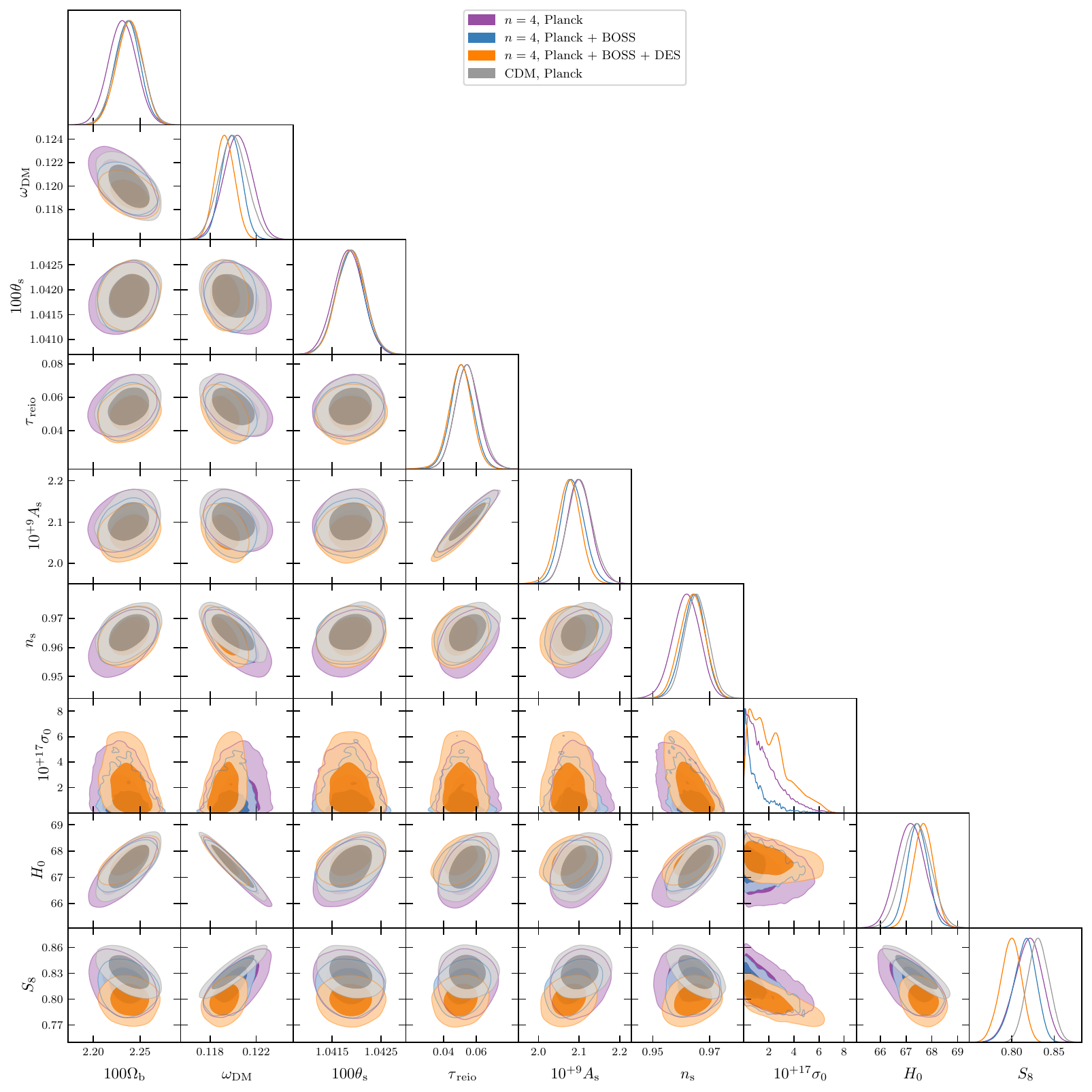}
\caption{\label{fig:n=4 triangle plot} 68\% and 95\% confidence level marginalized posterior distributions for the $n=4$ DM-baryon interacting model from different combinations of \textit{Planck}, BOSS, and DES data (colored), compared with posteriors for $\Lambda$CDM from \textit{Planck} (gray).}
\end{figure*}

\begin{figure*}[!htb]
\includegraphics[scale=0.575]{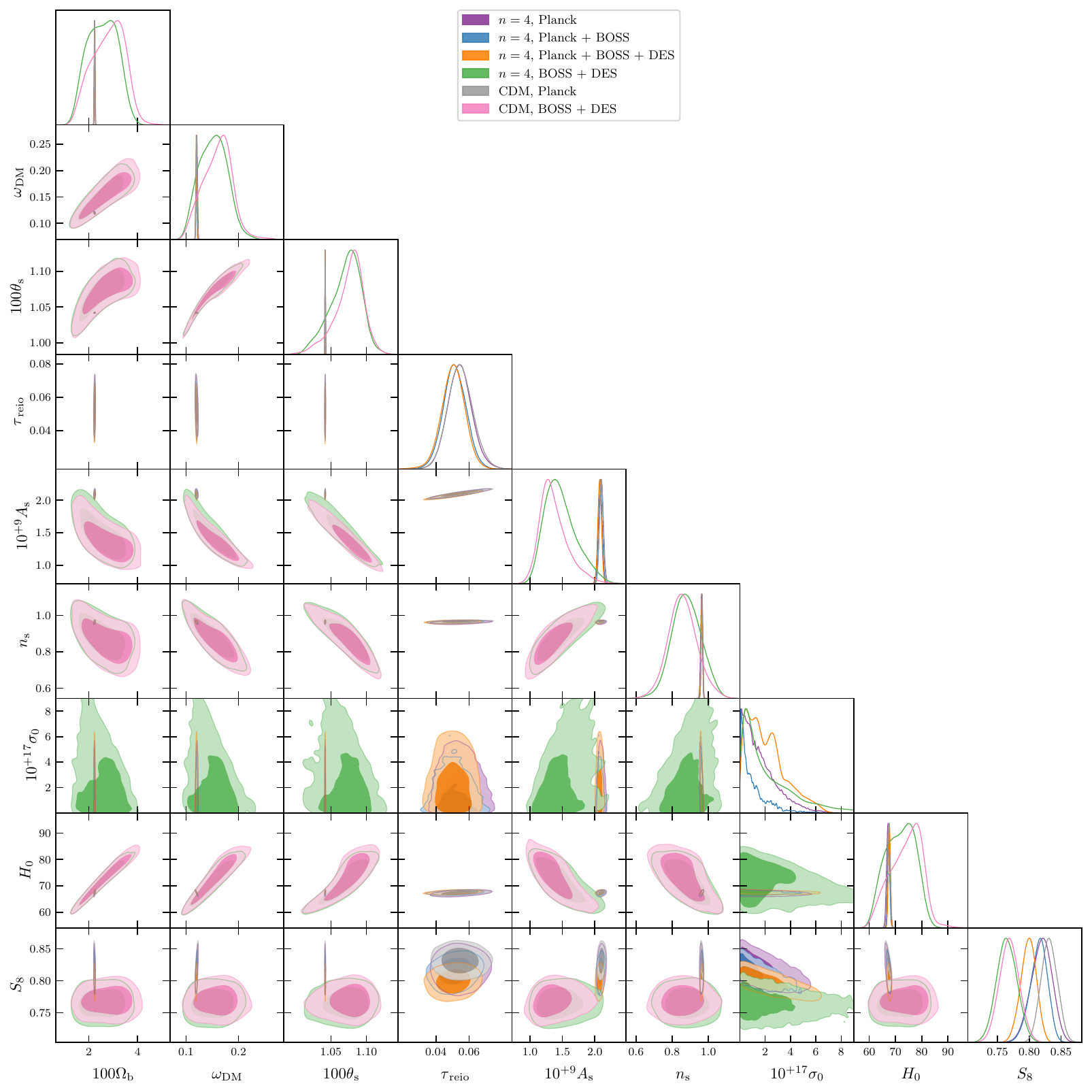}
\caption{\label{fig:n=4 triangle plot BOSS + DES} 68\% and 95\% confidence level marginalized posterior distributions for the $n=4$ DM-baryon interacting model from different combinations of \textit{Planck}, BOSS, and DES data (colored), compared with posteriors for $\Lambda$CDM from \textit{Planck} (gray). Same plot as Fig.~\ref{fig:n=4 triangle plot} but with BOSS + DES posteriors added.}
\end{figure*}

\begin{figure*}[!htb]
\includegraphics[scale=0.575]{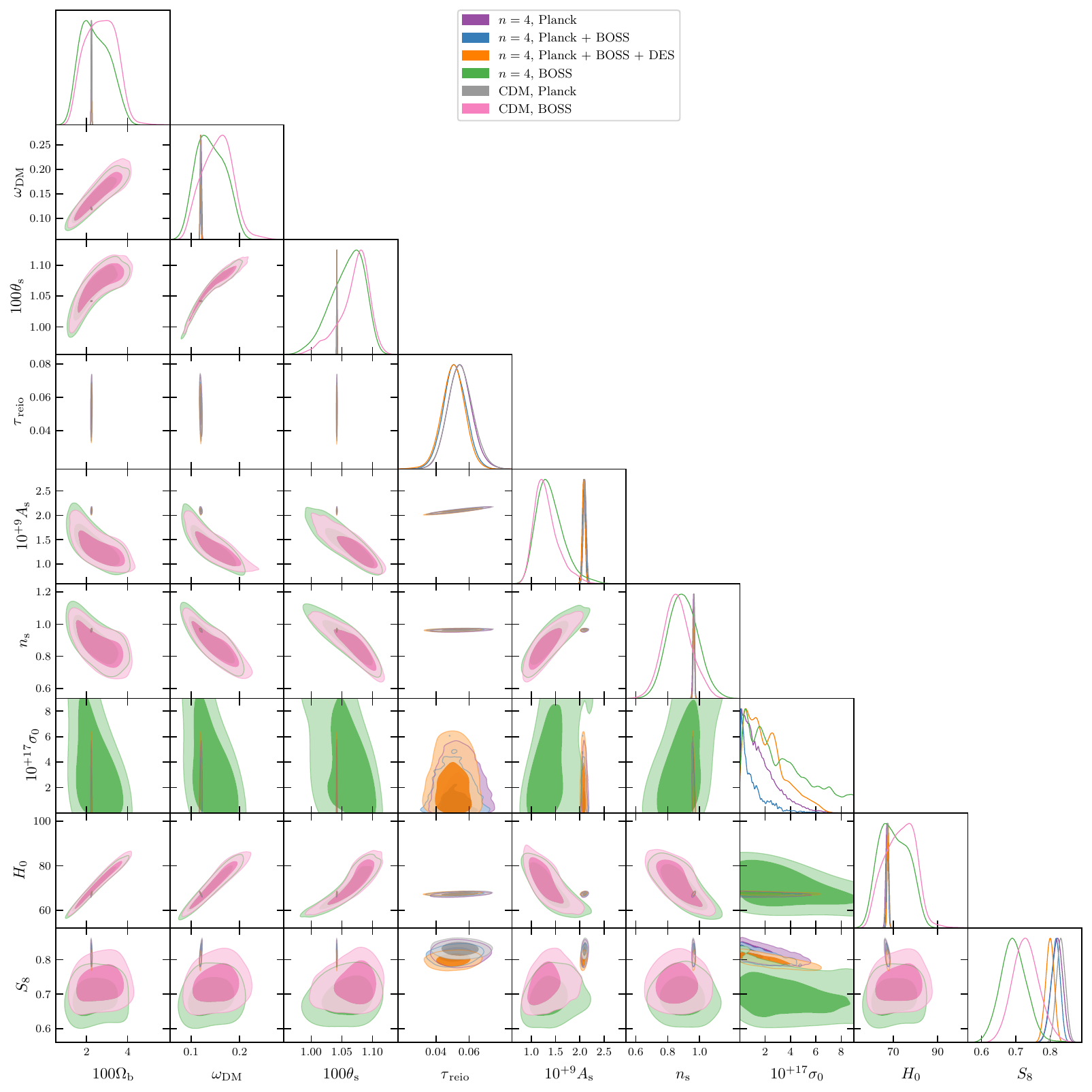}
\caption{\label{fig:n=4 triangle plot BOSS} 68\% and 95\% confidence level marginalized posterior distributions for the $n=4$ DM-baryon interacting model from different combinations of \textit{Planck}, BOSS, and DES data (colored), compared with posteriors for $\Lambda$CDM from \textit{Planck} (gray). Same plot as Fig.~\ref{fig:n=4 triangle plot} but with BOSS posteriors added.}
\end{figure*}

We display full marginalized posterior distributions for all relevant parameters in our analysis of the $n=4$, $f_\chi$ = 100\%, and $m_\chi$ = 1 MeV model in Fig.~\ref{fig:n=4 triangle plot}. We show the same posterior distributions along with the BOSS + DES posteriors in Fig.~\ref{fig:n=4 triangle plot BOSS + DES}, and the same posterior distributions along with the BOSS posteriors in Fig.~\ref{fig:n=4 triangle plot BOSS}.

\clearpage

\section{Posterior probability distributions for models in which a fraction of DM interacts}
\label{Appendix:posteriors_10}

\subsection{$n=-2$}

We display full marginalized posterior distributions for all relevant parameters in our analysis of the $n=-2$, $f_\chi$ = 10\%, and $m_\chi$ = 1 MeV, 1 GeV, and 10 GeV models in Fig.~\ref{fig:n=-2, m=1MeV triangle plot}, Fig.~\ref{fig:n=-2, m=1GeV triangle plot}, and Fig.~\ref{fig:n=-2, m=10 GeV triangle plot} respectively. 

\begin{figure*}[!htb]
\includegraphics[scale=0.575]{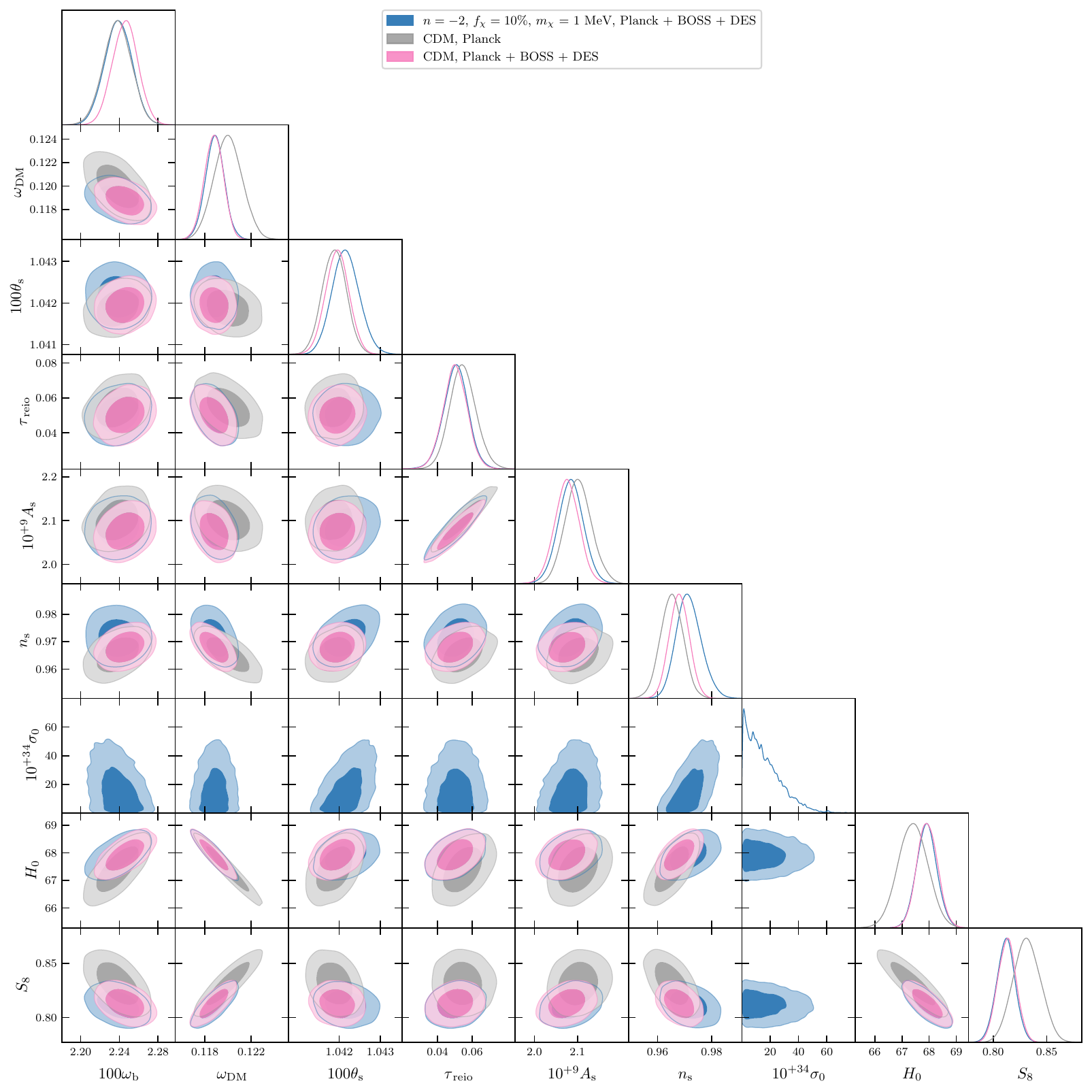}
\caption{\label{fig:n=-2, m=1MeV triangle plot} 68\% and 95\% confidence level marginalized posterior distributions for the $n=-2$, $f_\chi = 10\%$, and $m_\chi = 1$ MeV DM-baryon interacting model from a combined analysis of \textit{Planck}, BOSS, and DES data (blue), compared with posteriors for $\Lambda$CDM from a \textit{Planck}--only analysis (gray) and a combined \textit{Planck} + BOSS + DES analysis (pink).}
\end{figure*}

\begin{figure*}[!htb]
\includegraphics[scale=0.575]{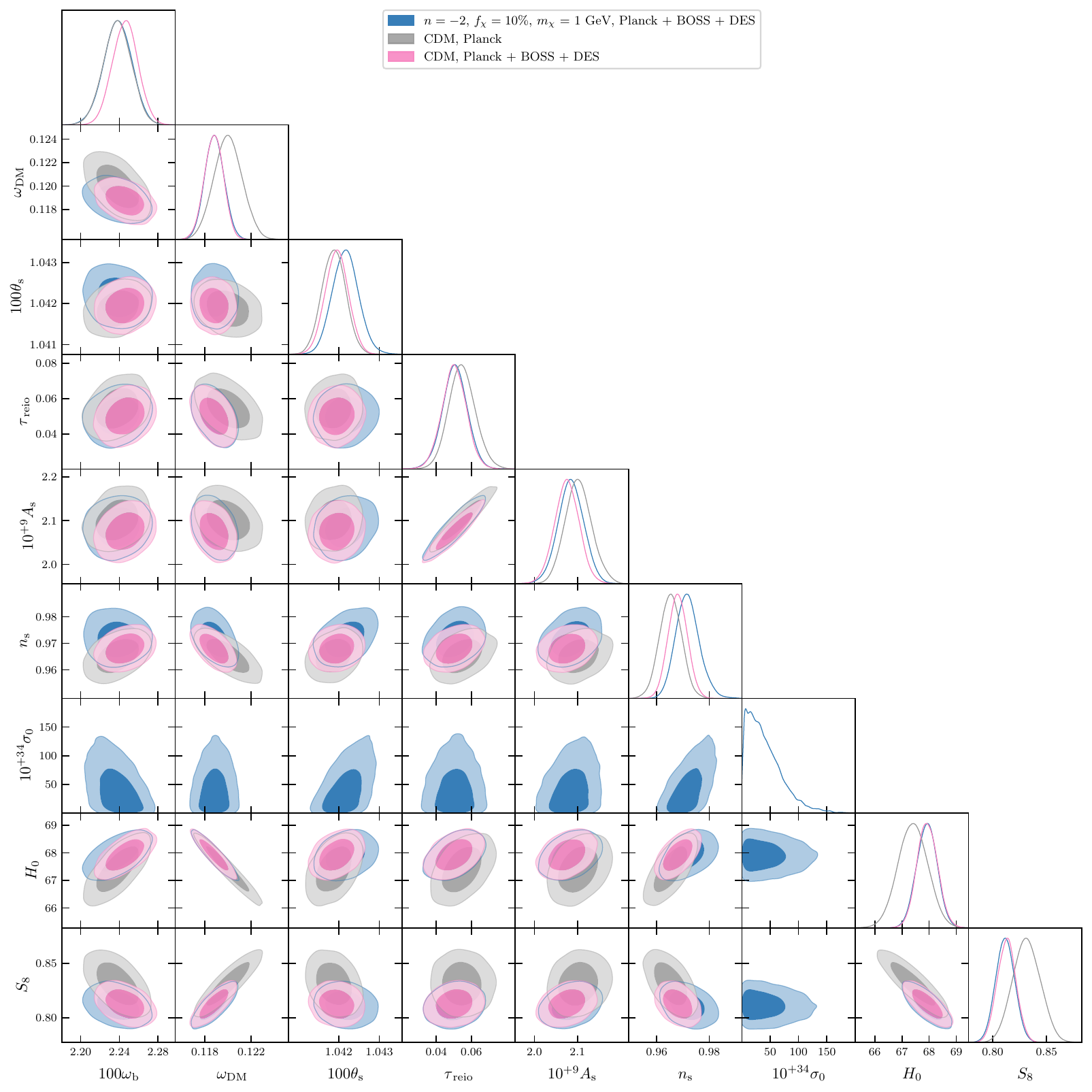}
\caption{\label{fig:n=-2, m=1GeV triangle plot} 68\% and 95\% confidence level marginalized posterior distributions for the $n=-2$, $f_\chi = 10\%$, and $m_\chi = 1$ GeV DM-baryon interacting model from a combined analysis of \textit{Planck}, BOSS, and DES data (blue), compared with posteriors for $\Lambda$CDM from a \textit{Planck}--only analysis (gray) and a combined \textit{Planck} + BOSS + DES analysis (pink).}
\end{figure*}

\begin{figure*}[!htb]
\includegraphics[scale=0.575]{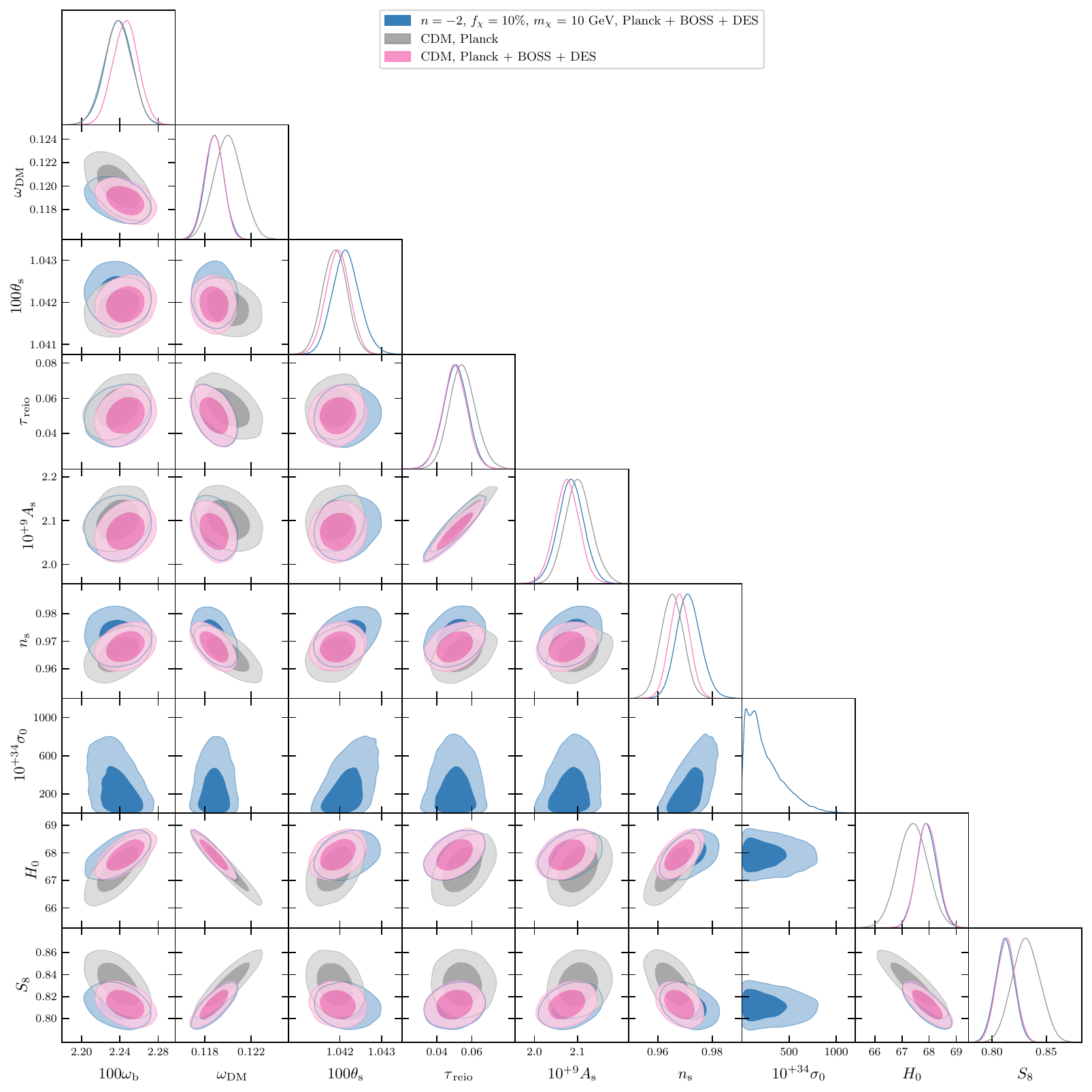}
\caption{\label{fig:n=-2, m=10 GeV triangle plot} 68\% and 95\% confidence level marginalized posterior distributions for the $n=-2$, $f_\chi = 10\%$, $m_\chi = 10$ GeV DM-baryon interacting model from a combined analysis of \textit{Planck}, BOSS, and DES data (blue), compared with posteriors for $\Lambda$CDM from a \textit{Planck}--only analysis (gray) and a combined \textit{Planck} + BOSS + DES analysis (pink).}
\end{figure*}

\clearpage

\subsection{$n=0$}

We display full marginalized posterior distributions for all relevant parameters in our analysis of the $n=0$, $f_\chi$ = 10\%, and $m_\chi$ = 1 MeV, 1 GeV, and 10 GeV models in Fig.~\ref{fig:n=0, m=1MeV triangle plot}, Fig.~\ref{fig:n=0, m=1GeV triangle plot}, and Fig.~\ref{fig:n=0, m=10 GeV triangle plot} respectively.

\begin{figure*}[!htb]
\includegraphics[scale=0.575]{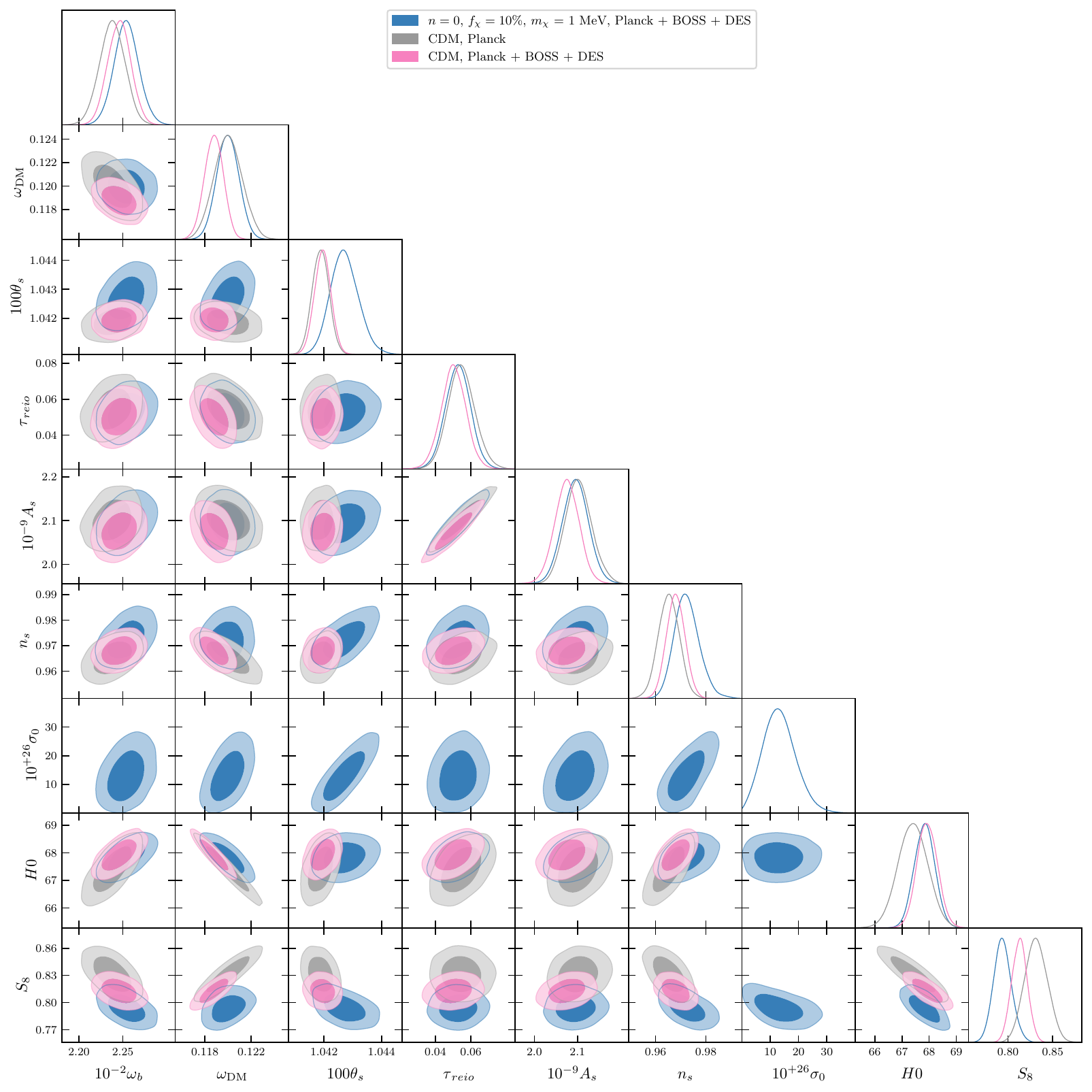}
\caption{\label{fig:n=0, m=1MeV triangle plot} 68\% and 95\% confidence level marginalized posterior distributions for the $n=0$, $f_\chi = 10\%$, and $m_\chi = 1$ MeV DM-baryon interacting model from a combined analysis of \textit{Planck}, BOSS, and DES data (blue), compared with posteriors for $\Lambda$CDM from a \textit{Planck}--only analysis (gray) and a combined \textit{Planck} + BOSS + DES analysis (pink).}
\end{figure*}

\begin{figure*}[!htb]
\includegraphics[scale=0.575]{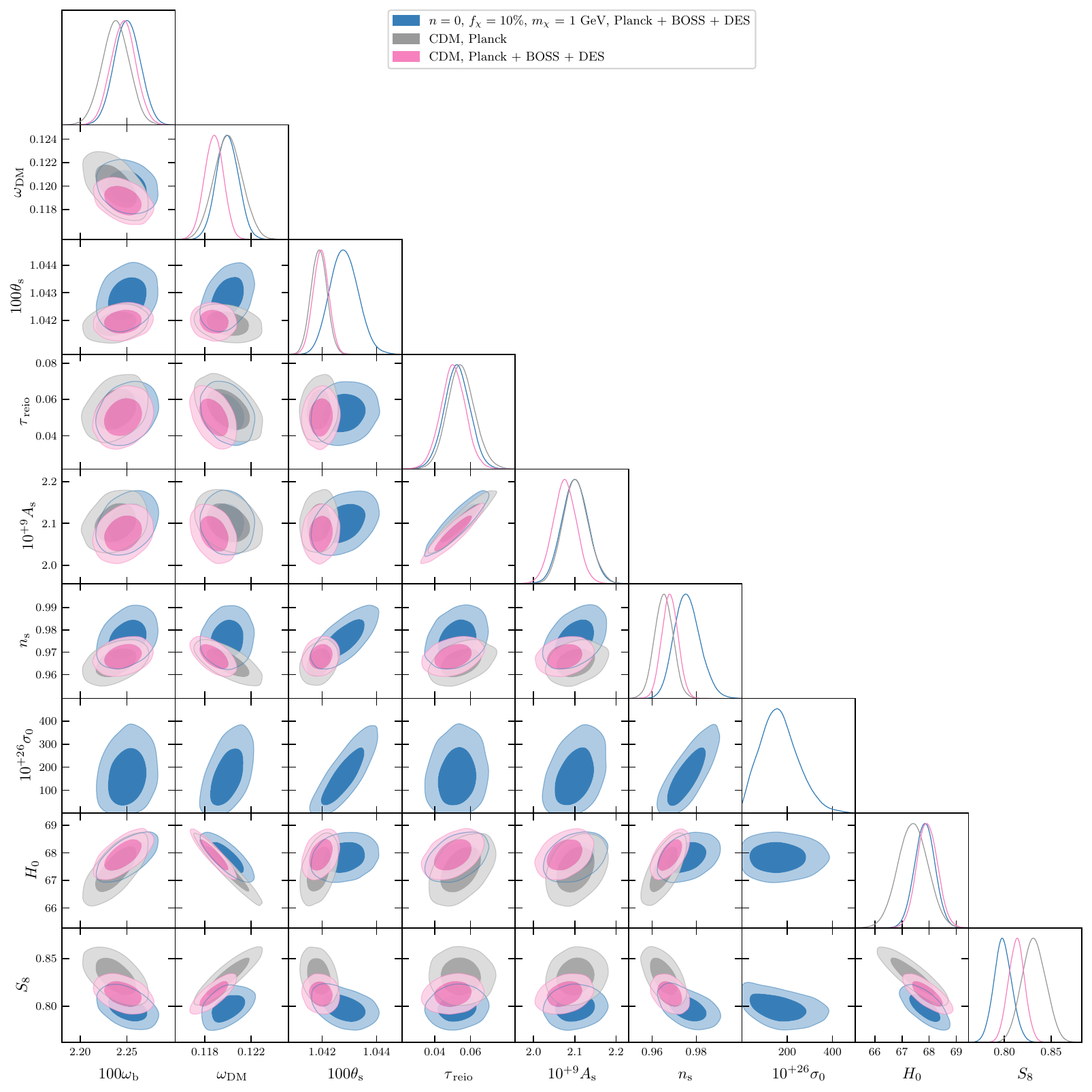}
\caption{\label{fig:n=0, m=1GeV triangle plot} 68\% and 95\% confidence level marginalized posterior distributions for the $n=0$, $f_\chi = 10\%$, and $m_\chi = 1$ GeV DM-baryon interacting model from a combined analysis of \textit{Planck}, BOSS, and DES data (blue), compared with posteriors for $\Lambda$CDM from a \textit{Planck}--only analysis (gray) and a combined \textit{Planck} + BOSS + DES analysis (pink).}
\end{figure*}

\begin{figure*}[!htb]
\includegraphics[scale=0.575]{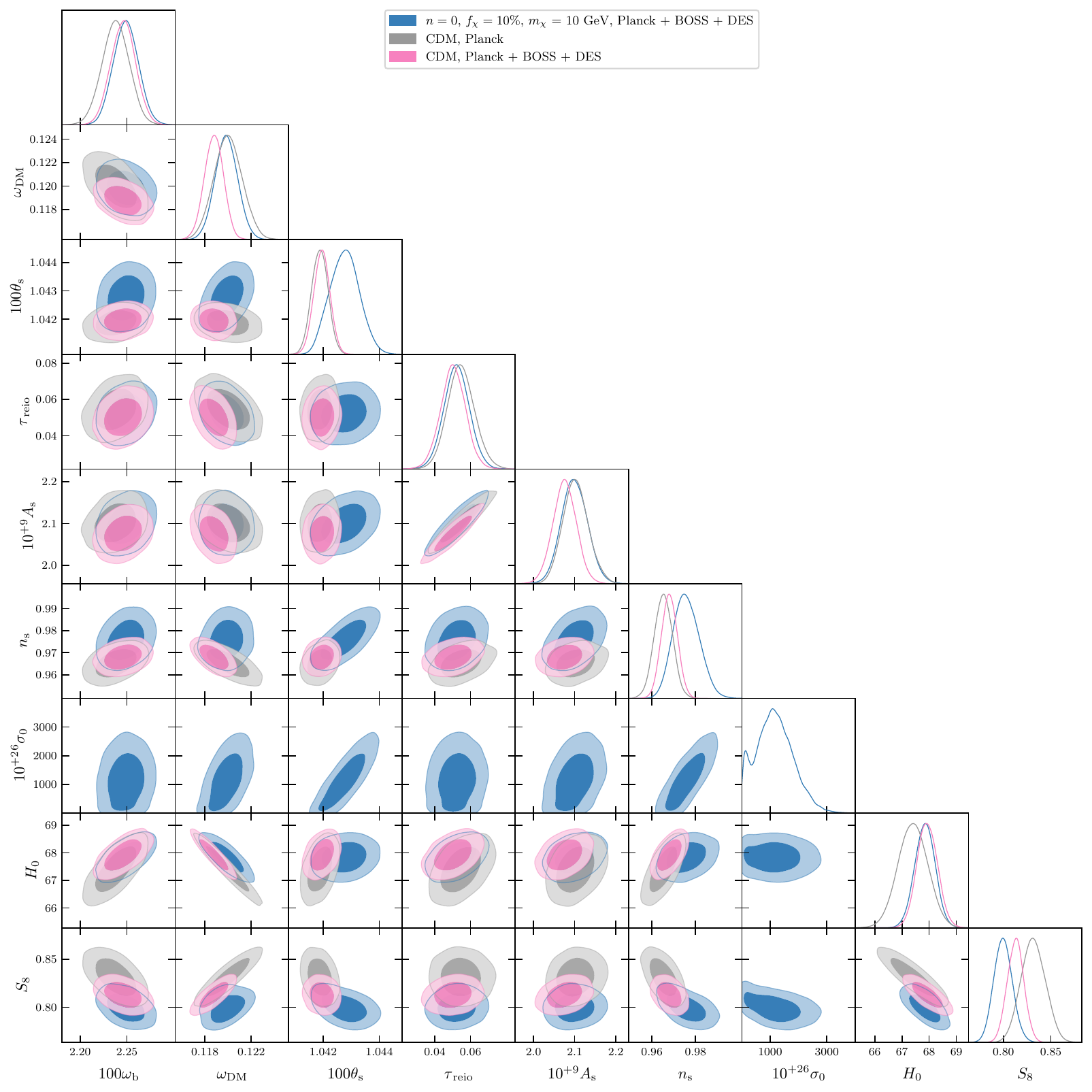}
\caption{\label{fig:n=0, m=10 GeV triangle plot} 68\% and 95\% confidence level marginalized posterior distributions for the $n=0$, $f_\chi = 10\%$, $m_\chi = 10$ GeV DM-baryon interacting model from a combined analysis of \textit{Planck}, BOSS, and DES data (blue), compared with posteriors for $\Lambda$CDM from a \textit{Planck}--only analysis (gray) and a combined \textit{Planck} + BOSS + DES analysis (pink).}
\end{figure*}

\clearpage

\subsection{$n=2$}

We display full marginalized posterior distributions for all relevant parameters in our analysis of the $n=2$, $f_\chi$ = 10\%, and $m_\chi$ = 1 MeV, 1 GeV, and 10 GeV models in Fig.~\ref{fig:n=2, m=1MeV triangle plot}, Fig.~\ref{fig:n=2, m=1GeV triangle plot}, and Fig.~\ref{fig:n=2, m=10 GeV triangle plot} respectively. 

\begin{figure*}[!htb]
\includegraphics[scale=0.575]{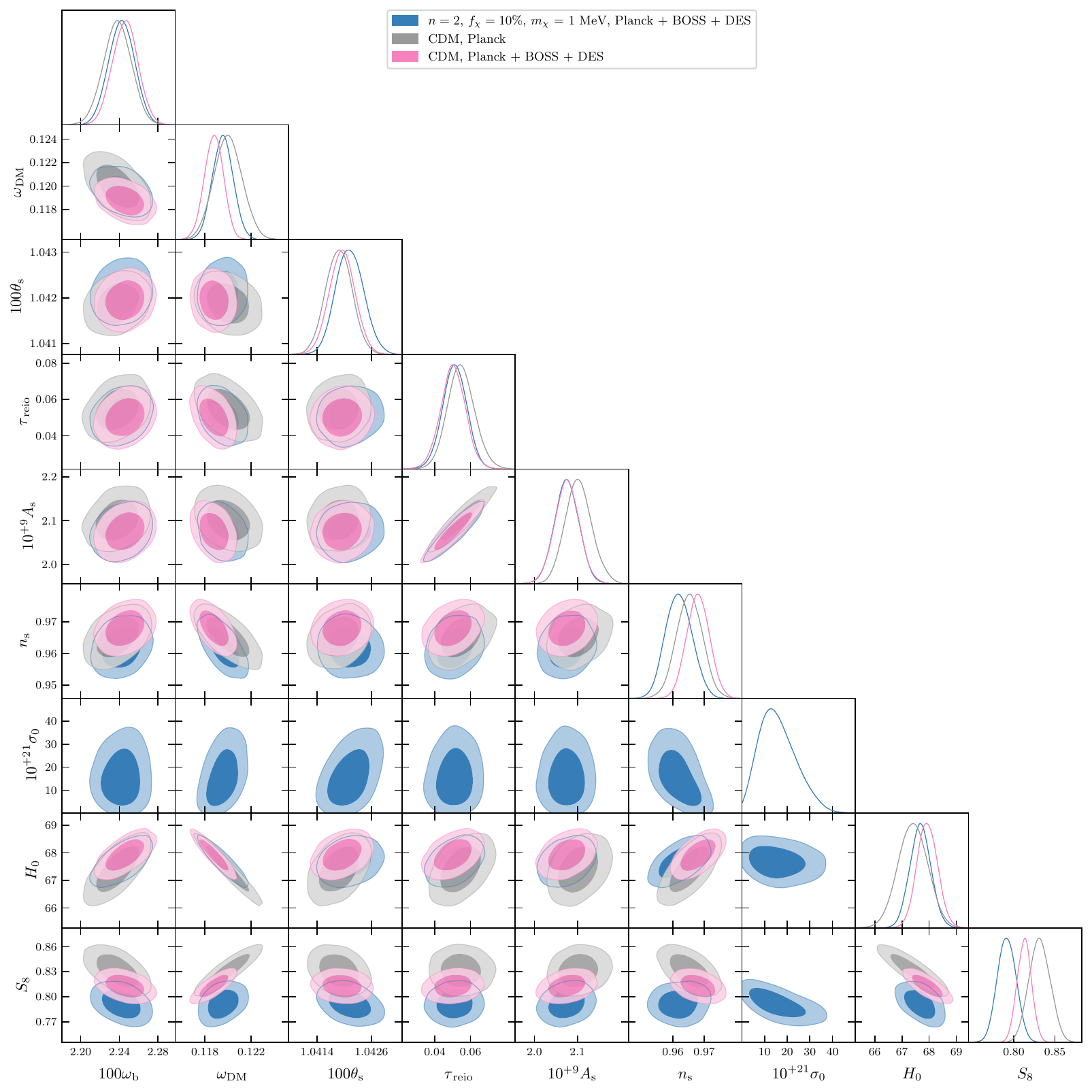}
\caption{\label{fig:n=2, m=1MeV triangle plot} 68\% and 95\% confidence level marginalized posterior distributions for the $n=2$, $f_\chi = 10\%$, and $m_\chi = 1$ MeV DM-baryon interacting model from a combined analysis of \textit{Planck}, BOSS, and DES data (blue), compared with posteriors for $\Lambda$CDM from a \textit{Planck}--only analysis (gray) and a combined \textit{Planck} + BOSS + DES analysis (pink).}
\end{figure*}

\begin{figure*}[!htb]
\includegraphics[scale=0.575]{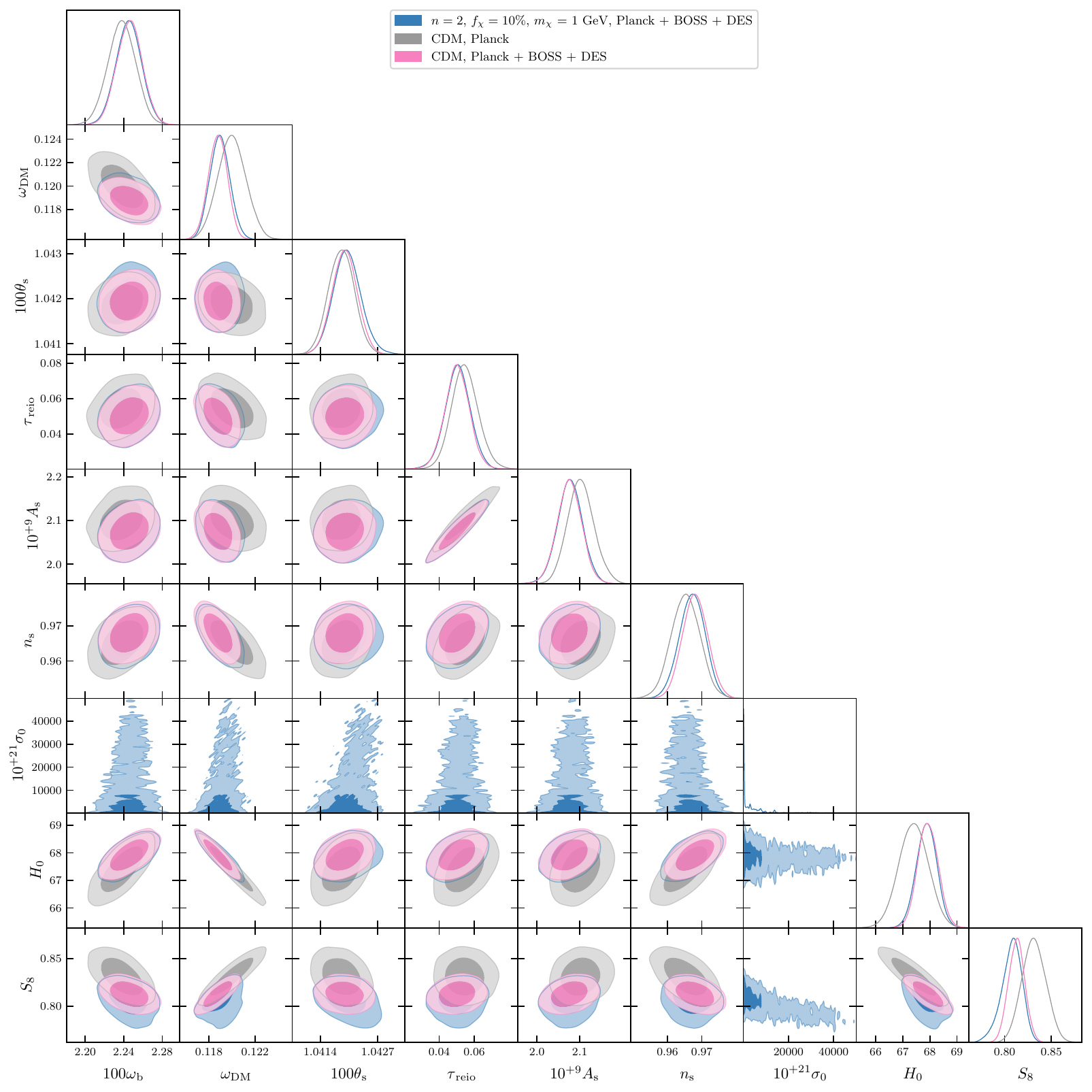}
\caption{\label{fig:n=2, m=1GeV triangle plot} 68\% and 95\% confidence level marginalized posterior distributions for the $n=2$, $f_\chi = 10\%$, and $m_\chi = 1$ GeV DM-baryon interacting model from a combined analysis of \textit{Planck}, BOSS, and DES data (blue), compared with posteriors for $\Lambda$CDM from a \textit{Planck}--only analysis (gray) and a combined \textit{Planck} + BOSS + DES analysis (pink).}
\end{figure*}

\begin{figure*}[!htb]
\includegraphics[scale=0.575]{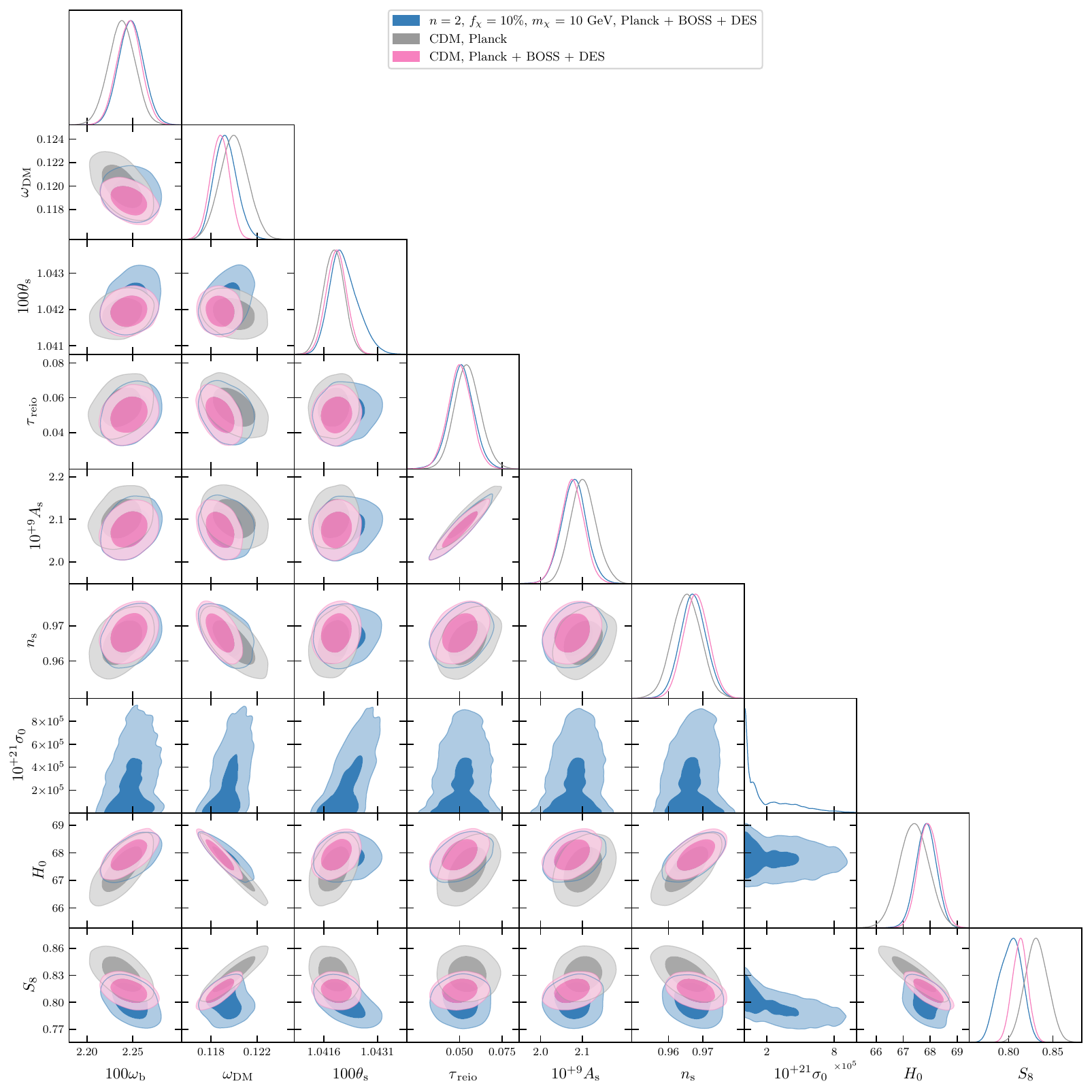}
\caption{\label{fig:n=2, m=10 GeV triangle plot} 68\% and 95\% confidence level marginalized posterior distributions for the $n=2$, $f_\chi = 10\%$, $m_\chi = 10$ GeV DM-baryon interacting model from a combined analysis of \textit{Planck}, BOSS, and DES data (blue), compared with posteriors for $\Lambda$CDM from a \textit{Planck}--only analysis (gray) and a combined \textit{Planck} + BOSS + DES analysis (pink).}
\end{figure*}

\clearpage

\subsection{$n=4$}

We display full marginalized posterior distributions for all relevant parameters in our analysis of the $n=4$, $f_\chi$ = 10\%, and $m_\chi$ = 1 MeV, 1 GeV, and 10 GeV models in Fig.~\ref{fig:n=4, m=1MeV triangle plot}, Fig.~\ref{fig:n=4, m=1GeV triangle plot}, and Fig.~\ref{fig:n=4, m=10 GeV triangle plot} respectively.

\begin{figure*}[!htb]
\includegraphics[scale=0.575]{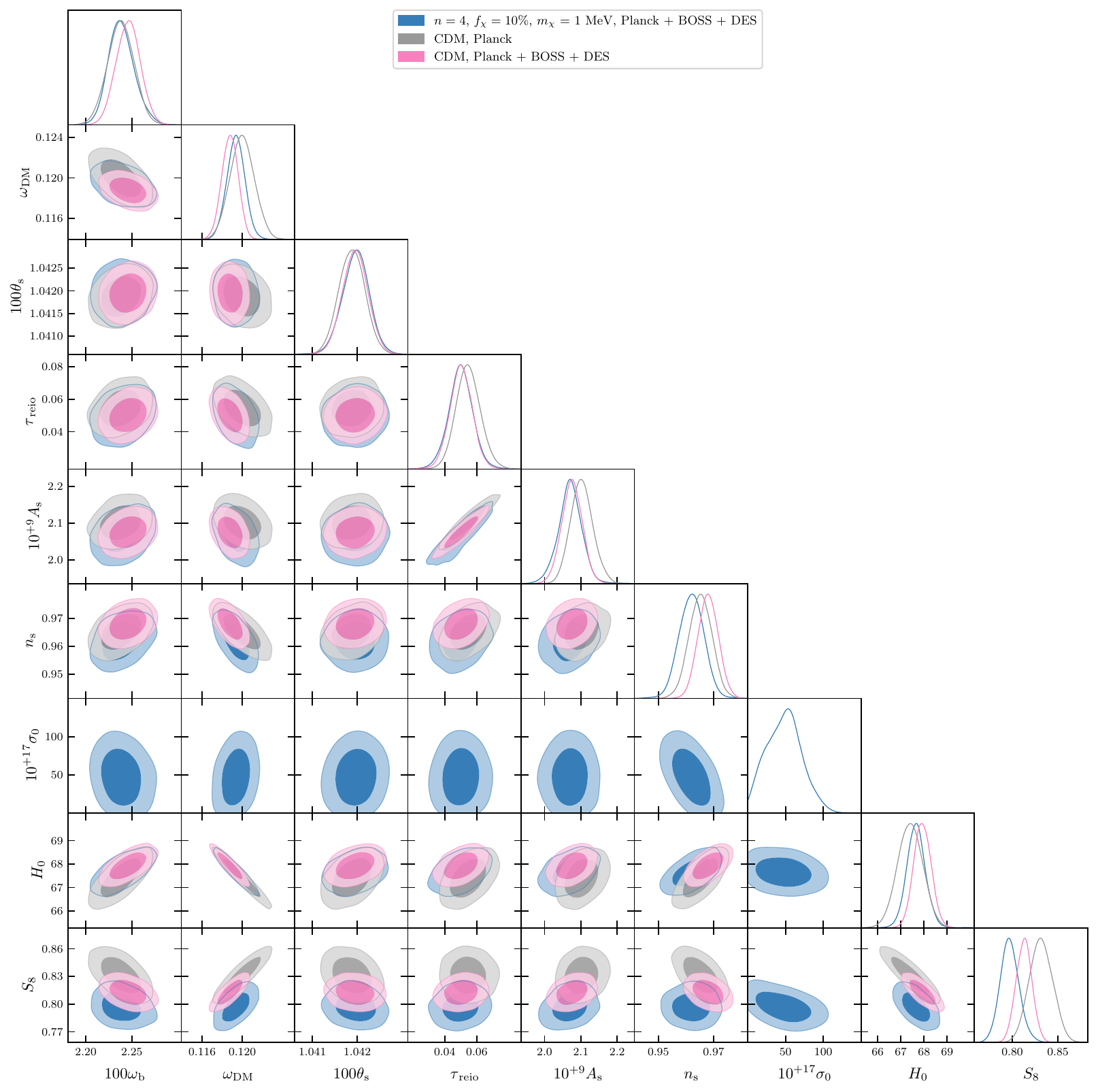}
\caption{\label{fig:n=4, m=1MeV triangle plot} 68\% and 95\% confidence level marginalized posterior distributions for the $n=4$, $f_\chi = 10\%$, and $m_\chi = 1$ MeV DM-baryon interacting model from a combined analysis of \textit{Planck}, BOSS, and DES data (blue), compared with posteriors for $\Lambda$CDM from a \textit{Planck}--only analysis (gray) and a combined \textit{Planck} + BOSS + DES analysis (pink).}
\end{figure*}

\begin{figure*}[!htb]
\includegraphics[scale=0.575]{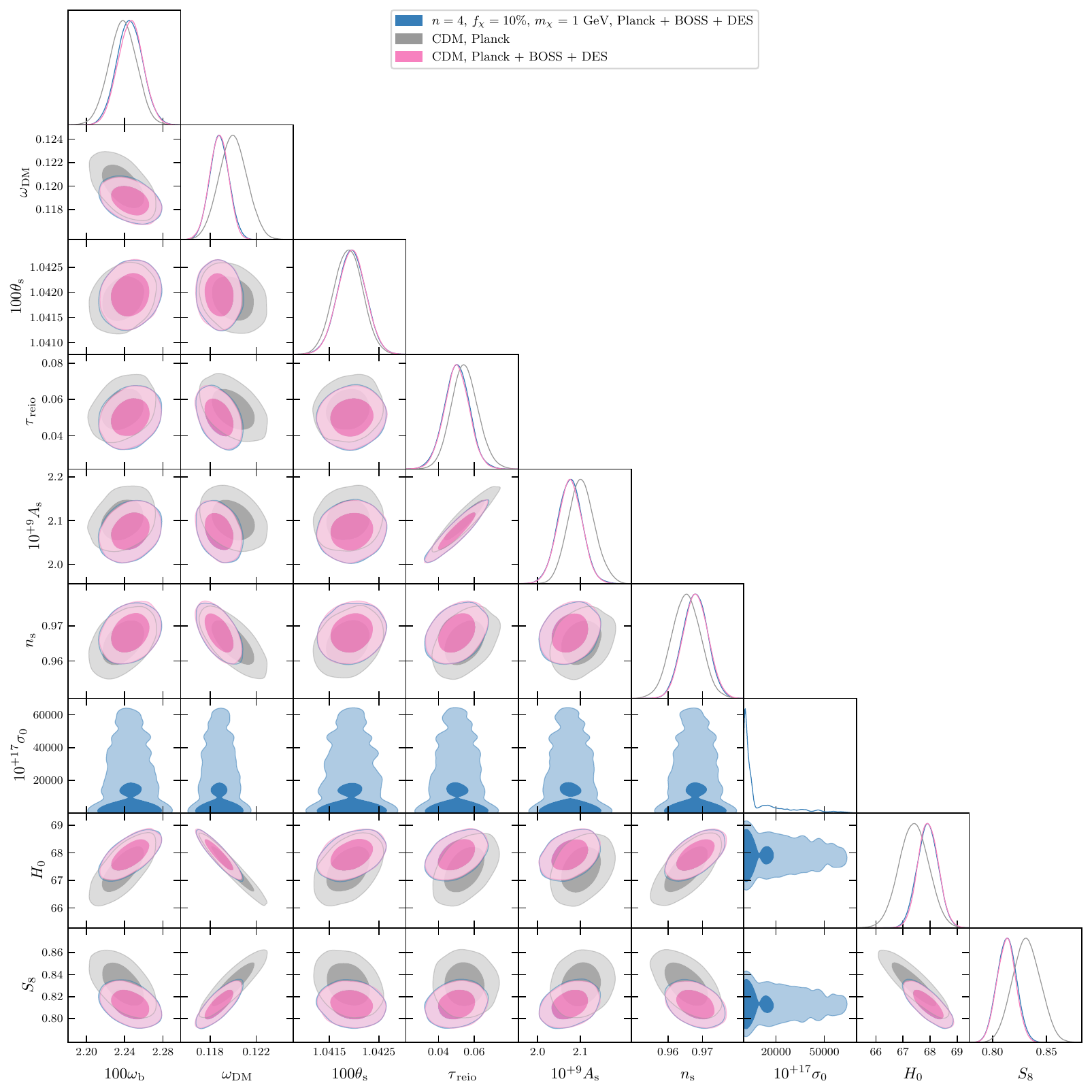}
\caption{\label{fig:n=4, m=1GeV triangle plot} 68\% and 95\% confidence level marginalized posterior distributions for the $n=4$, $f_\chi = 10\%$, and $m_\chi = 1$ GeV DM-baryon interacting model from a combined analysis of \textit{Planck}, BOSS, and DES data (blue), compared with posteriors for $\Lambda$CDM from a \textit{Planck}--only analysis (gray) and a combined \textit{Planck} + BOSS + DES analysis (pink).}
\end{figure*}

\begin{figure*}[!htb]
\includegraphics[scale=0.575]{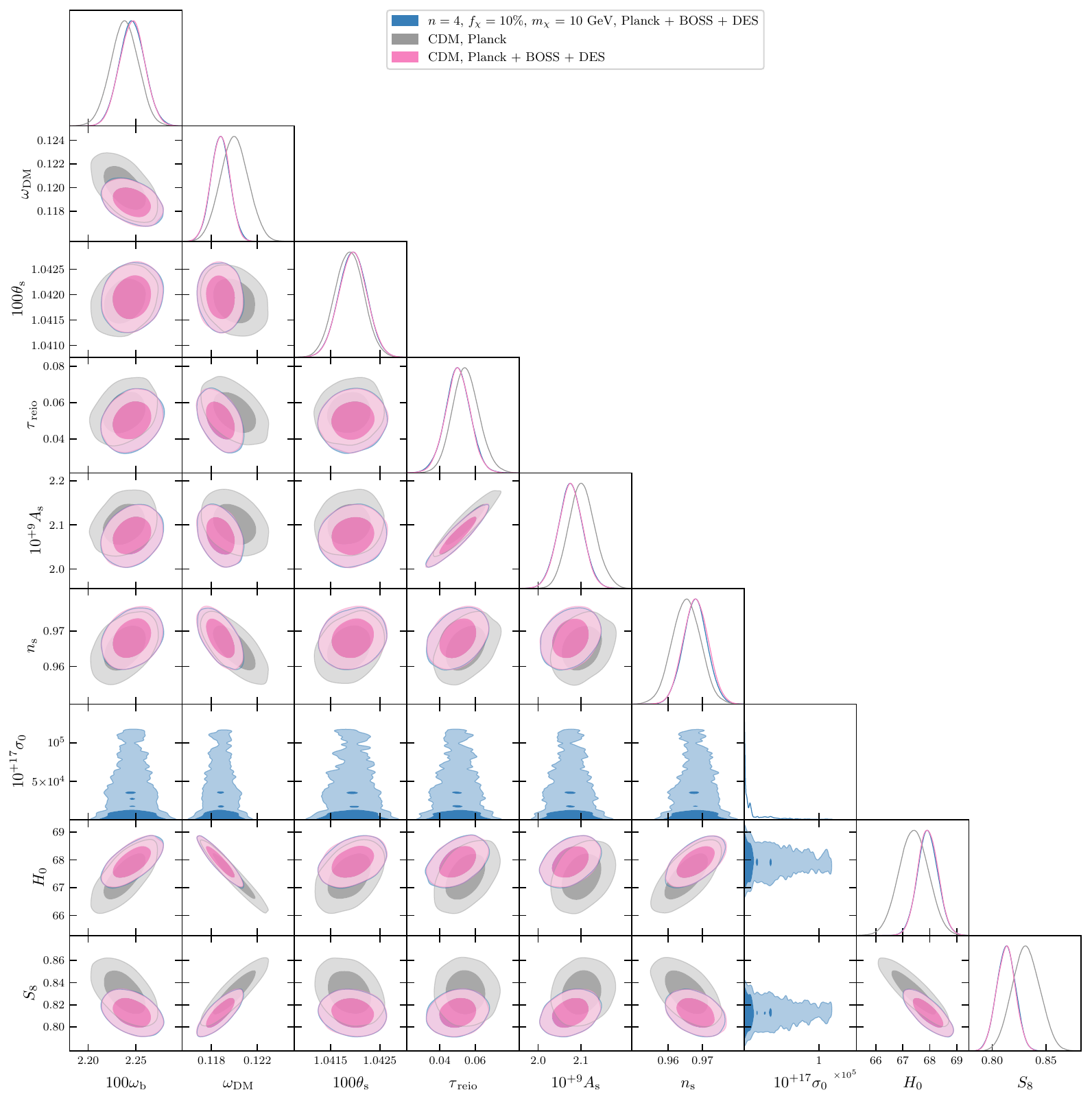}
\caption{\label{fig:n=4, m=10 GeV triangle plot} 68\% and 95\% confidence level marginalized posterior distributions for the $n=4$, $f_\chi = 10\%$, $m_\chi = 10$ GeV DM-baryon interacting model from a combined analysis of \textit{Planck}, BOSS, and DES data (blue), compared with posteriors for $\Lambda$CDM from a \textit{Planck}--only analysis (gray) and a combined \textit{Planck} + BOSS + DES analysis (pink).}
\end{figure*}

\clearpage

\section{Bias parameters}
\label{Appendix:bias}

\begin{figure*}[!htb]
\includegraphics[scale=0.325]{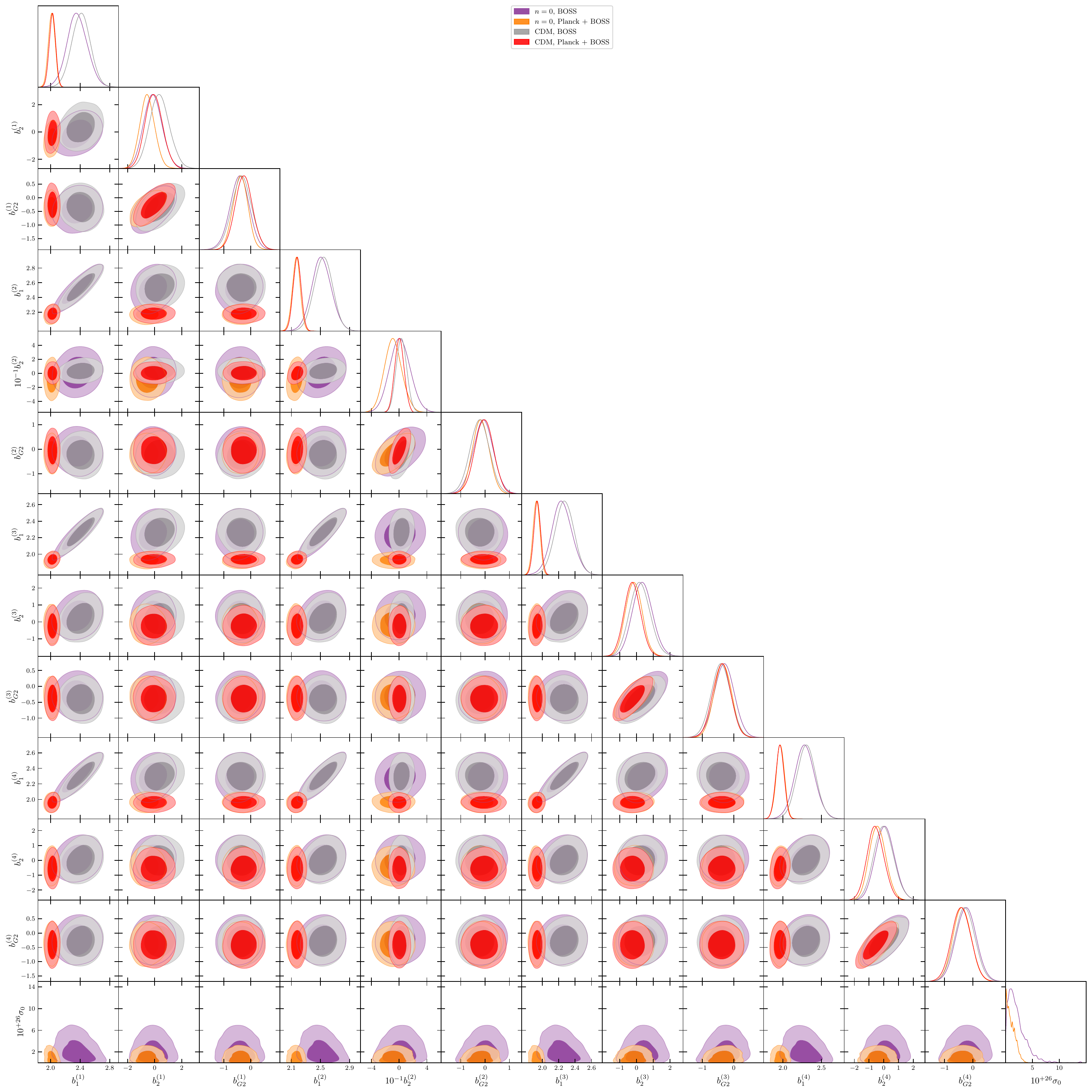}
\caption{\label{fig:n=0 bias parameters} Constraints on EFT bias parameters for the $n=0$, $f_\chi = 100\%$, $m_\chi = 1$ MeV DM-baryon interacting model from different combinations of \textit{Planck} and BOSS data, compared with constraints on EFT bias parameters for $\Lambda$CDM from these data.}
\end{figure*}

We show full marginalized posterior distributions for all EFT bias parameters in our analysis of the $n=0$, $f_\chi$ = 100\%, $m_\chi$ = 1 MeV model in Fig.~\ref{fig:n=0 bias parameters}. We do not show posterior distributions for all other powers of $n$, since we find that the posteriors for the EFT bias parameters are qualitatively the same for all $n$. We also find that the best-fit values for all EFT bias parameters in our analysis are similar to those found in $\Lambda$CDM.

\clearpage

\section{Tight-coupling approximation}
\label{Appendix:tight-coupling}

For positive powers of $n$, baryons and DM are coupled so strongly at early times that they experience a "baryon-DM slip", similar to the well-known photon-baryon slip. Thus, a tight-coupling approximation (TCA) must be implemented for these models so that \texttt{CLASS} may quickly and accurately evolve perturbations at high redshifts. We implement the TCA in a manner similar to \cite{Blas_2011}, starting with the last two formulas in Eq.~\ref{boltzmann}:
\begin{equation}\label{TCA1}
    \begin{split}
        \dot{\theta}_{\chi} +\frac{\dot{a}}{a}\theta_{\chi}-c^{2}_{\chi}k^{2}&\delta_{\chi} + R_{\chi}\Theta_{\chi \mathrm{b}} = 0, \\
        -\dot{\theta}_{\mathrm{b}} -\frac{\dot{a}}{a}\theta_{\mathrm{b}}+c^{2}_{\mathrm{b}}k^{2}\delta_{\mathrm{b}} + \frac{\rho_{\chi}}{\rho_{\mathrm{b}}}&R_{\chi}\Theta_{\chi \mathrm{b}}+ R_{\gamma}\left(\theta_{\gamma}-\theta_{\mathrm{b}}\right) = 0
    \end{split}
\end{equation}

where we have introduced the coupling term $\Theta_{\chi \mathrm{b}}=\theta_{\chi}-\theta_{\mathrm{b}}$. Adding these two equations, we get:
\begin{equation}\label{TCA2}
        \dot{\Theta}_{\chi \mathrm{b}} +\frac{\dot{a}}{a}\Theta_{\chi \mathrm{b}}+k^2\left(c^{2}_{\mathrm{b}}\delta_{\mathrm{b}} - c^{2}_{\chi}\delta_{\chi}\right) + R_{\chi}\Theta_{\chi \mathrm{b}}\left(1+\frac{\rho_\chi}{\rho_\mathrm{b}}\right) +R_{\gamma}\left(\theta_{\gamma}-\theta_{\mathrm{b}}\right) = 0
\end{equation}

Grouping terms and dividing by $R_\chi$, we have:
\begin{equation}\label{TCA3}
        \frac{1}{R_\chi}\dot{\Theta}_{\chi \mathrm{b}} + \left(1+\frac{\rho_\chi}{\rho_\mathrm{b}}+\frac{1}{R_\chi}\frac{\dot{a}}{a}\right)\Theta_{\chi \mathrm{b}} +  \frac{1}{R_\chi}\left(k^2\left(c^{2}_{\mathrm{b}}\delta_{\mathrm{b}} - c^{2}_{\chi}\delta_{\chi}\right)+R_{\gamma}\left(\theta_{\gamma}-\theta_{\mathrm{b}}\right)\right) = 0
\end{equation}

This formula is precisely of the form
\begin{equation}\label{TCA4}
    \epsilon y^\prime(t) + \frac{y(t)}{f(t)} + \epsilon g(t) = 0
\end{equation}

where $\epsilon$ (in our case, $R_\chi$) is a small parameter. The first-order approximation of the function $y$ is given by $\epsilon y_1$, where $y_1=-fg$. In our case, the function $y$ is equal to $\Theta_{\chi \mathrm{b}}$, and our first-order approximation is thus
\begin{equation}\label{TCA5}
        \Theta_{\chi \mathrm{b}} \sim -\frac{k^2\left(c^{2}_{\mathrm{b}}\delta_{\mathrm{b}} - c^{2}_{\chi}\delta_{\chi}\right)+R_{\gamma}\left(\theta_{\gamma}-\theta_{\mathrm{b}}\right)}{\frac{\dot{a}}{a}+R_\chi\left(1+\frac{\rho_\chi}{\rho_\mathrm{b}}\right)}
\end{equation}

We substitute this formula for $\Theta_{\chi \mathrm{b}}$ in the revised Boltzmann equations whenever $\tau_\chi H < 0.015$ and $\tau_\chi k < 0.01$, similar to how standard \texttt{CLASS} implements the TCA trigger for photon-baryon coupling. Here, $\tau_\chi$ is simply defined as $1/R_\chi$, and can be thought of as the conformal time scale of interaction between baryons and DM.
\clearpage

\section{IR Resummation for non-negative powers of $n$}
\label{Appendix:IR}

\begin{figure*}[ht!]
\includegraphics[scale=0.75]{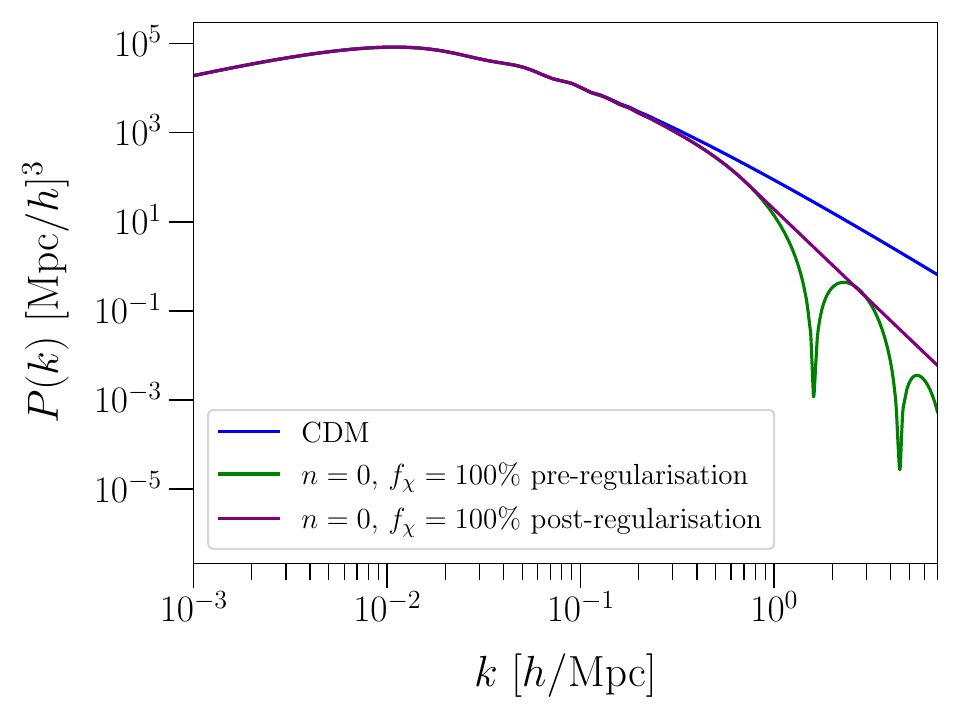}
\centering
\caption{Power spectra for the $n=0$, $f_\chi=100\%$, $m_\chi=1$ MeV model, with and without our regularisation procedure applied. For comparison, we display the CDM power spectrum for the same set of cosmological parameters.
\label{fig:regularised}}
\end{figure*}

In IDM scenarios where the power spectrum experiences steep suppression and vanishes, i.e. when the power law index $n\geq-2$ and $f_\chi=100\%$, the previous iteration of \texttt{CLASS-PT} was unable to perform infrared (IR) resummation~\cite{Senatore_2015,Baldauf:2015xfa,Blas_2016,Blas:2016sfa,Ivanov:2018gjr,Vasudevan:2019ewf}
successfully. Specifically, the prescription for splitting the real-space power spectrum into wiggly and non-wiggly components picks up noise from high ranges of $k$ that are irrelevant to the BAO wiggles IR resummation is attempting to capture. To address this, we implement a regularisation procedure that replaces the power spectrum with an analytic power law expression when the slope becomes too steep, i.e. when the spectral index $N\lesssim -4$ (where $P(k)\propto k^{N}$). We display sample power spectra for the $n=0$, $f_\chi=100\%$, $m_\chi=1$ MeV model, before and after regularisation, in Fig.~\ref{fig:regularised}.

\clearpage

\bibliographystyle{JHEP}
\bibliography{biblio.bib}

\end{document}